\documentclass[reprint,bibnotes,amsmath,amssymb,aps,prb,showpacs,floatfix,superscriptaddress,longbibliography]{revtex4-1}

\usepackage{amsmath}
\usepackage{amsfonts}
\usepackage{amssymb}
\usepackage{amsxtra}
\usepackage{xcolor}
\usepackage{graphicx}
\usepackage{dcolumn}
\usepackage{bm}
\usepackage[breaklinks=true,colorlinks,citecolor=blue,linkcolor=blue,urlcolor=blue]{hyperref}

\renewcommand{\BibitemShut}[1]{}
\def\be{\begin{equation}}
\def\ee{\end{equation}}
\def\nn{\nonumber}
\def\bea{\begin{eqnarray}}
\def\eea{\end{eqnarray}}
\newcommand\inv[1]{#1\raisebox{1.15ex}{$\scriptscriptstyle-\!1$}}

\begin{document}

\title{Berry curvature induced thermopower in type-I and type-II Weyl semimetals}
\author{Kamal Das}
\email{kamaldas@iitk.ac.in}
\affiliation{Department of Physics, Indian Institute of Technology Kanpur, Kanpur 208016, India}
\author{Amit Agarwal}
\email{amitag@iitk.com}
\affiliation{Department of Physics, Indian Institute of Technology Kanpur, Kanpur 208016, India}

\begin{abstract}
Berry curvature acts analogously to a magnetic field in the momentum-space, and it modifies the flow of charge carriers and entropy. This induces several  
intriguing magnetoelectric and magnetothermal transport phenomena in Weyl semimetals.  
Here, we explore the impact of the Berry curvature and orbital magnetization on the thermopower in tilted type-I and type-II Weyl semimetals, using semiclassical Boltzmann transport formalism. We analytically calculate the full magnetoconductivity matrix and use it to obtain the thermopower matrix for different orientations of the magnetic field $(B)$, with respect to the tilt axis. We find that the tilt of the Weyl nodes induces linear magnetic field  terms in the conductivity matrix, as well as in the thermopower matrix. The linear-$B$ term appears in the Seebeck coefficients, when the $B$-field is applied along the tilt axis. Applying the magnetic field 
in a plane perpendicular to the tilt axis results in a quadratic-$B$ planar Nernst effect, linear-$B$ out-of-plane Nernst effect and quadratic-$B$ correction in the Seebeck coefficient.\end{abstract}
\maketitle

\section{introduction}

Weyl semimetals (WSMs) host relativistic massless fermionic quasiparticles in the vicinity of the Weyl nodes which always come in pairs of opposite chirality \cite{Nielsen81,Ashvin18, Bansil16, Claudia_rev17}. 
Their existence has been demonstrated in several materials \cite{Kim13, xu15a, Xiong15, Lv15, xu15b} where either time reversal or space inversion symmetry is broken. Unlike their relativistic counterparts, in crystalline systems the Weyl quasiparticles can also break Lorentz invariance. Consequently, their dispersion can be tilted in a specific direction \cite{Soluyanov15, Yazyev16,  Chang16, Pengli17, Jiang17, Xu17, Pal18, Zhang18, Chang18}. Depending on the degree of the tilt, these WSMs can be classified as type-I or type-II. In a type-I WSM, the Fermi surface encloses only one kind of carriers, and has a vanishing density of states at the Weyl point. 
In contrast, a type-II WSM  has non-vanishing density of states at the Weyl point, and the Weyl point appears at the intersection of an electron and a hole pocket.

Interestingly, the Weyl nodes act as a source or sink of Berry curvature (BC), which in turn acts as a fictitious magnetic field in the momentum space \cite{Berry84, Xiao_Niu_rev10}. This leads to the possibility of several interesting transport phenomena in isotropic and tilted WSMs \cite{Nielsen83, Son12, Zyuzin_Burkov12, 
 Son_Spivak13, Hosur13, Kim14, Fiete14,  Zyuzin_A_A16, G_Sharma16, Wurff17, Nandini17, Sekine17, Ferreiros17, Steiner17, Watzman18}. 
Several of these have also been experimentally demonstrated\cite{Kim13, Xiong15, Desrat15, Huang15b, Hu16, Arnold16, Li16, Li17, Liang17, Sudesh17, Niemann17, Liang18, Nitesh18,  Noky18, Watzman18, Yang19}. 
For instance, negative magnetoresistivity (MR) \cite{Nielsen83, Son_Spivak13} has been observed in several WSM candidates including the TaAs family\cite{Huang15b, Hu16, Niemann17} and in WSMs induced magnetically from three dimensional Dirac semimetals\cite{Kim13, Xiong15, Li16}. 
The anomalous Hall effect \cite{Haldane04} predicted to exist in time reversal symmetry (TRS) broken WSMs\cite{Burkov14, Steiner17} has been recently seen 
in ZrTe$_{5}$ \cite{Liang18}. 
The corresponding effect in thermopower, the anomalous Nernst effect  in WSM\cite{ G_Sharma16, Ferreiros17} has been demonstrated 
in Cd${}_{3}$As${}_{2}$\cite{Liang17}, NbP\cite{Watzman18} and Ti${}_{2}$MnX\cite{Noky18}.
Chiral magnetic effect, a chiral anomaly induced phenomena\cite{Zyuzin_Burkov12, Son12, Wurff17} has been reported in ZrTe${}_{5}$\cite{Qiang_Li16}. 
The BC induced planar Hall effect, where the current response is measured in the plane of electric and magnetic field, has also been predicted in WSM\cite{Nandy17, Burkov17} and multi-WSM\cite{Dantas18, Nag18}
and experimentally demonstrated in WSM \cite{Nitesh18, Yang19}.
More recently, linear magnetic field dependence in both the MR and Hall responses is predicted to exist in tilted WSMs\cite{Zyuzin_V_A17, G_Sharma17a, Kamal19}.

Motivated by these recent studies, in this paper we explore the BC induced magnetothermopower in tilted type-I and type-II WSMs: the Seebeck 
and the Nernst coefficients (SCs and NCs, respectively). Our analytical calculations for the full conductivity and thermopower matrix are based on the  
Berry-connected-Boltzmann-transport formalism and include the effect of the orbital magnetic moment (OMM)\cite{Xiao_Niu_rev10}. 
The Seebeck effect captures the electric response along the temperature gradient while the Nernst effect captures the electric response  perpendicular to the temperature gradient. The conductivity and the thermopower are connected by the Mott relation [see Eq.~\eqref{thermopower_tensor}] even in the presence of the OMM correction \cite{Dong2018}. Thus the magnetothermopower broadly follows the 
magnetoelectric response, leading to the expectation of phenomena such as negative Seebeck effect and planar Nernst effects in WSMs. 
A similar kind of phenomenon is known to exist in ferromagnetic systems\cite{Bui14, Bui18} where spin dependent scattering induces a transverse velocity component in the charge carriers\cite{Wesenberg18}.

In this paper, we have calculated the electrical conductivity and thermopower matrix to explore the BC-induced magnetotransport in type-I and type-II WSMs. We have explicitly included the previously ignored impact of the OMM in all our calculations. 
Although the electrical conductivity matrix is well explored (excluding the OMM), the BC-induced thermopower and the impact of the tilt on it is relatively unexplored,  and this is the primary focus of this paper. 
In particular, we predict the following BC-induced phenomena: (1) linear-$B$ as well as quadratic-$B$ dependent SCs, (2) the existence of quadratic-$B$ planar Nernst as well as linear-$B$ out-of-plane Nernst response, (3) negative longitudinal (parallel electric and magnetic field) MR and positive perpendicular (perpendicular electric and magnetic field) MR in WSM. The rest of the paper is organized as follows: 
In Sec.~\ref{review1} we present the phenomenological equation of charge current and establish the relation between charge conductivity and thermopower. This is followed by a 
detailed discussion of the full magnetoconductivity matrix for type-I and type-II WSM in Sec.~\ref{effects of OMM}, and MR in Sec.~\ref{MR}.  
We discuss various aspects of the thermopower matrix in Secs.~\ref{Nernst},~\ref{asymptotic} and~\ref{internode}. We summarize our findings in Sec.~\ref{conclusions}.

\section{Thermopower in presence of Berry curvature} \label{review1}
Within the linear response theory, the phenomenological transport equation for the electrical current ${\bf j}^{e}$ is given by\cite{Ashcroft76}, 
\begin{eqnarray}\label{tetc1}
j_{i}^{e}=\sigma_{ij}E_{j}+\alpha_{ij}(-\nabla_{j} T)~.
\end{eqnarray}
Here, $E_j$ and ${\nabla_j}T$ are the external electric field and temperature gradient applied along the $j$th direction, $\sigma_{ij}$ denotes the elements of electrical conductivity matrix ($\tilde{\mathbf \sigma}$) and $\alpha_{ij}$ are the elements of thermoelectric conductivity matrix ($\tilde{\alpha}$).
These transport coefficients are calculated by doing a Brillouin zone sum over the relevant physical quantities, involving only the occupied states. In this paper, we use the semiclassical Boltzmann transport formalism to calculate the magnetoconductivity and magnetothermopower. 
The details of the Berry-connected-Boltzmann-transport formalism are discussed in Appendix~\ref{review}.  
The general expressions for BC-induced conductivity and thermopower are presented in Eqs.~\eqref{elec_cond}, and \eqref{therm_cond}, respectively.

The thermopower for an open circuit system is defined by setting $j^e_i = 0$ in Eq.~\eqref{tetc1}. In this scenario, 
the electric field generated by a temperature gradient is given by, 
\begin{equation} \label{nu1a}
E_i= \nu_{ij} \nabla_j T,~~{\rm where}~~ \tilde{\nu}\equiv\inv{\tilde \sigma}~{\tilde \alpha}~.
\end{equation}
The diagonal elements $\nu_{ii}$ denote the SCs whereas the off-diagonal elements $\nu_{ij}(i\neq j)$ are the NCs. 
It turns out that in the low temperature limit ($k_B T \ll \mu$), BC-induced thermopower can also be expressed in terms of the electrical conductivity 
using the Mott relation \cite{Dong2018}.  The Mott relation \cite{Ashcroft76, Xiao_Niu06} yields, 
\begin{equation}\label{mott1}
\alpha_{ij}=-\frac{\pi^{2}k_{B}^{2}T}{3e}~\left.\frac{\partial \sigma_{ij}}{\partial \epsilon}\right|_{\epsilon = \mu}~.
\end{equation}
Using Eq.~\eqref{mott1} in Eq.~\eqref{nu1a}, the thermopower matrix can be expressed solely in terms of the electrical conductivity matrix as\cite{Ashcroft76, Xiao_Niu06}
\begin{equation}\label{thermopower_tensor}
\tilde{\nu}=-\dfrac{\pi^{2}k_{B}^{2}T}{3e}~\inv{\tilde{\sigma}}~\left.\frac{\partial \tilde{\sigma}}{\partial \epsilon}\right|_{\epsilon = \mu}~.
\end{equation}

Further, to explicitly track the magnetic field dependence analytically, we express $\sigma_{ij}=\sigma_{ij}^{(0)}+ \sigma_{ij}^{(1)}+\sigma_{ij}^{(2)} + {\cal O}(B^3)$, and $\alpha_{ij}=\alpha_{ij}^{(0)}+ \alpha_{ij}^{(1)}+\alpha_{ij}^{(2)} + {\cal O}(B^3)$.  
Here the superscripts denote zeroth order (Drude and anomalous), linear and quadratic magnetic field terms, respectively. 
In the next section, we calculate the magnetoconductivity matrix for tilted WSM, including the OMM corrections. Since the tilt-axis (we consider $\hat{\bf z}$) breaks the TRS \emph{for each node}, the two cases in which $\bf B$ is applied parallel and perpendicular to $\hat{\bf z}$ result in different forms of the conductivity matrix.

\section{magnetoconductivity in type-I and type-II WSMs}
\label{effects of OMM}
The low energy Hamiltonian, for each of the chiral node of a tilted WSM is given by,\cite{Carbotte16}
\be\label{ham}
\mathcal{H}_{s}({\bf k}) = \hbar C_{s}k_{z}+s\hbar v_{F} {\bf \sigma}\cdot{\bf k}~.
\ee
Here, $s$ denotes chirality, $C_{s}$  ($v_F$) is the tilt (Fermi) velocity, and $\sigma=(\sigma_{x}, \sigma_{y}, \sigma_{z})$, are the Pauli matrices. 
In this paper, we consider a WSM with a pair of oppositely tilted nodes such that $C_{-s} = -C_s$. The degree of the tilt of the Weyl nodes can be quantified by the ratio of the 
tilt and the Fermi velocities: $R_{s}=C_{s}/v_{F}$. The above dispersion corresponds to the type-I class of WSMs in the regime $|R_s| <1$ and the type-II class in the regime $|R_{s}|>1$.

The BC can be easily calculated from Eq.~\eqref{BC} to be ${\bf \Omega}_{s}^{\lambda}= -\lambda s {\bf k}/({2k^{3}})$, where $\lambda = +1$ ($\lambda = -1$) 
denotes the conduction (valence) band. The OMM can be calculated from Eq.~\eqref{OMM} and can be expressed in terms of the BC\cite{Hayata17},
\be 
{\bf m}_{s}^{\lambda}= \lambda {e v_F k}{\bf \Omega}^{\lambda}_{s}~ = -s e v_F \frac{\bf k}{2k^2}~.
\ee
Both of the BC and the OMM are independent of the tilt velocity.  Furthermore, the OMM, and the resulting velocity correction are identical for both the bands. 

The impact of the tilt on the magnetoconductivity in type-I and type-II WSM was explored in Ref.~[\onlinecite{Kamal19}]. Here, we generalize those results to include the effect of the OMM. The conductivity matrix can be expressed as sum of the contributions in absence and presence of a magnetic field:
$\tilde{\sigma} = \tilde{\sigma}_{\rm D} + \tilde{\sigma}_{\bf B}$, such that $\tilde{\sigma}_{\bf B}$ vanishes as ${\bf B} \to 0$ and $\tilde{\sigma}_{\rm D}$ is the Drude 
conductivity. 
For the scenario in which ${\bf B} \perp \hat{\bf z}$, and the magnetic field is confined in the $x$-$y$ plane (planar geometry), we find that the conductivity matrix has this general form for both type-I and type-II WSMs, 
\be\label{charge_tensor_aniso}
\resizebox{\linewidth}{!}{$
\tilde{\sigma}_{\bf B} = 
\begin{pmatrix}
\sigma^{(2)}_{\perp}+\Delta\sigma^{(2)}\cos^{2}\phi & \Delta\sigma^{(2)} \sin(2\phi)/2 & \sigma_{\rm t}^{(1)} \cos\phi
\\
\Delta\sigma ^{(2)} \sin(2\phi)/2 & \sigma^{(2)}_{\perp} + \Delta \sigma^{(2)} \sin^{2}\phi &  \sigma_{\rm t}^{(1)} \sin\phi
\\
\sigma_{\rm t}^{(1)} \cos\phi & \sigma_{\rm t}^{(1)} \sin\phi & \sigma_{\rm z}^{(2)}
\end{pmatrix}. 
$}
\ee 
Here, $\phi$ is the angle of the magnetic field with respect to the $x$ axis. 
See Appendix \ref{Drude conductivities} for the details of calculation of $\tilde{\sigma}_{\rm D}$.
Here, $\sigma_{12} = \sigma_{xy}$ denotes the planar Hall conductivity 
and in addition there are new linear-$B$ terms such as $\sigma_{13} = \sigma_{xz}$ and $\sigma_{23} = \sigma_{yz}$, which were discussed in Ref.~[\onlinecite{Kamal19}].

For the other case of ${\bf B} \parallel \hat{\bf z}$, the general form of the conductivity matrix has a diagonal form \cite{Kamal19},  
\be\label{charge_tensor_aniso_para}
\tilde{\sigma}_{\bf B}  = 
\begin{pmatrix}
\sigma_{\rm l}^{(1)} + \sigma_{\rm l}^{(2)}  & 0 & 0
\\
0 &  \sigma_{\rm l}^{(1)}  + \sigma_{\rm l}^{(2)}   &  0
\\
0 & 0 &  \sigma_{\rm l z}^{(1)} + \sigma_{\rm l z}^{(2)}
\end{pmatrix}.
\ee
Here, the diagonal components have linear-$B$ dependence induced by the tilt. 
The analytical expression of the different conductivity components is presented in the subsections below where we have presented results including OMM. However, to explicitly highlight the impact of OMM, we have presented the general expressions in terms of $\gamma$, in Appendix \ref{expre_gamma}.

\subsection{Type-I WSMs}
\label{Type-I}

For the ${\bf B} \perp \hat{\bf z}$ case, the magnetoconductivity is given by Eq.~\eqref{charge_tensor_aniso}, where the conductivity coefficients 
$\sigma^{(2)}_{\perp}$, $\sigma_{\rm z}^{(2)}$ and $\Delta \sigma^{(2)}$ are proportional to $B^2$. In type-I WSM, including the OMM, these are explicitly given by 
\begin{gather}
\Delta \sigma^{(2)}  =  \sum_s \left(6 + 7R_{s}^{2}\right)\sigma_{0}~;~~~~~ \sigma^{(2)}_{\perp} = -\sum_s  2\sigma_{0}~,
\\ 
\sigma_{\rm z}^{(2)} =  -\sum_s  \left(2 - 8 R_s^2\right)\sigma_0~,
\end{gather}
where we have denoted the quadratic dependence as
\be 
\sigma_{0} \equiv \dfrac{e^2 \tau}{8 \pi^2}\dfrac{\hbar v_{F}^{3}}{15 \mu^2}\left(\dfrac{eB}{\hbar}\right)^2~.
\ee
These terms are finite even in the limit of $R_s \to 0$ and are even-function of $R_{s}$. Thus, the contributions from a pair of oppositely tilted nodes just adds up. 
Note the opposite sign of $\Delta\sigma^{(2)}$ and $\sigma_{\perp}^{(2)}$ and this will manifest in the perpendicular MR ($\phi = \pi/2$) being positive, and the longitudinal MR ($\phi =0$) being negative, 
as discussed in the next section.

In addition to these quadratic-$B$ terms, there are linear-$B$ dependent off-diagonal conductivity components as well ($\sigma_{xz}, \sigma_{yz} \propto \sigma_{\rm t}^{(1)} \propto B$).   
These terms arise solely due to the tilt of the Weyl nodes (which breaks the TRS for each node), and vanish in the limit of $R_s \to 0$. 
These linear-$B$ components are $\sigma_{xz}^{(1)}= \sigma_{yz}^{(1)}(\pi/2 -\phi)= \sigma_{\rm t}^{(1)}\cos\phi$, where $ \sigma_{\rm t}^{(1)}$ can be expressed as
\be \label{sigma_t}
\sigma_{\rm t}^{(1)} =   \sum_{s}\dfrac{s\sigma_{1}}{2R_s^2}\big[2R_{s} \left(1- 2R_s^2\right)  + \mathcal{F} \delta_{s}  \big]~. 
\ee
In the above equation, we have defined $\mathcal{F} \equiv 1- R_s^2$ and,
\be \nn
\sigma_{1}=\dfrac{e^{2}\tau}{(2\pi)^3}\dfrac{\pi v_F}{\hbar}\dfrac{eB}{\hbar}~,~~{\rm and}~~~~\delta_{s}=\ln \left(\dfrac{1-R_{s}}{1+R_{s}}\right)~.
\ee
Note that the contributions for the oppositely tilted nodes simply add up and the overall sign of this component depends on the details of tilt configuration. 

For the ${\bf B} \parallel \hat{\bf z}$ configuration, the conductivity matrix is given by Eq.~\eqref{charge_tensor_aniso_para}. 
As the matrix structure shows, in this case the longitudinal conductivities have a linear-$B$ dependence in addition to the quadratic-$B$ one. The quadratic-$B$ correction perpendicular to the tilt is  $\sigma_{\rm l}^{(2)}=\sigma^{(2)}_{\perp}=-\sum_s 2\sigma_{0}$, whereas along the tilt axis, it is given by
\be
\sigma_{\rm l z}^{(2)}=\sum_{s}\left(4 + 5R_s^2 \right)\sigma_{0}~. 
\ee
The linear-$B$ term in $\sigma_{xx} = \sigma_{yy}$ and $\sigma_{zz}$ is given by 
\be
\sigma_{\rm l}^{(1)} = - \sum_{s}\dfrac{s\sigma_{1}}{R_s^2}\left(2R_s + \delta_s \right);~~\sigma_{\rm l z}^{(1)} = -\sum_{s}s\sigma_1 \left(2R_s\right)~.
\ee
We emphasize that all the linear-$B$ conductivity discussed in this section is $\propto \sigma_1$, in which there is no explicit $\mu$ dependence. 
This is primarily a consequence of $\Omega_{\bf k} \propto 1/k^2$ in WSMs. The only $\mu$ dependence of $\sigma_1$ arises from the energy dependence of the scattering timescale $\tau$.

\subsection{Type-II WSMs}
The low energy model Hamiltonian of Eq.~\eqref{ham} corresponds to the type-II class in the regime $|R_{s}|> 1$. 
In this regime, the Fermi surface of the  Weyl node comprises of ``unbounded" electron and hole pockets. Hence both the bands take part in transport. 
And to truncate the ``unbounded sea" of the charge carriers, we need to introduce a cutoff in the momentum space along the radial direction ($\Lambda_{k}$). 
In real materials, this is akin to the bandwidth of the system. For simplicity, we present all the conductivity terms only upto first order in $k_F/\Lambda_k \equiv 1/\tilde{\Lambda}_{k}$, and assume $\mu > 0$ without the loss of generality.

First, we will consider the planar geometry. 
In this case the form of the conductivity matrix is given by Eq.~\eqref{charge_tensor_aniso}, with various elements are given by 
\bea
\Delta \sigma^{(2)}&=&2 \sum_s \mathcal{K} \left(30R_s^8 + 35 R_s^6 +50 R_s^4 - 9R_s^2 -2   \right), 
\\
\sigma^{(2)}_{\perp}&=& \sum_s \mathcal{K}\left(5R_s^6 - 60R_s^4 + 25R_s^2 -2 \right),
\\ 
\sigma_{\rm z}^{(2)}&=&2  \sum_s \mathcal{K} \left(15R_s^8 + 65 R_s^6 -35 R_s^4 - R_s^2 + 4\right),
\eea
where $\mathcal{K} \equiv \frac{\sigma_{0}}{16|R_{s}|^{5}}$. Note that the tilt induced corrections occur as even powers of $R_s$, implying the addition of contributions from the oppositely tilted nodes.
The linear-$B$ correction in the out-of-plane off-diagonal conductivities can be written as $\sigma_{xz}^{(1)}=\sigma_{yz}^{(1)}(\pi/2 -\phi)=\sigma_{\rm t}^{(1)}\cos\phi$. Here,
\be 
\sigma_{\rm t}^{(1)} = \sum_{s}\dfrac{s\sigma_1}{2R_s^4}{\rm sgn}(R_{s})\left[2 + R_s^2 - 5R_s^4 - \mathcal{F}R_s^2\delta_{s}^{1} \right]~, 
\ee
and we have defined 
\be
\delta_{s}^{1}=\ln (R_{s}^{2}-1) + 2 \ln\tilde{\Lambda}_{k}~.
\ee

Now, we consider a magnetic field along the direction of the tilt (${\bf B} \parallel \hat {\bf z}$). 
The linear-$B$ correction to the longitudinal component in the $x-y$ plane, $\sigma_{xx} = \sigma_{yy}$, is given by 
\be
\sigma_{\rm l}^{(1)} = -\sum_{s}\dfrac{s\sigma_{1}}{R_{s}^{4}}{\rm sgn}(R_{s})\left[  3R_s^2-1-\delta_{s}^{1}  R_s^2\right]~.
\ee
The linear-$B$ correction to $\sigma_{zz}$ is given by 
\be
\sigma_{\rm l z}^{(1)} = -2\sum_{s}\dfrac{s\sigma_{1}}{R_{s}^{4}}{\rm sgn}(R_{s})\left(2R_s^4 - 2R_s^2 + 1\right)~.
\ee
The quadratic-$B$ correction to $\sigma_{xx}$ and $\sigma_{yy}$ is given by 
\be
\sigma_{\rm l}^{(2)}=\sum_{s}\dfrac{\sigma_{0}}{8|R_{s}|^{5}}
\left(4 - 25R_{s}^{2}+5R_{s}^{6}\right)~.
\ee
The corresponding term for the $\sigma_{zz}$ component is given by 
\be
\sigma_{\rm l z}^{(2)} = \sum_{s}\dfrac{\sigma_{0}}{2|R_{s}|^{5}}
\left(20R_s^6 - 5R_s^4 + 5R_s^2 -2\right)~.
\ee

Having obtained the full conductivity matrix for tilted WSM, now we discuss tilt and OMM dependence of the MR -- the quantity generally probed in experiments.

\begin{figure}[t]
\includegraphics[width=0.5\textwidth]{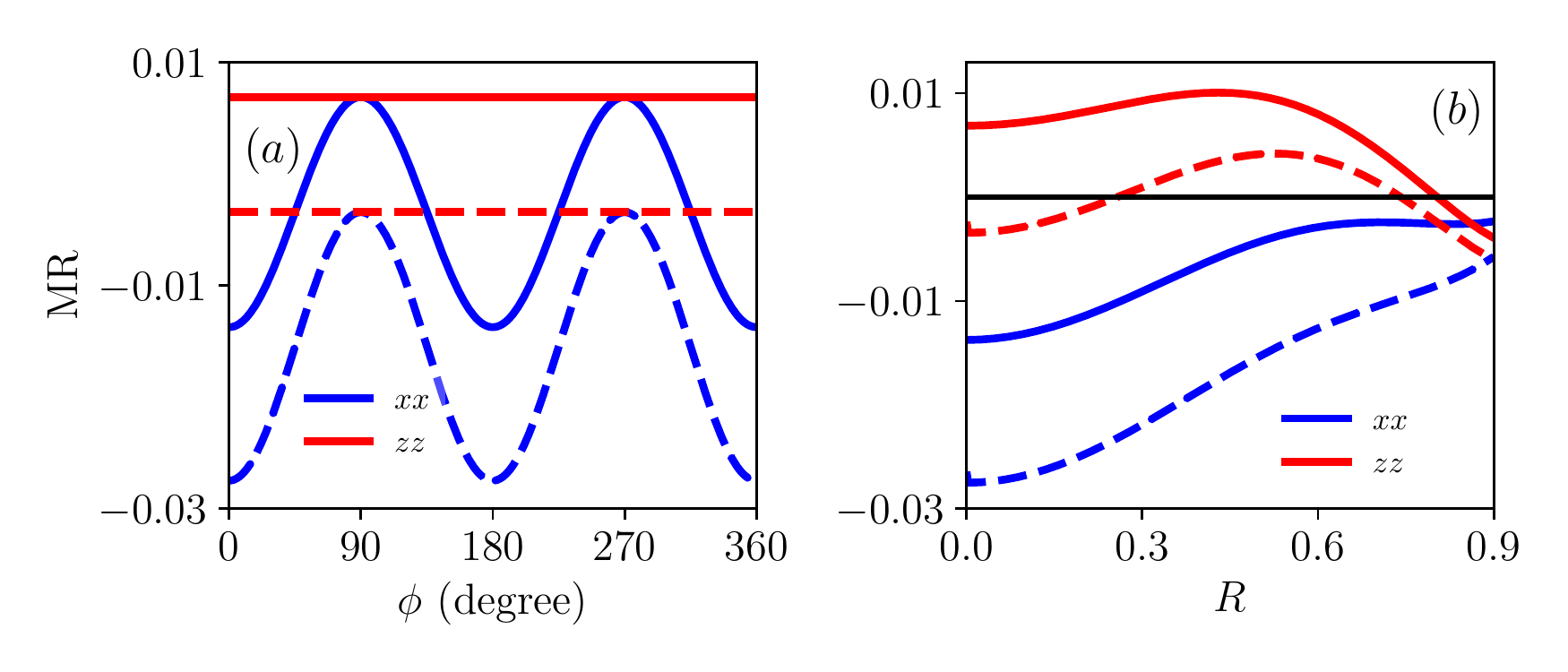}
\caption{a) MR for type-I WSM ($R = 0$) as a function of the angle between ${\bf E}$ and ${\bf B}$ for the planar geometry. The planar MR (MR$_{xx}(\phi)$) varies as $\cos^2 \phi$ (blue lines). The longitudinal MR (MR$_{xx}(\phi=0)$) is negative irrespective of OMM correction. The perpendicular MR (MR$_{zz}$) becomes positive on including the OMM (solid red line). (b) MR as a function of tilt ($R$) for the configuration $R_- = -R_+ = R$. The longitudinal MR remains negative (blue lines) while the perpendicular MR changes sign at a certain critical $R$ value, beyond which it remains negative (red lines). 
Here, we have used the following parameters: $\mu=0.1$ eV, $v_{F}=10^{6}$ m/s and $B=4$ T.}
\label{fig_11}
\end{figure}

\section{Magnetoresistivity}
\label{MR}

The resistivity matrix is obtained by inverting the conductivity matrix. The corresponding MR is given by MR$_{ii} = \rho_{ii}(B)/\rho_{ii}(0)-1$. Below we discuss the longitudinal and perpendicular MR for the two cases of ${\bf B} \perp \hat{\bf z}$ and ${\bf B} \parallel \hat{\bf z}$.

For the case of planar geometry, using Eq.~\eqref{charge_tensor_aniso} we obtain the planar resistivity to be 
\be\label{Eq.33}
\rho_{xx}=\rho_{\rm D} - \rho^{(2)}_{\perp} + \left[ \left(\rho_{\rm t}^{(1)}\right)^{2}\rho_{\rm D}^{\rm z}[\rho_{\rm D}]^{-2} - \Delta \rho^{(2)}\right]\cos^{2}\phi~.
\ee
Here, we have defined the Drude resistivity in the $x$-$y$ plane as $\rho_{\rm D}=1/\sigma_{\rm D}$, and along the $z$ axis as $\rho_{\rm D}^{\rm z}=1/\sigma_{\rm D}^{z}$. 
Additionally, we have defined the following: $\rho^{(2)}_{\perp} = \sigma^{(2)}_{\perp}/{\sigma_{\rm D}}^2$, $\rho_{\rm t}^{(1)} =\sigma_{\rm t}^{(1)}/{\sigma_{\rm D}}^2$ and $\Delta \rho^{(2)} = \Delta \sigma^{(2)}/{\sigma_{\rm D}}^2$. 
It is evident from Eq. \eqref{Eq.33} that the planar MR [MR$_{xx} (\phi)$] is anisotropic and varies as $\cos^2\phi$ on changing the planar $\bf B$ direction with respect to the $x$ axis. 
In Fig. \ref{fig_11} we have plotted the MR with $\phi$ and tilt factor. We have used dotted lines for conductivities without the contribution of OMM ($\gamma =0$) in our plots. Note that
the longitudinal MR [MR$_{xx}(\phi =0)$] is negative irrespective of inclusion or exclusion of the OMM and the degree of the tilt of the WSM. 
However, the perpendicular MR [MR$_{xx}(\phi =\pi/2$)] becomes positive on including the OMM terms $(\gamma = 1)$ for isotropic WSMs shown in Fig. \ref{fig_11}(a).  
For the out-of-plane perpendicular MR (MR$_{zz}$), we obtain the resistivity to be 
\be 
\rho_{zz} = \rho_{\rm D}^{\rm z}-\rho_{\rm z}^{(2)} + \left( \rho_{\rm t z}^{(1)}\right)^{2}\rho_{\rm D}[\rho_{\rm D}^{\rm z}]^{-2}~.
\ee
Here, we have defined $\rho_{\rm z}^{(2)}=\sigma_{\rm z}^{(2)}/{\sigma_{\rm D}^{\rm z}}^2$ and correction due to the linear-$B$ Hall conductivity component as $ \rho_{\rm tz}^{(1)}= \sigma_{\rm t}^{(1)}/{\sigma_{\rm D}^{\rm z}}^2$.
The OMM correction forces the out-of-plane perpendicular MR to be positive (solid red line) for WSM with small tilt  -- as shown in panels (a) and (b) of Fig.~\ref{fig_11}.

\begin{figure}[t]
\includegraphics[width=0.5\textwidth]{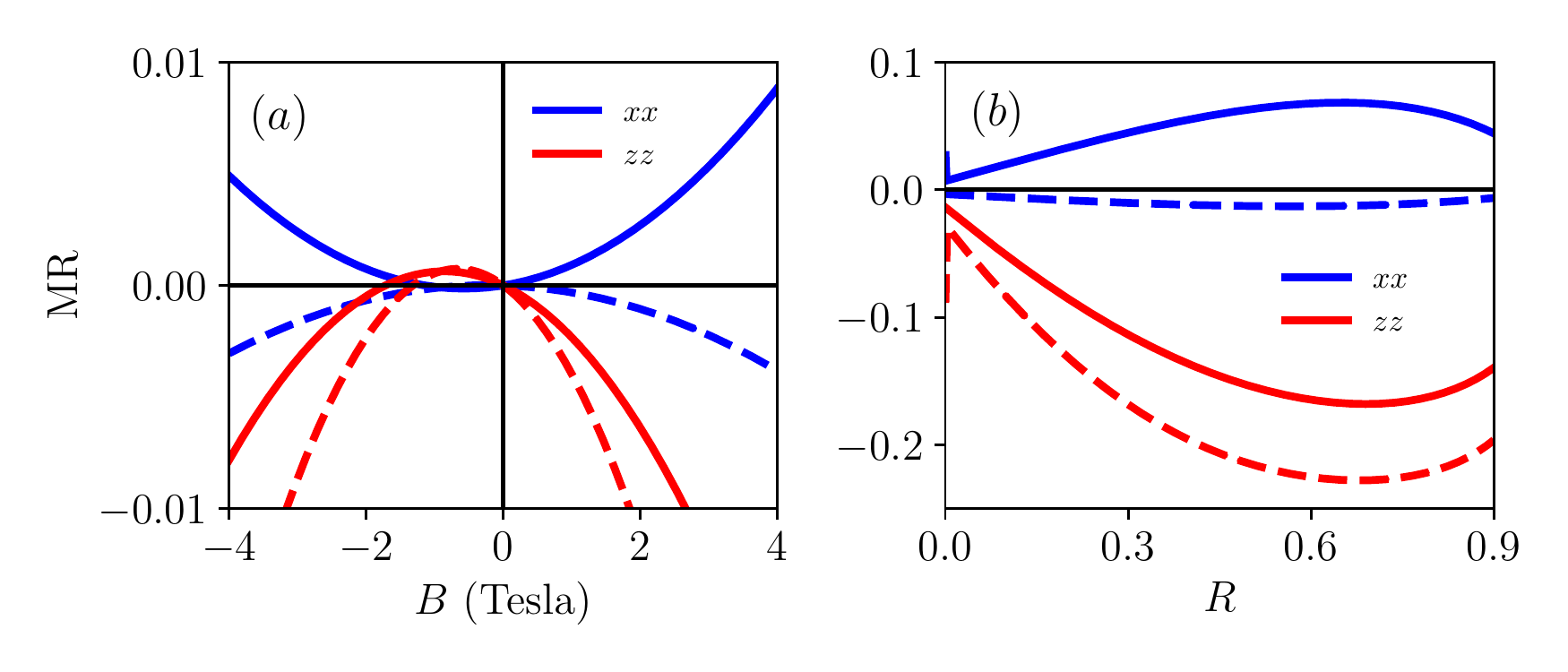}
\caption{MR of type-I WSM for the case of ${\bf B} \parallel {\bf R}$ with tilt configuration of Fig.~\ref{fig_11}. (a) The magnetic field dependence of the MR. Note that MR$_{xx}$ and MR$_{zz}$ have a linear-$B$ contribution for a finite tilt (here $R = .015$), leading to an asymmetry in the MR curves about the $B=0$ line.   
(b) Variation of MR as a function of the tilt parameter $R$ at $B=4$ T. Note that 
the inclusion of OMM correction forces perpendicular MR (MR$_{xx}$) to be positive (solid blue line), while longitudinal MR (MR$_{zz}$) remains negative with or without OMM correction (solid or dashed red line, respectively). Here the parameters used are identical to those of Fig.~\ref{fig_11}. 
}
\label{fig_22}
\end{figure}

For the other case of ${\bf B} \parallel \hat{\bf z}$, we calculate resistivity along the tilt direction from Eq. \eqref{charge_tensor_aniso_para}, and it is given by 
\be 
\rho_{zz} = \rho_{\rm D}^{\rm z}-\rho_{\rm l z}^{(1)} + \left(\rho_{\rm l z}^{(1)}\right)^{2}[\rho_{\rm D}^{\rm z}]^{-1}-\rho_{\rm l z}^{(2)}~.
\ee
Here we have defined $ \rho_{\rm l z}^{(1)} = \sigma_{\rm l z}^{(1)}/(\sigma_{\rm D}^{\rm z})^{2}$ and $\rho_{\rm{lz}}^{(2)}=\sigma_{\rm{lz}}^{(2)}/(\sigma_{\rm D}^{\rm z})^{2}$.
Evidently, in this case the longitudinal MR, MR$_{zz}$, will have linear-$B$ contribution for a tilted WSM, while its absolute value depends on the degree of the tilt, starting with a negative value for an isotropic WSM. This linear-$B$ part gives rise to an asymmetry in the MR curve as $B$ goes from negative to positive -- see Fig.~\ref{fig_22}(a).  
Note that the inclusion of OMM does not change the sign of longitudinal MR. 
The expression for $\rho_{xx}$ is given by 
\be 
\rho_{xx}=\rho_{\rm D} - \rho_{\rm l}^{(1)} + \left(\rho_{\rm l}^{(1)}\right)^{2}[\rho_{\rm D}]^{-1} - \rho_{\rm l}^{(2)}~.
\ee
Here, we have defined $\rho_{\rm l}^{(1)} = \sigma_{\rm l}^{(1)}/{\sigma_{\rm D}}^2$ and $\rho_{\rm l}^{(2)}=\sigma_{\rm l}^{(2)}/{\sigma_{\rm D}}^2$. Similar to the case of $\rho_{zz}$, $\rho_{xx}$ also has linear-$B$ contributions leading to asymmetric MR curves around the $B=0$ line shown in Fig.~\ref{fig_22}(a). 
However, unlike the case of longitudinal MR, the perpendicular MR, MR$_{xx}$, changes sign on including the OMM and reverses from negative to positive as shown in Fig.~\ref{fig_22}(b).

Our findings for isotropic WSM that the longitudinal MR is negative, while the perpendicular MR is positive, are consistent with the experimental MR results reported in Dirac semimetals \cite{Li16, Xiong15} and isotropic WSMs\cite{Huang15b, Niemann17}. 
We emphasize that the inclusion of OMM is crucial to capture the correct sign of the perpendicular MR.

\section{Thermopower in Weyl semimetal}
\label{Nernst}

In this section we calculate the magnetic field dependent thermopower at low temperature using the Mott relation \cite{Dong2018}.  
Let us first consider the case ${\bf B} \perp \hat{\bf z}$. Since $\tilde{\alpha} \propto \partial_\mu \tilde{\sigma}$, 
the thermoelectric conductivity matrix retains the form of Eq.~\eqref{charge_tensor_aniso}, and it is given by 
\be\label{thermoelectric_tensor_aniso}
\resizebox{\linewidth}{!}{$
\tilde{\alpha}_{\rm \bf B} =
\begin{pmatrix}
\alpha^{(2)}_{\perp} + \Delta\alpha^{(2)} \cos^2\phi & \Delta\alpha^{(2)}\sin(2\phi)/2 &  \alpha_{\rm t}^{(1)} \cos \phi
\\
\Delta\alpha^{(2)} \sin(2\phi)/2& \alpha^{(2)}_{\perp} + \Delta\alpha^{(2)} \sin^2\phi &  \alpha_{\rm t}^{(1)} \sin \phi
\\
\alpha_{\rm t}^{(1)} \cos \phi &  \alpha_{\rm t}^{(1)} \sin \phi &  \alpha_{\rm z}^{(2)}
\end{pmatrix}.
$}
\ee
The different thermoelectric conductivity elements in the matrix are connected to the corresponding elements in the conductivity matrix of Eq.~\eqref{charge_tensor_aniso} via the Mott relation [Eq.~\eqref{mott1}]. 
At a first glance it seems that the out-of-plane Hall components ($\alpha_{xz}$ and $\alpha_{yz} \propto \alpha_{\rm t}^{(1)}$) are zero for type-I WSM as the corresponding elements in the electrical conductivity matrix are independent of the Fermi energy. However, the scattering timescale is generally dependent on the Fermi energy, and this would lead to a finite linear-$B$ term in the 
thermoelectric conductivity matrix as well. Another possibility is that the deviations from the linear model, for example in a lattice model, can also lead to finite linear-$B$ contribution.  
Similar physics is seen in the case of the finite anomalous Nernst response in a tight-binding model of WSMs\cite{G_Sharma16}, 
even though the anomalous Hall coefficient is independent of the Fermi energy in the isotropic low energy model of WSM\cite{Fiete14}.

The thermopower matrix can now be calculated by using Eqs.~\eqref{charge_tensor_aniso} and ~\eqref{thermoelectric_tensor_aniso} in Eq.~\eqref{nu1a}.
The SC in the planar configuration can be expressed in the form $\nu_{yy}=\nu_{xx}(\pi/2 - \phi)$, where
\be \label{Seebeck_xx}
\nu_{xx} - \nu_{\rm D}=\nu_{\perp}^{(2)}+\Delta\nu^{(2)} \cos^{2}\phi~.
\ee
Here, we have defined $\nu_{\rm D}=\alpha_{\rm D}/\sigma_{\rm D}$ as the usual Drude coefficient calculated in Appendix \ref{Drude conductivities} and the magnetic field dependent coefficients are given by
%
\begin{gather}
\nu^{(2)}_{\perp } = \sigma_{\rm D}^{-2}\left(\sigma_{\rm D}\alpha^{(2)}_{\perp}-\alpha_{\rm D}\sigma^{(2)}_{\perp}\right),
\\
\resizebox{\linewidth}{!}{$\Delta\nu^{(2)} = \dfrac{1}{\sigma_{\rm D}^{2}}\left[(\sigma_{\rm D}\Delta \alpha^{(2)} - \alpha_{\rm D} \Delta \sigma^{(2)} + 
\left(\sigma_{\rm t}^{(1)}\alpha_{\rm D}- \alpha_{\rm t}^{(1)} \sigma_{\rm D} \right)\dfrac{\sigma_{\rm t}^{(1)}}{\sigma_{\rm D}^{\rm z}}\right]$}.
\end{gather}%
The out-of-plane SC (along the $z$ axis) can be expressed as 
$\nu_{zz}=\nu_{\rm D}^{\rm z}+ \nu_{\rm z}^{(2)}$,
where $\nu_{\rm D}^{\rm z} \equiv \alpha_{\rm D}^{\rm z}/\sigma_{\rm D}^{\rm z}$ is the Drude contribution along the tilt axis and the corresponding quadratic-$B$ correction 
is given by 
\be \label{Eq32}
\resizebox{\linewidth}{!}{$
\nu_{\rm z}^{(2)}=\frac{1}{(\sigma_{\rm D}^{\rm z})^{2}}\left[\sigma_{\rm D}^{\rm z} \alpha_{\rm z}^{(2)} - \alpha_{\rm D}^{\rm z} \sigma_{\rm z}^{(2)} + 
\left(\sigma_{\rm t}^{(1)} \alpha_{\rm D}^{\rm z}- \alpha _{\rm t}^{(1)}\sigma_{\rm D}^{\rm z} \right)\dfrac{ \sigma_{\rm t}^{(1)}}{\sigma_{\rm D}}\right]$}.
\ee
%

For the planar configuration, we obtain the coefficient for the planar Nernst effect,  
\be \label{Nernst_yx}
\nu_{yx}=\Delta \nu^{(2)} \sin \phi \cos \phi~.
\ee
This has an identical angular dependence on the planar angle between ${\bf E}$ and ${\bf B}$ to that of the planar Hall effect. In addition to the planar Nernst effect, we find the out-of-plane linear-$B$ NCs, and are given by $\nu_{xz}=\nu_{\rm t}^{(1)}\cos\phi=\nu_{yz}(\pi/2 - \phi)$,
with
\be 
\nu_{\rm t}^{(1)}=\frac{1}{\sigma_{\rm D}^{\rm z}}\left(\frac{ \alpha_{\rm t}^{(1)} \sigma_{\rm D}^{\rm z} -  \sigma_{\rm t}^{(1)} \alpha_{\rm D}^{\rm z}}{\sigma_{\rm D}}\right)~.
\ee

\begin{figure}[t!]
\includegraphics[width=0.5\textwidth]{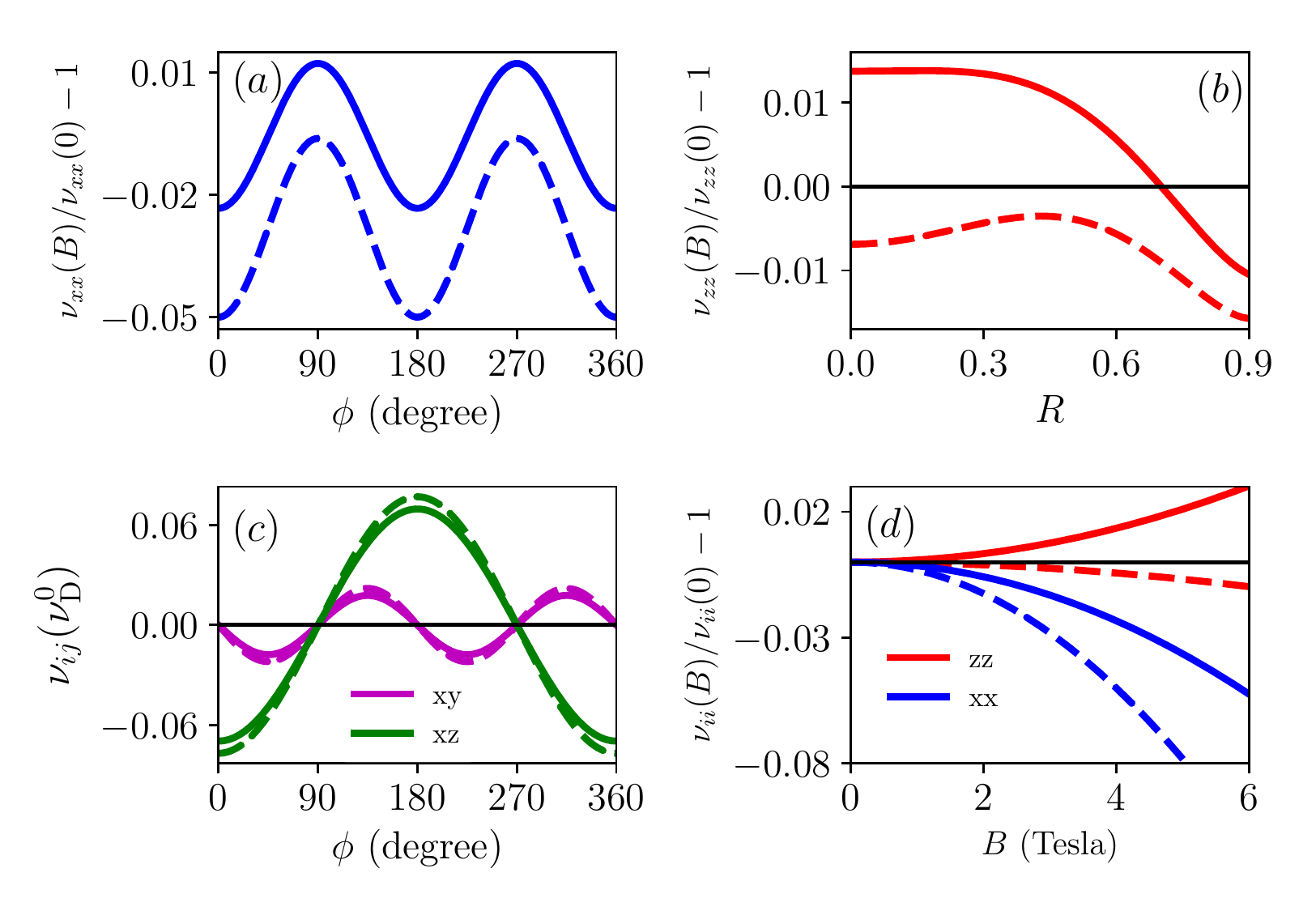}
\caption{Various components of thermopower in the planar geometry for the type-I class with tilt configuration $R_- =-R_+ = R$. (a) The $\cos^2 \phi$ dependence of the planar SC including (solid line) and excluding (dashed line) the OMM correction. (b) The dependence of the out-of-plane SC with the tilt parameter. Note that the inclusion of OMM correction (solid line) 
changes the sign of the $B$ dependent contribution from negative to positive, up to a critical $ R$. (c) The angular dependence of the planar NC ($\nu_{xy}\propto \sin 2\phi$) and the out-of-plane NC ($\nu_{xz} \propto \cos \phi$). (d) The $B$ dependence of the longitudinal and the out-of-plane transverse SC, which results in a negative and a positive Seebeck effect, respectively. We have used the parameters of Fig.~\ref{fig_11} and $R=0.3$.}
\label{fig.33}
\end{figure}

\begin{figure}[t!]
\includegraphics[width=0.5\textwidth]{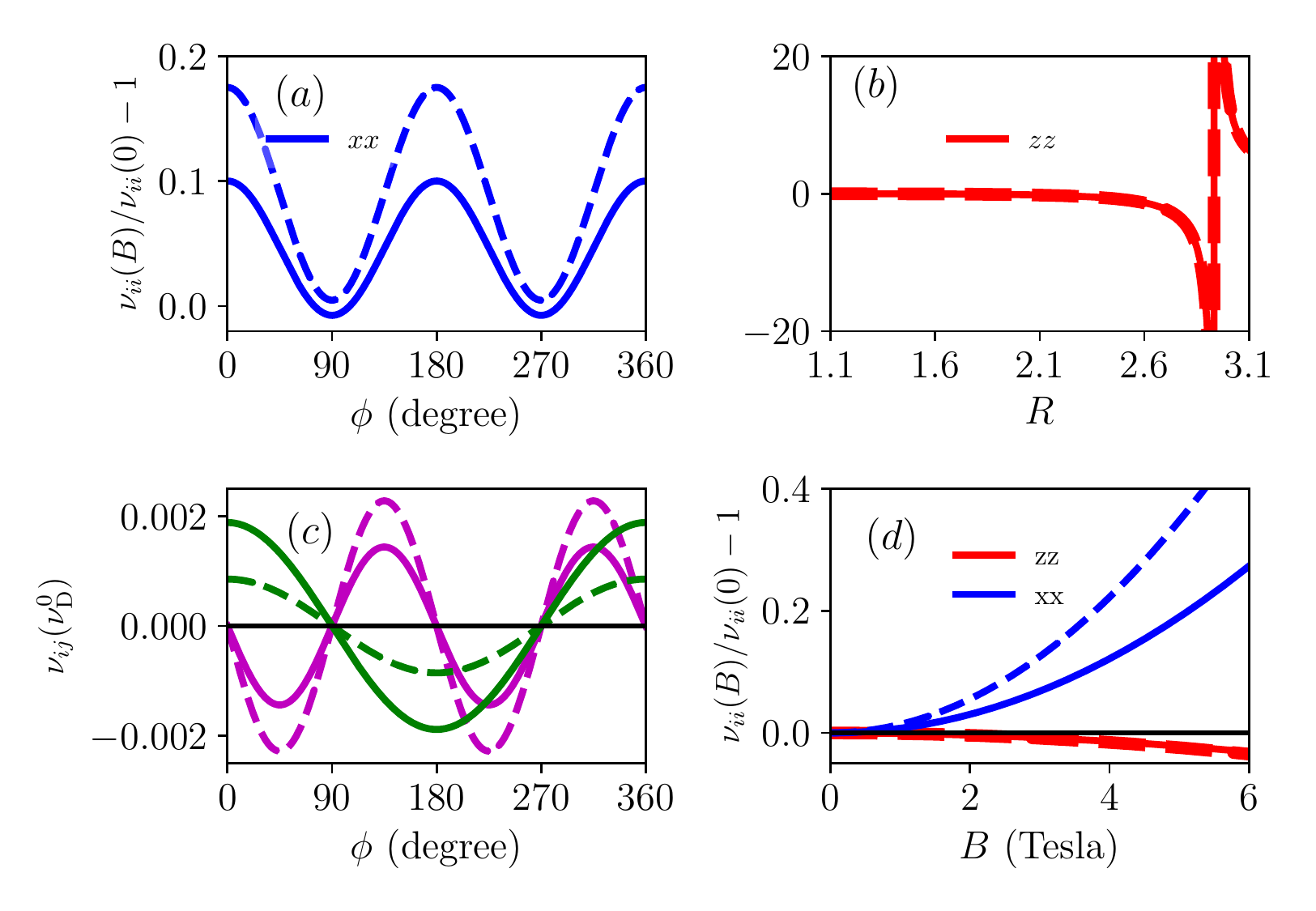}
\caption{Same as Fig.~\ref{fig.33}, but for a type-II WSM. (a) The $\phi$ dependence of $\nu_{xx}$. (b) The tilt dependence of $\nu_{zz}$ has contributions from electrons as well as holes. Here the sign reversal in 
$\nu_{zz}$ from negative to positive arises from the sign of the Drude component reversing at large $R$ (see Fig.~\ref{fig.drude}). This is a direct consequence of the hole carriers dominating the transport on increasing the WSM tilt. (c) The $\phi$ dependence of planar ($\nu_{xy}$ - purple curve) and out-of-plane ($\nu_{xz}$ - green curve) NCs. Note the phase difference of $\pi$ in the $\nu_{xz}$ response between a type-I and a type-II WSM. (d) The $B$ induced part of the SC has opposite signs for the longitudinal $\nu_{xx}(\phi=0)$, and the out-of-plane transverse $\nu_{zz}$ components. Here, we have used the parameters of Fig.~\ref{fig_11} and the tilt parameter $R=1.5$ and cutoff $\tilde{\Lambda}_k=10$.}
\label{fig.44}
\end{figure}

The angular dependence of the planar SC ($\nu_{xx}\propto \cos^2 \phi$), is shown in Fig.~\ref{fig.33}(a) for type-I WSMs and in Fig.~\ref{fig.44}(a) for type-II WSMs. The relative phase difference between the two classes is due to the opposite sign of Drude conductivity shown in Appendix \ref{Drude conductivities}. The  planar NC ($\nu_{xy} \propto \sin 2\phi$) and the out-of-plane NC ($\nu_{xz} \propto \cos\phi$) are highlighted in Figs.~\ref{fig.33} (c), and~\ref{fig.44} (c), for type-I and type-II WSMs, respectively. Again we find a relative phase difference in the linear-$B$ NC between the two classes. However this is not due to the Drude conductivity but due to the `tilted over' nature of the type-II WSM.
The inclusion of an OMM has a significant impact on the perpendicular SCs (magnetic field perpendicular to temperature gradient). It reverses the sign of the $B$-induced contribution in the $\nu_{zz}$ for the type-I WSM from negative to positive upto a critical tilt parameter, beyond which it retains its negative value [see Fig.~\ref{fig.33}(b)]. This is reminiscent of the sign change also seen in the perpendicular MR in Fig.~\ref{fig_11} (b).
Note that the sign reversal of $\nu_{zz}(B)/\nu_{zz}(0)-1$ in Fig.~\ref{fig.44}(b) for $R \approx 3.1$ arises from the corresponding sign change in the Drude component, $\nu_{\rm D}^{\rm z}$ as shown in Fig.~\ref{fig.drude}. 

In the presence of BC, the magnetic field suppresses the longitudinal SC [$\nu_{xx} (\phi=0)$] resulting in what is termed as a negative Seebeck effect. At the same time, it enhances the perpendicular SC ($\nu_{zz}$) for a type-I WSM, as shown in Fig.~\ref{fig.33}(d). This kind of negative longitudinal Seebeck effect and positive perpendicular Seebeck effect has been experimentally observed in the magnetically induced isotropic WSM phase in Cd${}_{3}$As${}_{2}$\cite{Jia16} and NbP\cite{Stockert17}. 
Our calculations predict that for a type-II WSM, the sign of both the longitudinal and the perpendicular SC change as compared to the type-I class,  as indicated in 
Fig.~\ref{fig.44}(d). This is because of the sign change of the corresponding Drude components as discussed in Appendix \ref{Drude conductivities}. 

For the case of ${\bf B} \parallel \hat{\bf z}$, the thermoelectric conductivity matrix can be written as 
\be\label{thermo_tensor_aniso_para}
\tilde{\alpha}-\tilde{\alpha}_{\rm D}=
\begin{pmatrix}
 \alpha_{\rm l}^{(1)} + \alpha_{\rm l}^{(2)}  & 0 & 0
\\
0 &   \alpha_{\rm l}^{(1)}  + \alpha_{\rm l}^{(2)}   &  0
\\
0 & 0 & \alpha_{\rm lz }^{(1)} + \alpha_{\rm l z}^{(2)}
\end{pmatrix}.
\ee
Using Eqs.~\eqref{charge_tensor_aniso_para} and \eqref{thermo_tensor_aniso_para} in Eq.~\eqref{nu1a} yields the thermopower matrix. 
For this configuration, since both $\tilde{\sigma}$ and $\tilde{\alpha}$ are diagonal, the thermopower matrix has no off-diagonal terms {i.e.}, no Nernst response. The diagonal components are given by $\nu_{xx}=\nu_{yy}$ and 
\bea \label{oop}
\nu_{xx}=\nu_{\rm D} + \nu_{\rm l}^{(1)} + \nu_{\rm l}^{(2)}~,\\
\nu_{zz}=\nu_{\rm D}^{\rm z} + \nu_{\rm l z}^{(1)} + \nu_{\rm l z}^{(2)}~.
\eea
Here, we have defined the linear-$B$ correction along the $x$ axis to be 
\be 
 \nu_{\rm l}^{(1)} = \dfrac{\sigma_{\rm D} \alpha_{\rm l}^{(1)} -\alpha_{\rm D} \sigma_{\rm l}^{(1)}}{\sigma_{\rm D}^{2}}~,
\ee
and the quadratic-$B$ correction in Eq.~\eqref{oop} reads as 
\be 
\nu^{(2)}_{\rm l} = \dfrac{1}{\sigma_{\rm D}^{2}}\left(\alpha_{\rm l}^{(2)}\sigma_{\rm D} - \alpha_{\rm D}\sigma_{\rm l}^{(2)} 
+ 
\left(\sigma_{\rm l}^{(1)} \alpha_{\rm D}- \alpha_{\rm l}^{(1)} \sigma_{\rm D}\right)\dfrac{\sigma_{\rm l}^{(1)}}{\sigma_{\rm D}}\right).
\ee
The linear and quadratic-$B$ correction along the $z$ direction can be generated from the above two equations simply by replacing the $x$ component of $\sigma$'s and $\alpha$'s by their $z$ components.

\begin{figure}[t]
\includegraphics[width=0.5\textwidth]{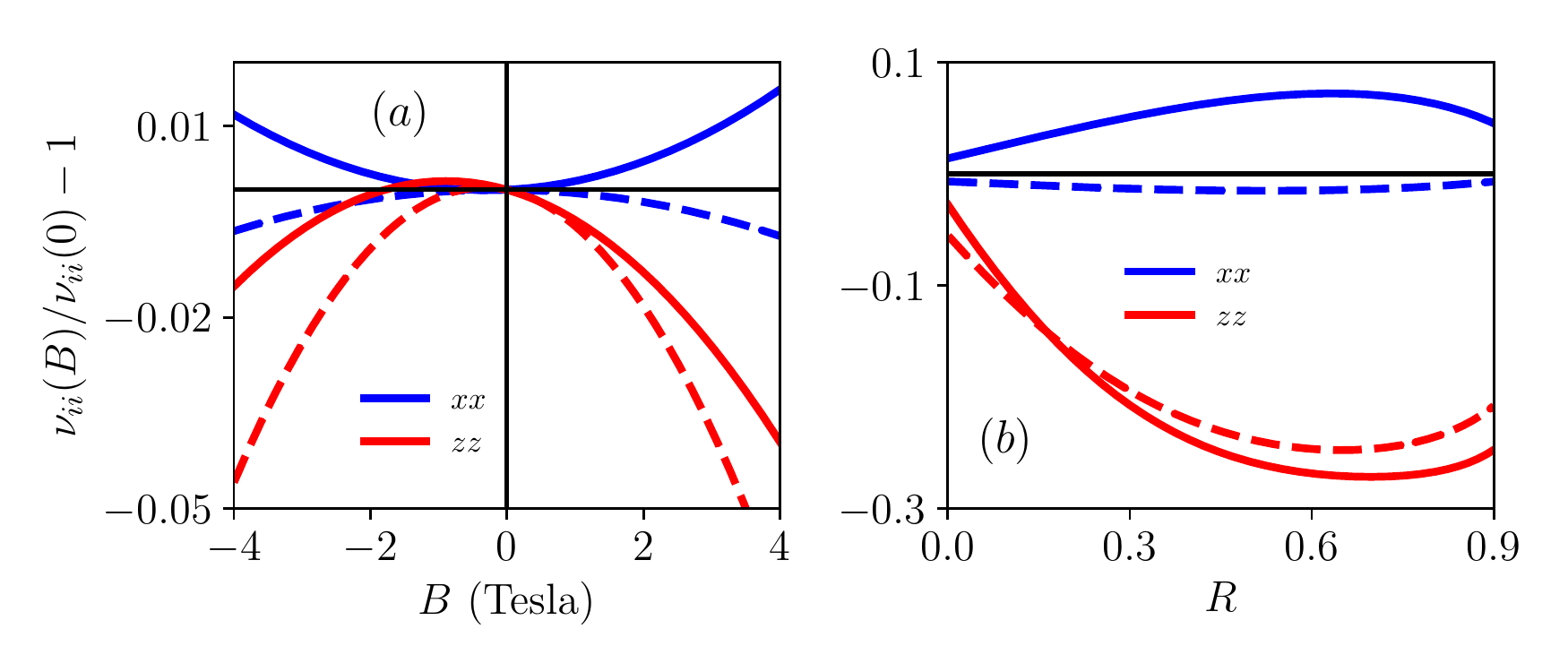}
\caption{SCs of type-I WSM for ${\bf B} \parallel {\bf R}$ with tilt configuration $R_- =-R_+=R$. (a) The $B$ dependence of the SCs at $R=.015$. The linear-$B$ terms in the $\nu_{xx}$ and $\nu_{zz}$ expressions lead to the asymmetry in the SC curves as $B$ changes from positive to negative. 
(b) The tilt dependence of the SCs at $B=4$ T. The longitudinal SC ($\nu_{zz}$) is negative irrespective of OMM correction (red lines). Note that the inclusion of the OMM correction has a 
significant impact on perpendicular SC ($\nu_{xx}$) as evident from the difference between the dashed (excluding OMM) and the solid blue lines (including OMM). 
We have used the parameters of Fig \ref{fig_11}.}
\label{fig.55}
\end{figure}

\begin{figure}[t]
\includegraphics[width=0.5\textwidth]{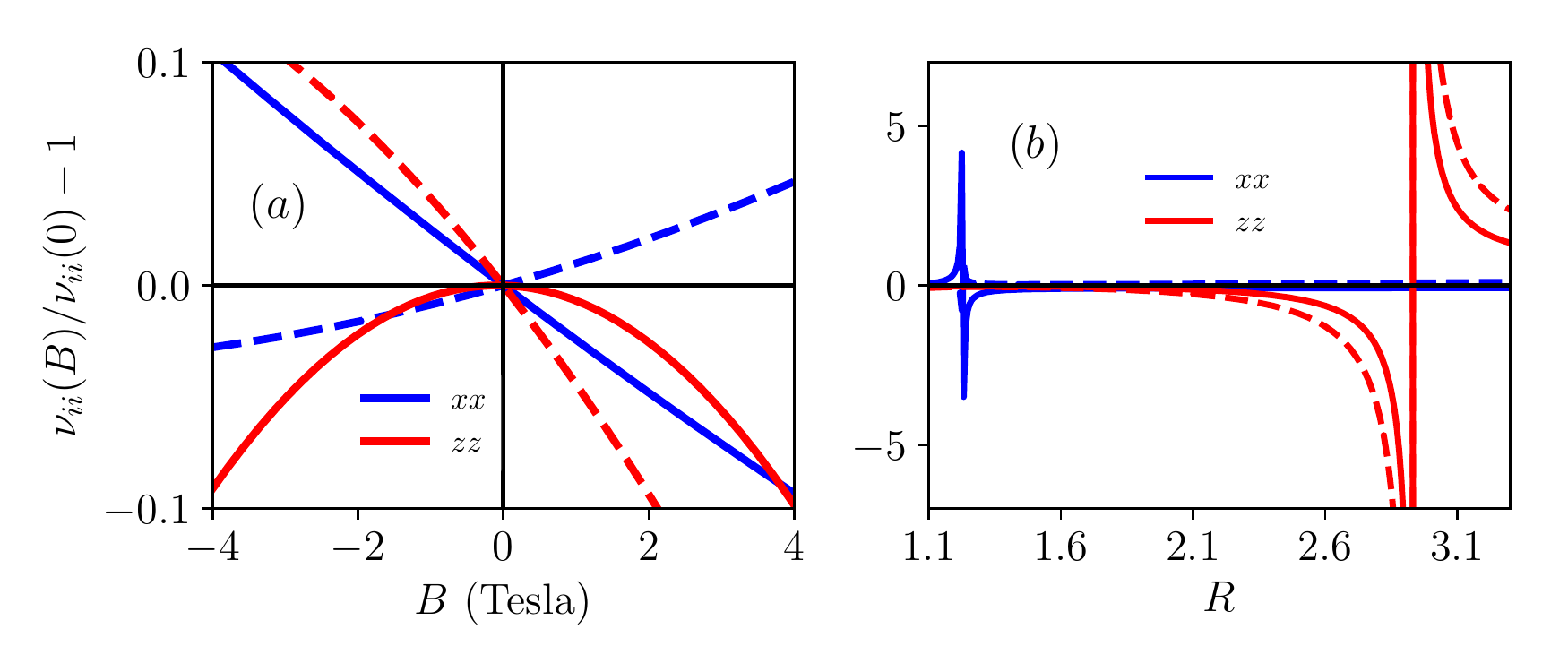}
\caption{Same as Fig.~\ref{fig.55}, but for type-II WSM. (a) $B$ dependence of the SCs at $R=2$. The linear-$B$ terms in $\nu_{xx}$, dominates its behavior for small $B$ with a negative slope, as shown (solid blue line -- including OMM). 
(b) The tilt dependence of the SC at $B=4$ T. Note that for a given $R$, the signs of $\Delta \nu_{xx}(B)/\nu_{xx}(0)$ and  $\Delta \nu_{zz}(B)/\nu_{zz}(0)$ are opposite. The sign reversal in each of them is a consequence of the corresponding Drude components flipping sign. This in turn occurs as different carriers start dominating the transport as shown in Fig.~\ref{fig.drude}. We have used the parameters of Fig.~ref{fig.44}.}
\label{fig.66}
\end{figure}

Interestingly, the SCs have a linear-$B$ dependence, arising from TRS breaking tilt. This is reminiscent of linear-$B$ terms also appearing in MR. 
The tilt and $B$ dependence of the longitudinal SCs for type-I WSM is shown in Fig.~\ref{fig.55}, while the same for type-II is shown in Fig.~\ref{fig.66}. 
Evidently the OMM plays an important role, reversing the sign of the perpendicular SC ($\nu_{xx}$) in type-I as well as type-II WSMs. 
Furthermore, in the case of a type-II WSM the linear-$B$ component of $\nu_{xx}$ dominates for small $B$, and the corresponding curve for 
$\nu_{xx}(B)/\nu_{xx}(0) - 1$ is almost linear in Fig.~\ref{fig.66}(a), with a negative slope.

\begin{table*}
\caption{The Berry curvature, OMM ($\gamma=1$) and tilt induced $B$-linear correction to thermopower. Only nonzero corrections are listed below.  
We have defined the dimensionless thermopower, $\nu^{(1)}_{ij} =  \nu_1 \tilde{\nu}^{(1)}_{ij}$, and $\nu_1$ is defined in Eq.~\eqref{nu_1}. We have neglected terms of the order of $\frac{x}{f(x)^2}$ for the type-III class, and $\frac{x^\prime}{g(x^\prime)}$ and $\frac{1}{g(x^\prime)^2}$ for the type-II class to obtain a  simpler form of thermopower. \label{T1}
}
\begin{tabular}{cccc}
\hline \hline
\rule{0pt}{5ex} 
\begin{tabular}{c}  $\tilde{\nu}^{(1)}_{ij} = \tilde{\nu}^{(1)}_{ji}$ \end{tabular}  
& 
\begin{tabular}{c} Type-I  [$R \to 0+|R|]$ \\ ${\cal{O}}(R)$ \end{tabular}
& 
 \begin{tabular}{c} Type-III [$|R|\to 1-x]$ \\${\cal O}(x)$\end{tabular} 
& 
\begin{tabular}{c}Type-II  $[|R|\to 1 + x']$\\$ {\cal O}(x')$\end{tabular}\\[3ex]
\hline
\rule{0pt}{6ex} 
\begin{tabular}{c} $({\bf B}\parallel \hat{\bf z})$\\ \end{tabular}
& 
\begin{tabular}{c} $~\tilde{\nu}_{\rm lz}^{(1)}\approx -2R$ \\[1ex]
$\tilde{\nu}_{\rm l}^{(1)}\approx \frac{2}{3}R~$
\end{tabular}
&
\begin{tabular}{c}  $\tilde{\nu}_{\rm lz}^{(1)}\approx -\frac{2}{3f(x)}\left(2 -  8x\right)$ \\[1ex] 
$\tilde{\nu}_{\rm l}^{(1)}\approx \frac{4}{3}f(x)x$;~~ ~$f(x)\equiv \log\frac{2}{x}-2$
\end{tabular}  
& 
\begin{tabular}{c} $~~~~\tilde{\nu}_{\rm lz}^{(1)}=-\frac{4}{3g(x^{\prime})};~~~ g(x^\prime) \equiv 2 + \ln \left(2 x^\prime \tilde \Lambda_k^2\right)$ 
\\[1ex]
$\tilde{\nu}^{(1)}_{\rm l}\approx \frac{4}{3}\left[g(x^\prime)-3\right]x^{\prime}$ \end{tabular} \\[7ex]
\begin{tabular}{c} 
$({\bf B \perp \hat{\bf z}})$\end{tabular}
& 
\begin{tabular}{c}  $\tilde{\nu}_{\rm t}^{(1)}\approx -\frac{4}{3}R$ \end{tabular}
& 
\begin{tabular}{c} $\tilde{\nu}_{\rm t}^{(1)}\approx -\frac{4}{3}x$ 
\end{tabular}  
&
\begin{tabular}{c} $\tilde{\nu}_{\rm t}\approx \frac{4}{3}x^\prime$ \end{tabular}\\[3ex]
\hline \hline
\end{tabular}
\end{table*}

\section{Limiting cases: $R \to 0$ and $R \to 1$}
\label{asymptotic}

In this section, we summarize our results for different components of thermopower in the asymptotic limit of no tilt, $R \to 0$, and critical tilt, $R \to 1$, 
which is called a type-III WSM, and serves as the boundary between type-I and type-II. 
To be specific, we work with the tilt configuration $R_- = - R_+ = R$ with $R>0$, though the results are similar for the other configuration as well. 
We will consider three specific cases: (a) vanishing tilt, $R \to 0$, (b) tilt tending to $R \to 1-0^+$ from below, and (c) tilt tending to $R \to 1+ 0^+$ from above.
We present linear-$B$ results in Table \ref{T1} and quadratic-$B$ results in Table \ref{T2}. 

The linear-$B$ correction to the thermopower is of the order of $-\frac{\pi^2}{3 e} \left(k_B^2 T\right)\nu_1$, where we have defined,
\be \label{nu_1}
\nu_1 \equiv \dfrac{\sigma_1 }{\sigma_{\rm D}^0} \dfrac{\alpha_{\rm D}^0}{\sigma_{\rm D}^0} 
= \frac{3}{\mu} \dfrac{\hbar^2 v_F^2}{2 \mu^2}\dfrac{e B}{\hbar}~.
\ee
The quadratic-$B$ correction is of the order of $-\frac{\pi^2}{3 e} \left(k_B^2 T\right) \nu_0$, where $\nu_0$ is defined as
\be \label{nu_0}
\nu_0 \equiv \dfrac{\sigma_{0} \alpha_{\rm D}^0 - \alpha_{0}  \sigma_{\rm D}^0  }{\left(\sigma_{\rm D}^0\right)^2}
=
-\dfrac{4}{5\mu} \left(\dfrac{\hbar^2 v_F^2}{2 \mu^2} \dfrac{e B}{\hbar}\right)^2~.
\ee
%
%
%
Interestingly, we find that all the $B$-linear terms tabulated in Table~\ref{T1}, vanish as $R \to 0$ as well as $R \to 1$.

\section{effect of chiral anomaly}	
\label{internode}

So far in this paper, we have discussed the thermopower due to {\it intranode scattering} and the effect of BC. In this section, we estimate the effect of {\it internode scattering} as the origin of non-trivial thermopower in type-I WSM. 
Internode scattering stabilizes the chiral anomaly in WSMs leading to different chemical potential in different Weyl nodes. \cite{Son_Spivak13,PhysRevB.97.035403,Kamal19} 
For calculating charge conductivity and thermoelectric coefficient due to internode scattering of a tilted WSMs, we borrow the formalism from Ref.~[\onlinecite{ Zyuzin_V_A17}]. 

For the case of ${\bf B} \perp \hat{\bf z}$, we calculate the charge conductivity 
matrix due to chiral anomaly to be 
\be \label{chrg_cond_CA}
\resizebox{\linewidth}{!}{$
\tilde{\sigma}_{\bf B} = 
\begin{pmatrix}
\sigma^{(2)}_{\rm ca}\cos^{2}\phi & \sigma^{(2)}_{\rm ca} \sin\phi \cos\phi & \frac{1}{9}\sigma_{\rm ca}^{(1)} \cos\phi
\\
\sigma_{\rm ca}^{(2)} \sin \phi \cos \phi &   \sigma^{(2)}_{\rm ca} \sin^{2}\phi &  \frac{1}{9}\sigma_{\rm ca}^{(1)} \sin\phi
\\
\frac{1}{4}\sigma_{\rm ca}^{(1)} \cos\phi & \frac{1}{4}\sigma_{\rm ca}^{(1)} \sin\phi &  0
\end{pmatrix}. 
$}
\ee 
Note the difference in the matrix structure in Eq.~\eqref{chrg_cond_CA}, from the intranode contribution given in Eq. \eqref{charge_tensor_aniso}. 
%
For the case of ${\bf B} \parallel \hat{\bf z}$, the only non zero component is $\sigma_{zz}$  and it is given by
\be \label{xxxx}
\sigma_{zz} =\left(\frac{1}{4}+\frac{1}{9}\right)\sigma_{\rm ca }^{(1)} + \sigma_{\rm ca}^{(2)}~.
\ee
%
In Eqs.~\eqref{chrg_cond_CA} and \eqref{xxxx}, we have defined (for $\gamma = 1$), 
\be \label{chrg_CA}
\sigma^{(2)}_{\rm ca} = \dfrac{e^2 \tau_v}{18 \pi^2 \hbar} \dfrac{e^2v_F^3}{\mu^2} B^2,~~~{\rm and}~~~ \sigma_{\rm ca}^{(1)} = - 2\dfrac{ e^2 \tau_v}{\pi^2 \hbar} \dfrac{e v_F}{\hbar} R B~.
\ee
Note that both these coefficients are proportional to the internode scattering time $\tau_v$ as expected. Furthermore, while the $\sigma^{(2)}_{\rm ca}$ term solely arises from the chiral anomaly inducing ${\bf E} \cdot {\bf B}$ term, 
the $\sigma_{\rm ca}^{(1)}$ term primarily arises from the tilted nature of the WSM and it vanishes as $R \to 0$. 
The Fermi energy dependence of both the linear and quadratic terms in Eq.~\eqref{thermo_tensor_aniso_para}, is identical to the corresponding intranode contributions. As a consistency check, we note that if we ignore the OMM correction in 
$\sigma^{(2)}_{\rm ca}$, 
then its numerical prefactor changes from $1/18$ to $1/4$, and  $\sigma^{(2)}_{\rm ca}$ becomes identical to Eq.~(17) in Ref.~[\onlinecite{Son_Spivak13}.] 

Using these coefficients and the Mott relation [valid in the limit $\mu/(k_B T) \gg 1$], we can calculate the correction to thermopower due to the chiral anomaly. For  ${\bf B} \perp \hat{\bf z}$, the thermopower matrix is given by
\be
\resizebox{\linewidth}{!}{$
\tilde{\nu}_{\bf B} = 
\begin{pmatrix}
\nu^{(2)}_{\rm ca}\cos^{2}\phi & \nu^{(2)}_{\rm ca} \sin\phi \cos \phi & \frac{1}{9}\nu_{\rm ca}^{(1)} \cos\phi
\\
\nu^{(2)}_{\rm ca} \sin\phi \cos\phi &  \nu^{(2)}_{\rm ca} \sin^{2}\phi &  \frac{1}{9}\nu_{\rm ca}^{(1)} \sin\phi
\\
\frac{1}{4}\nu_{\rm ca}^{(1)} \cos\phi & \frac{1}{4}\nu_{\rm ca}^{(1)} \sin\phi &  \nu_{\rm ca,z}^{(2)}
\end{pmatrix}. 
$}
\ee
Similarly, for ${\bf B} \parallel \hat{\bf z}$, the only non zero component of thermopower is given by
\be 
\nu_{zz} =\left(\frac{1}{4}+\frac{1}{9}\right)\nu_{\rm ca }^{(1)} + \nu_{\rm ca,zz}^{(2)}
\ee
%
The quadratic-$B$ correction to the thermopower matrix due to internode scattering is given by can be written, in units of $ -\frac{\pi^2}{3e}\left(k^2_B T\right)$, with $\zeta \equiv\frac{\hbar^2 v_F^2}{2 \mu^2} \frac{e B}{\hbar}$, as
\begin{gather} \label{nu_ca_2}
\nu_{\rm ca}^{(2)} = -\dfrac{8}{3\mu} \zeta^2\dfrac{\tau_v}{\tau} \left(1- 3\dfrac{\tau_v}{\tau} R^2 \right),~
\nu_{\rm ca,z}^{(2)} = \dfrac{8}{\mu} \zeta^2\left(\dfrac{\tau_v}{\tau}R\right)^2.
\\
\nu_{\rm ca,zz}^{(2)} = -\dfrac{2}{9\mu} \zeta^2\dfrac{\tau_v}{\tau} \left(4 - 169\dfrac{\tau_v}{\tau} R^2 \right)~.
\end{gather}
Similarly, the linear-$B$ correction, in units of $ -\frac{\pi^2}{3e}\left(k^2_B T\right)$ can be expressed as 
\be \label{nu_ca_1}
\nu_{\rm ca}^{(1)}=\frac{24}{\mu} \zeta \frac{\tau_v}{\tau}R~.
\ee
We emphasize that this linear-$B$ correction in thermopower, is one of the significant findings of this paper. It primarily arises due to the tilted nature of the WSM and vanishes as $R \to 0$.  

Given that the internode scattering involves relatively large momentum transfer as compared to the intranode scattering timescale \cite{Son_Spivak13, Zyuzin_V_A17}, generally we have $\tau_v \gg \tau$.  
Additionally, since the internode scattering terms [Eqs.~\eqref{nu_ca_2} and~\eqref{nu_ca_1}] are $\sim \tau_v/\tau$ times than the intranode scattering terms, the contribution of the internode scattering terms will dominate in the thermopower as well as in the electrical conductivity.

\begin{table*}
\caption{The Berry curvature, OMM, and tilt induced quadratic-$B$ correction to the thermopower. Only nonzero corrections are listed below. 
We have defined the dimensionless thermopower, $\nu^{(2)}_{ij} =  \nu_0 \tilde{\nu}^{(2)}_{ij}$, and $\nu_0$ is defined in Eq.~\eqref{nu_0}.  We have neglected terms of the order of $\frac{x}{f(x)^2}$ for type-III class, and $\frac{x^\prime}{g(x^\prime)}$ and $\frac{1}{g(x^\prime)^2}$ for type-II class to obtain a  simpler form of thermopower.
\label{T2}}
\begin{tabular}{cccc}
\hline \hline
\rule{0pt}{5ex} 
\begin{tabular}{c}  $\tilde{\nu}^{(2)}_{ij}=\tilde{\nu}^{(2)}_{ji}$ \end{tabular}  
& 
\begin{tabular}{c} Type-I  [$R \to 0+|R|]$ \\ ${\cal{O}}(R)$ \end{tabular}
& 
 \begin{tabular}{c} Type-III [$|R|\to 1-x]$ \\${\cal O}(x)$\end{tabular} 
& 
\begin{tabular}{c}Type-II  $[|R|\to 1 + x']$\\$ {\cal O}(x')$\end{tabular}\\[3ex]
\hline
\rule{0pt}{6ex} 
\begin{tabular}{c}$({\bf B}\parallel \hat{\bf z})$
\end{tabular}
& 
\begin{tabular}{c}  
$\tilde{\nu}_{\rm l}^{(2)} = 2;~~\tilde{\nu}_{\rm lz}^{(2)}=-4$ 
\end{tabular}
&
\begin{tabular}{c} $\tilde{\nu}_{\rm l}^{(2)}=\frac{8}{3}x$
\\
$~~~~\nu_{\rm lz}^{(2)}=-\frac{4}{f(x)^2}\left[\frac{3}{2}f(x)\nu_0 - \frac{4}{9}\frac{\nu_1^2}{\nu_{\rm D}^0}\right]+ \frac{74}{3f(x)}\nu_0 x~~~~$  
\end{tabular}  
& 
\begin{tabular}{c}$\tilde{\nu}^{(2)}_{\rm l}\approx \frac{8}{3} x^\prime$
\\
$\tilde{\nu}_{\rm lz}^{(2)}\approx -\frac{6}{g(x^\prime)}$ \end{tabular} 
\\[7ex]
\begin{tabular}{c}
$({\bf B}\perp \hat{\bf z})$
\end{tabular} 
&
\begin{tabular}{c}
$\tilde{\nu}_{\rm z}^{(2)}\approx 2$ \\ [0.5ex]
$\Delta \tilde{\nu}^{(2)}\approx -6;\tilde{\nu}_{\perp}^{(2)} \approx 2$ 
\end{tabular} 
&
\begin{tabular}{c}
$\nu_{\rm z}^{(2)}\approx -\frac{4}{3f(x)} \left(3\nu_0  - \left[17 \nu_0 + \frac{2}{3} \frac{\nu_1^2}{\nu_{\rm D}^0}\right]x \right)$ \\[1ex]
$\Delta \nu^{(2)}\approx -\frac{4}{9}\left[39\nu_0 -\frac{\nu_1^2}{\nu_{\rm D}^0}\frac{2}{f(x})  \right]x;~~\tilde{\nu}_{\perp}^{(2)} \approx \frac{8}{3}x$ \end{tabular} 
& 
\begin{tabular}{c}
$\tilde \nu_{\rm z}^{(2)} \approx -\frac{4}{g(x^\prime)}$~\\[1ex]$~\Delta \tilde \nu^{(2)}\approx-\frac{4}{3}13 x^\prime;$ ~~$\tilde{\nu}_{\perp}^{(2)}= \frac{8}{3} x^\prime$
\end{tabular}\\[6ex]
\hline \hline
\end{tabular}
\end{table*}

\section{Conclusions}
\label{conclusions}
The presence of the BC and OMM in WSMs influences the flow of charge carriers as well as entropy in the presence of a magnetic field. This manifests as several interesting magnetoelectric and magnetothermal transport properties in WSM. 
{Since the Weyl nodes always come in pairs in a WSM, both the intranode and internode scattering play an important role in determining the electrical conductivity and thermopower. 
In this paper, we have primarily focused on the impact of the BC and OMM on the thermopower due to intranode scattering in a tilted WSM, and briefly discussed the effect of the  internode scattering timescale.} Our analytical calculations of the full conductivity and thermopower matrix, are based on the BC-connected semiclassical Boltzmann transport formalism, and explicitly include the effects of the OMM. The latter modifies the energy-dispersion of the Bloch electrons which also manifests in the modified velocity of carriers, as well as in the Fermi function. However, the Mott relation connecting the conductivity matrix to the thermopower matrix remains intact on including the effects of the OMM. 

We find that the OMM has a significant impact on the perpendicular MR in WSMs. Consistent with experiments, our calculations show that the longitudinal MR (${\bf B} \parallel {\bf E}$) in isotropic WSMs is always negative, while the perpendicular MR (${\bf B} \perp {\bf E}$) is positive on including the effect of the OMM. However, in tilted WSMs, the perpendicular MR can also flip sign to become negative for the large tilt parameter [see Fig.~\ref{fig_11} (b)].

In a type-I WSM, for the case of ${\bf B} \perp \hat{\bf z}$, we find that increasing the magnetic field reduces the longitudinal SC, giving rise to 
 a  negative Seebeck effect in analogy with negative MR [see Fig.~\ref{fig.33}(d)]. 
For the perpendicular SC we find it to be positive for small tilt parameters, but it reverses sign for large tilt parameters. 
Analogous to the planar Hall effect, we also find the existence of a planar Nernst effect, which has an angular dependence $\nu_{xy} \propto \sin(2 \phi)$. 
Additionally, we also find a linear-$B$ out-of-plane Nernst response in WSMs with a finite tilt. 
For the other case of ${\bf B} \parallel \hat{\bf z} $, we find the conductivity and the thermopower matrix to be diagonal, with tilt induced linear-$B$ terms in the 
longitudinal as well as perpendicular components. This manifests in an asymmetry in the MR and SC curve around the $B=0$ line, as shown in Figs.~\ref{fig.55} and \ref{fig.66}.

For the case of a type-II WSM, the scene is a bit mixed up, owing to the contributions of both electron and hole carriers for all energies. 
We find that even in the absence of a magnetic field, the Drude SC can be positive or negative depending on the tilt (see Fig.~\ref{fig.drude}). 
For the case of ${\bf B} \perp \hat{\bf z}$ in a type-II WSM, we find that in contrast to the case of a type-I WSM, the longitudinal SC is positive while the perpendicular SC is negative. The angular dependence of the planar ($\nu_{xy} \propto \sin 2 \phi$) and the out-of-plane Nernst effect ($\nu_{xz} \propto \cos \phi$) is the same for type-I and type-II WSMs. For the other case of ${\bf B} \parallel \hat{\bf z}$, we find that the linear-$B$ terms dominate the $\nu_{xx}$ for small magnetic fields. 
We expect similar effects (such as planar Peltier effect and linear-$B$ out-of-plane Peltier effect, among others) to also arise in the diagonal and the off-diagonal coefficients corresponding to the Peltier effect.  

Additionally, we have also explored the impact of intranode scattering and chiral anomaly on the electrical conductivity and thermopower matrix in tilted WSMs. Remarkably, we find that the intranode scattering and chiral anomaly in tilted WSMs also lead to $B$-linear terms in the electrical conductivity as well as in the thermopower matrix. Furthermore, as the conductivity and thermopower matrix $\propto \tau_v$ and since $\tau_v \gg \tau$, the internode contribution dominates.

\acknowledgements
A. A. acknowledges funding support by Dept. of Science and Technology, Government of India, via DST grant no. DST/NM/NS/2018/103(G), and from SERB grant number CRG/2018/002440. 
K. D. acknowledges Indian Institute of Technology Kanpur for PhD fellowship. 

\appendix

\section{Berry-connected Boltzmann transport formalism} \label{review}

{The Boltzmann transport formalism for magnetotransport works well for relatively small magnetic fields where the effects of Landau quantization can be ignored. 
The equations of motion (EOM) approach works well in the regime where several Landau levels are occupied: $\hbar \omega_c \ll \mu$, with $\mu$ denoting the chemical potential, and $\omega_c$ is the cyclotron frequency. In addition, the relaxation time approximation for the non-equilibrium distribution function (NDF) works well in the regime $v_F\tau \ll l$, where $v_F$ is the Fermi velocity, $\tau$ is the relaxation time scale and $l\equiv \sqrt{\hbar/eB}$ is the magnetic length for cyclotron motion\cite{G_Sharma17a} with $B$ as the magnetic field. For WSM, the Fermi velocity is found to be in the range of $10^5$-$10^6$ m/s\cite{Watzman18}. The Fermi energy and scattering time are found to be of the order of a few meV and $0.1$ ps, respectively\cite{Zhang17}.

\subsection{Semiclassical transport with Berry curvature and orbital magnetic moment}

The EOM describing the dynamics of the center of the carrier wave-packet (location at $\bf r$, and having the Bloch wave-vector $\bf k$) in a given band is
 given by \cite{Xiao_Niu_rev10,Morimoto16,Marder10}
\begin{eqnarray}\label{eom_r}
\dot{\bf r}& = &D_{\bf k}\left[\tilde{{\bf v}}_{\bf{k}}+\frac{e}{\hbar}({\bf E}\times {\bf \Omega}_{\bf k})+\frac{e}{\hbar}(\tilde{{\bf v}}_{\bf{k}}\cdot {\bf \Omega}_{\bf k}){\bf B}\right],
\\\label{eom_k}
\hbar\dot{\bf k }&=& D_{\bf k}\left[-e{\bf E} - e(\tilde{{\bf v}}_{\bf{k}}\times {\bf B})-\frac{e^{2}}{\hbar}({\bf E}\cdot{\bf B}){\bf  \Omega_{\bf{k}}}\right].
\end{eqnarray}
Here $-e$ is the electronic charge and and we have defined $D_{\bf k}\equiv [1+\frac{e}{\hbar}({\bf B}\cdot{\bf \Omega}_{\bf{k}})]^{-1}$.  The band velocity is given by $\hbar \tilde{{\bf v}}_{\bf k} =\nabla_{\bf k}\tilde{\epsilon}_{\bf k}$, where  ${\tilde\epsilon}_{\bf k}=\epsilon_{{\bf k}}-{\bf m}_{\bf k}\cdot {\bf B}$ is the electronic dispersion modified by the intrinsic OMM. The modified band velocity can now be expressed as $\tilde{{\bf v}}_{{\bf k}}={\bf v}_{{\bf k}}-\gamma{\bf v}_{{\bf k}}^{m}$, where ${\bf v}_{{\bf k}}^{m} = \frac{1}{\hbar}{\bf \nabla}_{{\bf k}}({\bf m}_{\bf k}\cdot {\bf B})$, and the factor of $\gamma= 0/1$ is introduced to keep track of the OMM dependent corrections.

The BC modified group velocity in Eq.~\eqref{eom_r} has two interesting effects: 
the ${\bf E}\times {\bf \Omega}_{\bf k}$ term gives rise to the intrinsic anomalous Hall effect \cite{Sinitsyn08, Haldane04},
while the $(\tilde{{\bf v}}_{\bf{k}}\cdot {\bf \Omega}_{\bf k}){\bf B}$ term gives rise to the chiral magnetic effect in the presence of non-zero chiral chemical potential\cite{Kim14}. 
In Eq.~\eqref{eom_k}, the first two terms are the well known Lorentz force, whereas the third $({\bf E}\cdot{\bf B}){\bf  \Omega_{\bf{k}}}$ term manifests the effect of the chiral anomaly leading to negative MR\cite{Son_Spivak13} in WSMs. The modified EOM also changes the phase space volume by a factor $D_{\bf k}$ . 
%
To compensate for this, such that the number of states in the phase-space volume element is preserved, we have   
$d{\bf k} \to d{\bf k}/D_{\bf k}$. 
This factor needs to be incorporated 
whenever the wave-vector summation is converted in an integral over the Brillouin zone in the presence of the BC\cite{Xiao_Niu05, Duval06}.

The three-component BC and the intrinsic OMM can be obtained from their respective tensors via the relation: $A_{a}=\varepsilon_{a b c}A^{b c}$, where $\varepsilon_{abc}$ is the anti-symmetric Levi-Civita symbol. The corresponding Berry tensor is given by\cite{Chang96, Dai17}
%
%
%
%
%
%
\be \label{BC}
{\bf \Omega}_{n}^{a b}=-2~\dfrac{{\rm Im}
\left[\langle n|\partial_{k_{a}}\mathcal{H}|n^{\prime}\rangle
\langle n^{\prime}|\partial_{k_{b}}\mathcal{H}|n\rangle\right]}
{(\epsilon_{n}-\epsilon_{n^{\prime}})^2}~,
\ee
where $n$ is the band index with $\mathcal{H}|n\rangle = \epsilon_{n}|n\rangle$.
Similarly, the OMM tensor is given by\cite{Chang96, Dai17}
\be  \label{OMM}
{\bf m}_{n}^{a b}=-\dfrac{e}{\hbar}~\dfrac{{\rm Im}\left[\langle n|\partial_{k_{a}}\mathcal{H}|n^{\prime}\rangle 
\langle n^{\prime}|\partial_{k_{b}}\mathcal{H}|n\rangle\right]}
{\epsilon_{n}-\epsilon_{n^{\prime}}}~.
\ee

The dynamics of the NDF $g_{\bf r, k}$, is described by the Boltzmann kinetic equation.  
In the steady state the NDF kinetic equation for each node is given by \cite{Ashcroft76}
\be\label{bte_2}
\dot{\bf{r}}\cdot {\bf \nabla}_{\bf{r}} ~g_{\bf{r},\bf{k}} +\dot{\bf{k}}\cdot{\bf\nabla}_{\bf{k}}~g_{\bf{r},\bf{k}} =I_{\rm coll}\{g_{\bf{r},\bf{k}}\}~,
\ee
where the right hand side is the collision integral. In the relaxation time approximation,
$
I_{\rm coll}\{g_{\bf{r},\bf{k}}\}=-\dfrac{g_{\bf{r},\bf{k}}-f_{\rm eq}}{\tau_{\bf{k}}}~,
$ 
where $f_{\rm eq} \equiv f_{\rm eq}(\tilde{\epsilon}_{\bf k},\mu,T) = (e^{\beta(\tilde{\epsilon}_{\bf k} - \mu)} + 1)^{-1}$ is the equilibrium Fermi-Dirac distribution function with $\beta^{-1} \equiv k_{B}T$.
The scattering timescale $\tau_{\bf k}$ is the effective intra-node relaxation time which we consider to be constant ($\tau_{\bf k}\to \tau$) for simplicity. 
Note that in an anisotropic tilted WSM, the scattering timescale should be anisotropic. However, for simplicity, we will consider the 
scattering timescale to be isotropic, and the anisotropy of the band structure will appear only in the modified anisotropic velocities, and the anisotropic Fermi surface.

%
Substituting Eqs.~\eqref{eom_r}-and \eqref{eom_k} in Eq.~\eqref{bte_2}, we obtain an approximate steady state NDF, upto  first order in ${\bf E}$ and $\nabla T$:  
\begin{widetext}
\be \label{ndf}
g_{{\bf r, k}}=f_{\rm eq} + 
\left[D_{\bf k}\tau \left(-e{\bf E} - \frac{(\tilde{\epsilon}_{{\bf k}}-\mu)}{T}~{\bf \nabla_{\bf r}}T\right) 
\cdot
\left(\tilde{{\bf v}}_{{\bf k}} + \frac{e {\bf B}(\tilde{{\bf v}}_{{\bf k}}\cdot{\bf \Omega}_{{\bf k}})}{\hbar}\right)
%
\right]
\bigg(-\dfrac{\partial f_{\rm eq}}{\partial \tilde{\epsilon}_{{\bf k}}}\bigg)~.
\ee
\end{widetext}
%
{Note that in this paper, our primary focus is on the BC connected conductivity and we have not included the impact of the Lorentz force terms in modifying the NDF\citep{Kim14, Jacoboni10} in Eq.~\eqref{ndf}. 
%
%
The Lorentz force contribution to conductivity proportional to $\frac{eB}{\mu}\tau v_F^2$ and its effect is more prominent in scenarios when the magnetic field is perpendicular to the transport direction. The corresponding BC contribution is proportional to $\frac{eB} 
{\mu^2}\hbar v_F^2$ (for intranode scattering) in the electrical conductivity, and its impact is 
more when electric and magnetic fields are parallel.  A direct comparison between the Lorentz force terms and the BC induced terms is not feasible, as far as the MR is concerned. We refer the reader to Ref.~[\onlinecite{PhysRevB.98.205139}] for an excellent discussion on this issue, and proceed below with the discussion on the BC induced conductivity. }


Armed with the equation of motion and the NDF, we now proceed to calculate current. In the presence of a finite OMM, the total local current can be expressed as \cite{Xiao_Niu06}
\be \label{eq:10}
{\bf j}^{\rm loc}=-e\int [d{\bf k}]\inv{D}\dot{{\bf r}}~g_{{\bf r},{\bf k}} + {\bf \nabla}_{{\bf r}} \times \int [d{\bf k}] \inv{D}{\bf m}_{\bf k}~f_{\rm eq}.
\ee
Here we have used the shorthand $[d{\bf k}] = d{\bf k}/({2 \pi})^d$, with $d$ being the dimension of the system. The additional second term arises from the intrinsic OMM of individual carriers, and can be physically attributed to the rotating dynamics of the finite width Bloch wave-packet.  However, the `magnetization current' is not observable in conventional transport measurement. Consequently, the transport current 
is defined as\cite{Cooper97, Xiao_Niu06}
\be \label{current}
{\bf j}^{\rm tr} ={\bf j}^{\rm loc} - { \bf \nabla}_{{\bf r}} \times {\bf M}({\bf r})~,
\ee
where $ {\bf M}({\bf r})$ is the total orbital magnetization in real space.
%
%
%
The magnetization for a given chemical potential ($\mu$) and $T$ is given by ${\bf M}=-\partial F/\partial {\textbf{B}}|_{\mu,T}$, where $F$ is the grand-canonical potential defined as \cite{Xiao_Niu06} 
\be \label{Free energy}
F=-\frac{1}{\beta}\int [d{\bf k}]\left(1+\frac{e}{\hbar}{\bf B}\cdot{\bf \Omega}_{\bf{k}}\right) \ln[1+e^{-\beta({\tilde\epsilon}_{\bf k}-\mu)}]~.
\ee
Note that in Eq.~\eqref{current}, the curl in the real space will involve temperature gradients, and the second term gives rise to the anomalous thermo-electric Hall effect. 

\subsection{Electric and thermoelectric conductivity}

Using Eqs.~\eqref{eom_r},~\eqref{ndf}, and~\eqref{Free energy} in Eq.~\eqref{current}, yields the following general expression for the BC dependent part of the electrical conductivity tensor 
\begin{widetext}
\begin{multline}\label{elec_cond}
\sigma_{ij}^{\rm total} =-\frac{e^{2}}{\hbar}\int[d{\bf k}]~\epsilon_{ijl}\Omega_{{\bf k}}^{l}~f_{\rm eq}+e^{2}\tau\int [d{\bf k}] D_{\bf k}
\bigg[\tilde{v}_{i}  + \dfrac{eB_{i}}{\hbar}(\tilde{{\bf v}}_{{\bf k}}\cdot {\bf \Omega}_{{\bf k}})\bigg]
\bigg[\tilde{v}_{j}  + \dfrac{eB_{j}}{\hbar}(\tilde{{\bf v}}_{{\bf k}}\cdot {\bf \Omega}_{\bf{k}})\bigg]\bigg(-\dfrac{\partial f_{\rm eq}}{\partial \tilde{\epsilon}_{{\bf k}}}\bigg)~.
\end{multline}
Here $\tilde{v}_{j}$ denotes the $j$th component of $\tilde{{\bf v}}_{\bf k}$, and $\epsilon_{ijl}$ is the Levi-Civita antisymmetric tensor. 
Similarly, the BC dependent part of the thermoelectric conductivity tensor can be explicitly obtained to be
\begin{multline}\label{therm_cond}
\alpha_{ij}^{\rm total}=\dfrac{k_{B}e}{\hbar}\int[d{\bf k}]~\epsilon_{ijl}\Omega_{{\bf k}}^{l}~\xi_{{\bf k}}-e\tau\int [d{\bf k}] D_{\bf k}~\frac{(\tilde{\epsilon}_{{\bf k}}-\mu)}{T}
\bigg[\tilde{v}_{i}  + \dfrac{eB_{i}}{\hbar}(\tilde{{\bf v}}_{{\bf k}}\cdot {\bf \Omega}_{{\bf k}})\bigg]
\bigg[\tilde{v}_{j}  + \dfrac{eB_{j}}{\hbar}(\tilde{{\bf v}}_{{\bf k}}\cdot {\bf \Omega}_{\bf{k}})\bigg]\bigg(-\dfrac{\partial f_{\rm eq}}{\partial \tilde{\epsilon}_{{\bf k}}}\bigg)~. 
\end{multline}
\end{widetext}
In Eq.~\eqref{therm_cond} we have defined,  
\begin{equation}\label{entropy}
\xi_{{\bf k}}=\beta(\tilde{\epsilon}_{{\bf k}}-\mu)f_{\rm eq}+\ln[1+e^{-\beta(\tilde{\epsilon}_{{\bf k}}-\mu)}]~.
\end{equation}
While Eq.~\eqref{therm_cond} can be evaluated separately, in the low temperature limit ($k_BT \ll \mu$) it can also be obtained from Eq.~\eqref{elec_cond} by using the Mott relations \cite{Ashcroft76,Xiao_Niu06} which also hold in the presence of BC and the OMM. In fact the validity of the Mott relation including the OMM correction has also been proved recently, in a more general setting, in Ref.~[\onlinecite{Dong2018}].

The first term on the right hand side of Eqs.~\eqref{elec_cond} and \eqref{therm_cond} denote the  anomalous Hall effect \cite{Burkov14, Goswami13, Haldane04} and the anomalous thermoelectric effect\cite{Xiao_Niu06,  Fiete14}, respectively. 
In WSM, the anomalous Hall conductivity, $\sigma_{xy}^{A}$ has been shown to be linearly proportional to the internode separation\cite{Fiete14}.
The anomalous thermoelectric conductivity $\alpha_{xy}^{A}$ was shown to be zero\cite{Fiete14} in a linearized model but finite for a lattice model\cite{G_Sharma16, Gorbar17}. The finite contribution in $\alpha_{xy}^{A}$ in a lattice model originates from band curvature effects beyond the linear dispersion. 
In the case of a tilted WSM, described by a linear dispersion, $\alpha_{xy}^{A}$ is finite for both the type-I and the type-II class of WSMs\cite{Ferreiros17}.

In the last term in Eqs.~\eqref{elec_cond} and \eqref{therm_cond}, one of the anomalous velocity terms arises from the ${\bf E} \cdot {\bf B}$ term in Eq.~\eqref{eom_k}, and the other from the NDF. 
For parallel electric and magnetic fields,  this is what leads to NMR\cite{Nielsen83, Son_Spivak13}, which is quadratic in the magnetic field, and is a relatively well established transport signature\cite{Kim13, Huang15b}. 
This term also leads to the planar Hall effect\citep{Burkov17, Nandy17}, in which a Hall voltage is generated in the plane of the electric and magnetic fields, as long as they are not parallel or perpendicular to each other. 

Expanding Eqs.~\eqref{elec_cond} and \eqref{therm_cond} in powers of $B$ (expansion of Fermi function\cite{Morimoto16}), the zeroth order, linear and quadratic-$B$ components of the transport coefficients can be expressed as \cite{Ashcroft76}: 
$\sigma^{(o)}_{ij} \equiv \mathcal{L}^{0{(o)}}_{ij}$ and $\alpha^{(o)}_{ij}\equiv -\frac{1}{eT}\mathcal{L}^{1{(o)}}_{ij}$, where $o=\{0,1,2\}$ refers to the order of magnetic field. For the first-order terms we find 
%
%
\begin{widetext}
\begin{gather} \label{first_order}
\mathcal{L}_{ij}^{p (1)}=e^{2}\tau\int [d{\bf k}]\bigg[\left(\epsilon-\mu\right)^{p}
\bigg(\Big[\frac{e}{\hbar}\left(v_{i}B_{j} +v_{j}B_{i}\right)({\bf v}\cdot {\bf \Omega})
-\frac{e}{\hbar}{\bf \Omega} \cdot{\bf B}~v_{i}v_{j}
\\\nn
-\gamma \left(v_{i}v_{j}^{m}+v_{j} v_{i}^{m}\right)\Big]
\left(-f_{0}^{\prime}\right)
-\gamma~v_{i}v_{j} \left({\bf m}\cdot {\bf B}\right)\left(-f_{0}^{\prime \prime}\right) \bigg)
- \delta(p-1)\left(\gamma ~{\bf m} \cdot {\bf B} \right)^{p}v_{i}v_{j}\left(-f_{0}^{\prime}\right)\bigg]~.
\end{gather}
Here $p =$ 0 (or 1) for the electric (or thermoelectric) conductivity. 
%
%
Similarly, the quadratic terms can be expressed as\cite{Morimoto16} 
\begin{multline}\label{second_order}
\mathcal{L}_{ij}^{p (2)} = e^{2}\tau\int [d{\bf k}]\Bigg[\left(\epsilon-\mu\right)^{p}
\bigg(\Big[v_{i} v_{j}\left(\dfrac{e}{\hbar} {\bf \Omega}\cdot {\bf B}\right)^{2}
-\dfrac{e}{\hbar} {\bf \Omega} \cdot {\bf B} \left\{
\dfrac{e}{\hbar}\left(v_{i}B_{j}+v_{j}B_{i}\right){\bf v}\cdot {\bf \Omega}-\gamma \left(v_{i}v_{j}^{m}+v_{j}v_{i}^{m}\right)\right\}
\\
+\left(\frac{eB_{i}}{\hbar}\frac{eB_{j}}{\hbar}\left({\bf v}\cdot{\bf \Omega}\right)^{2}
-\gamma \left\{\frac{e}{\hbar}~\left(v_{i}^{m}B_{j}+v_{j}^{m}B_{i}\right)({\bf v}\cdot {\bf \Omega})
-\frac{e}{\hbar}~\left(v_{i}B_{j}+v_{j}B_{i}\right)({\bf v}^{m} \cdot{\bf \Omega})
+v_{i}^{m} v_{j}^{m}\right\}\right)\Big]
\left(-f_{0}^{\prime}\right)
\\
- \gamma \left[\dfrac{e}{\hbar}\left(v_{i}B_{j}+v_{j}B_{i}\right)({\bf v} \cdot {\bf \Omega})
-\left(v_{i}v_{j}^{m}+v_{j}v_{i}^{m}\right)
-v_{i}v_{j}\dfrac{e}{\hbar}  {\bf \Omega} \cdot {\bf B}\right]\left({\bf m}\cdot{\bf B}\right)\left(-f_{0}^{\prime \prime}\right)
~+~\frac{\gamma}{2} v_{i}v_{j}\left({\bf m}\cdot{\bf B}\right)^{2}\left(-f_{0}^{\prime \prime \prime}\right) \bigg)
\\
- \delta(p-1)\left(\gamma {\bf m}\cdot{\bf B}\right)^{p}\left(\left[\frac{e}{\hbar}\left(v_{i}B_{j}+v_{j}B_{i}\right)({\bf v}\cdot {\bf \Omega})
- \left(v_{i}v_{j}^{m}+v_{j} v_{i}^{m}\right)
-v_{i}v_{j}\frac{e}{\hbar}{\bf \Omega} \cdot{\bf B}~\right]
\left(-f_{0}^{\prime}\right)
-~v_{i}v_{j} \left({\bf m}\cdot {\bf B}\right)\left(-f_{0}^{\prime \prime}\right)\right) \Bigg].
\end{multline}
\end{widetext}
The last term in both Eqs.~\eqref{first_order} and \eqref{second_order} only contributes to $\alpha_{ij}$ ($p=1$). 
As an additional consistency check, it is straight forward to derive the Mott relations separately for linear-$B$ and quadratic-$B$ terms using Eqs.~\eqref{first_order} and \eqref{second_order}.

\section{Drude conductivities}
\label{Drude conductivities}
In this section we calculate Drude conductivities of tilted WSMs\cite{Carbotte16, Kamal19}.
The diagonal components of conductivity in the absence of a magnetic field are called the Drude conductivities. For the type-I class, Drude conductivity is given by $\sigma_{xx}^{(0)}=\sigma_{yy}^{(0)}$, where
\be\label{sigma_xx_0_I}
\sigma_{xx}^{(0)}=\sum_s \dfrac{3\sigma_{\rm D}^{0}}{4R_s^{3}}\left[\dfrac{2R_s}{(1-R_s^{2})} + \ln\left(\dfrac{1-R_s}{1+R_s}\right)\right]~.
\ee
Drude conductivity along the $z$ direction is given by
\be \label{sigma_zz_0_I}
\sigma_{zz}^{(0)}=\sum_s\dfrac{3\sigma_{\rm D}^{0}}{2R_s^{3}}\left[-2R_s -\ln\left(\dfrac{1-R_s}{1+R_s}\right) \right]~.
\ee
In the limit $R_{s} \to 0$ (ideal WSM), the Drude conductivity is equal in all three directions and is given by
\be \label{drude_iso}
\sigma_{\rm D}^0=\frac{4 \pi}{3}\dfrac{e^{2}}{h} \dfrac{\mu^{2} \tau}{h^{2} v_{F}} ~.
\ee
For the type-II class a finite cutoff in momentum space ($\Lambda_{k}$) is unavoidable to calculate the conductivities. This determines  the Drude conductivity along the $x$ direction as
\be\label{sigma_xx_0_II}
\sigma_{xx}^{(0)}=\sum_{s}\dfrac{3\sigma_{\rm D}^{0}}{4|R_{s}|^{3}}\left[\dfrac{3-R_{s}^{2}}{R_{s}^{2}-1}
+ \left( R_{s}^{2} -1\right) \tilde{\Lambda}_k^{2} -\delta_{s}^{1}
\right]~.
\ee
The same along the $z$ direction is given by
\be\label{sigma_zz_0_II}
\sigma_{zz}^{(0)}=\sum_{s}\dfrac{3\sigma_{\rm D}^{0}}{2|R_{s}|^{3}}\left[3-R_{s}^{2}
+  \left( R_{s}^{2} -1\right)^2\tilde{\Lambda}_k^{2} +\delta_{s}^{1} \right]~.
\ee
These expressions of Drude conductivities are exact as there are no approximations due to large cutoff.

\begin{figure}[t]
\includegraphics[width=.95\linewidth]{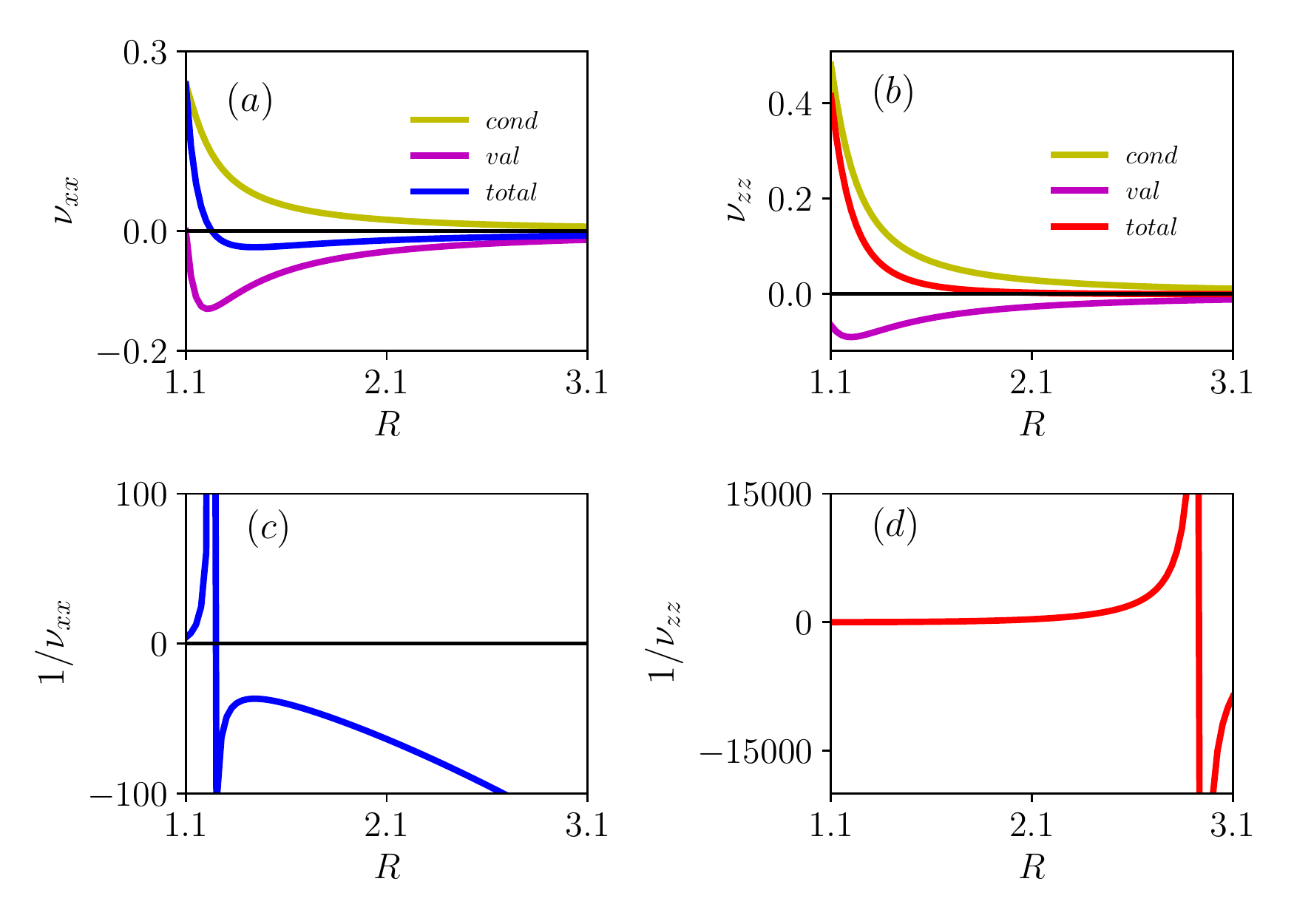}
\caption{(a) The Drude ($B=0$) component of the SC ($\nu_{xx}$), showing the contribution of the conduction and the valance bands separately. 
It reverses sign from positive to negative, as the contribution from the holes (valance band states) starts to dominate. 
(b) The Drude component of the SC ($\nu_{zz}$) highlighting the contribution from the different bands. All the components are scaled by the isotropic Drude counterpart, $\nu_{\rm D}$. (c) and (d) show the flip in sign of the inverse of the Drude components $1/\nu_{xx}$, and $1/\nu_{xx}$, which is also reflected in Fig.~\ref{fig.66}(b).} 
\label{fig.drude}
\end{figure}

Now we discuss the Drude thermopower. The Drude thermopower for isotropic WSM, using Eqs. \eqref{thermopower_tensor} and \eqref{drude_iso}, calculated to be
\be
\nu_{\rm D}^0 = -\dfrac{2 \pi^2}{3} \dfrac{k_B }{e} \dfrac{k_B T}{\mu}~.
\ee
Note that for a constant relaxation time the Drude SC is scattering time independent. It is evident that for $\mu>0$, the Drude coefficient is negative and for $\mu<0$ it is positive, showing the electron and hole type of carriers respectively. 
For the tilted type-I class, the Drude coefficient is identical to the isotropic one 
but for the type-II class, since the Fermi energy comes along the cutoff, we find considerable effect due to tilt.
As in type-II WSMs both the bands contributes and the contributions are opposite in nature, we expect the Drude SC to be zero for equal contribution, which we get in our calculation for large tilt factor limit as shown in Fig.~\ref{fig.drude}. We have shown the contribution of both the bands. For the $xx$ components, the valence band dominates in flow of entropy [see Fig.~\ref{fig.drude}(a)] whereas for $zz$ components the conduction band dominates [see Fig.~\ref{fig.drude}(b)].

\section{Expressions with $\gamma$}
\label{expre_gamma}
\textit{Type-I WSM.--} For ${\bf B} \perp \hat{\bf z}$, the conductivities for the type-I class are given by
\bea
\sigma^{(2)}_{\perp}&=& \sum_s  (1-3\gamma)\sigma_{0}~,\\ 
\Delta \sigma^{(2)} & = & \sum_s \left[7+13R_{s}^{2}-\gamma\left(1 + 6R_{s}^{2}\right)\right]\sigma_{0}~, \\ 
\sigma_{\rm z}^{(2)}&=& \sum_s  \left[1+7R_{s}^{2}-\gamma\left(3-R_{s}^{2}\right)\right]\sigma_{0}~.
\eea
Here, the factor $\gamma = 1~ (0)$ explicitly keeps track of the terms arising from the presence (absence) of the OMM\cite{Marco15}. The corrections due to the OMM (the $\gamma$ dependent terms) tend to suppress the conductivities. Most importantly the inclusion of the OMM in the conductivity changes the sign of $\sigma_{\perp}^{(2)}$. The linear-$B$ correction to the  transverse conductivity is given by
\bea \label{sigma_t}
\sigma_{\rm t}^{(1)} &=&   \sum_{s}\dfrac{s\sigma_{1}}{6R_{s}^{4}}\big[2R_{s} \left\{(3 - 2 R_s^2)(\gamma-1) + 3 R_s^2 (1- 2R_{s}^{2})\right\} \nn \\
& &  + 3\mathcal{F} \delta_{s}  (\gamma-1 + R_s^2) \big]~. 
\eea

For ${\bf B} \parallel \hat{\bf z}$, the quadratic corrections are given by 
\be 
\sigma_{\rm l}^{(2)} = \sum_s \left(1 - 3\gamma \right)\sigma_0; ~~\sigma_{\rm lz}^{(2)} = \sum_s \left[8+\gamma \left( 5R_s^2 -4\right) \right]\sigma_0~.
\ee
The linear-$B$ term in $\sigma_{xx} = \sigma_{yy}$ is given by 
\bea\nn
\sigma_{\rm l}^{(1)} &=& \sum_{s}\dfrac{s\sigma_{1}}{6R_{s}^{4}}\Big[2R_s\left\{\left(3-2R_s^2\right)(\gamma-1) -6\gamma R_s^2\right\}
\\
&&-3\delta_s\left\{\left(R_s^2 -1 \right)(\gamma -1) + 2\gamma R_s^2 \right\}\Big]~.
\eea
For $\sigma_{zz}$, the linear-$B$ correction is given by 
\bea\nn
\sigma_{\rm l z}^{(1)} &=& \sum_{s}\dfrac{s\sigma_1}{3R_s^4}\Big[ 2R_s\left\{\left(3 - 5R_s^2\right)(1 - \gamma) - 3R_s^4\right\}
\\
&&+3\mathcal{F}^2\delta_{s}(1 - \gamma)\Big]~.
\eea

\textit{Type-II WSM.--} First, we will consider the planar geometry (${\bf B} \perp \hat{\bf z}$).
In this case the form of the conductivity matrix is given by Eq.~\eqref{charge_tensor_aniso}, and the elements of the conductivity matrix are given by 
\bea
\Delta \sigma^{(2)}&=& \sum_s \mathcal{K} \left(\mathcal{A}_{\mathcal{R}}-\gamma \mathcal{A}_{\mathcal{M}}\right)~, 
\\
\sigma^{(2)}_{\perp}&=& \sum_s \mathcal{K}\left(\mathcal{B}_{\mathcal{R}}-\gamma \mathcal{B}_{\mathcal{M}}\right)~,
\\ 
\sigma_{\rm z}^{(2)}&=&2  \sum_s \mathcal{K} \left(\mathcal{D}_{\mathcal{R}}-\gamma \mathcal{D}_{\mathcal{M}}\right)~.
\eea
where $\mathcal{K} \equiv \frac{\sigma_{0}}{16|R_{s}|^{5}}$.
For the planar components (responses in the $x$-$y$ plane) of the conductivity, we have defined the following polynomials of $R_s$:
\bea 
\mathcal{A}_{\mathcal{R}}&=&2\left(1-R_{s}^{2}+5R_{s}^{4}+125R_{s}^{6}+30R_{s}^{8}\right),
\\
\mathcal{A}_{\mathcal{M}}&=&2\left(3+8R_{s}^{2}-45R_{s}^{4}+90R_{s}^{6}\right),
\\
\mathcal{B}_{\mathcal{R}}&=&\left( 1-5R_{s}^{2}+15R_{s}^{4}+5R_{s}^{6}\right),
\\
\mathcal{B}_{\mathcal{M}}&=&3\left(1-10R_{s}^{2}+25R_{s}^{4}\right).
\eea
For the $\sigma_{zz}$ component, we have defined the following polynomials of $R_s$:
\bea
\mathcal{D}_{\mathcal{R}}&=&\left(-2+11R_{s}^{2}-25R_{s}^{4}+65R_{s}^{6}+15R_{s}^{8}\right),\\
\mathcal{D}_{\mathcal{M}}&=&2\left(-3+6R_{s}^{2}+5R_{s}^{4}\right).
\eea
The linear-$B$ correction in the out-of-plane off-diagonal conductivities can be written as $\sigma_{xz}^{(1)}=\sigma_{yz}^{(1)}(\pi/2 -\phi)=\sigma_{\rm t}^{(1)}\cos\phi$. Here, 
\bea \nn
\sigma_{\rm t}^{(1)} &=& \sum_{s}\dfrac{s\sigma_1}{6R_s^4}{\rm sgn}(R_{s})\Big[\left(11-24R_s^2 + 21R_s^4\right)(\gamma - 1) 
\\
& + & 3\gamma\left(2 + R_s^2 - 5R_s^4\right) - 3\mathcal{F}\delta_{s}^{1}(\gamma  -1 + R_s^2)\Big]~. 
\eea
Now, we consider a magnetic field along the direction of the tilt (${\bf B} \parallel \hat{\bf z}$). 
The linear-$B$ correction to the longitudinal component in the $x/y$ plane, $\sigma_{xx} = \sigma_{yy}$, is given by 
\bea \nn
\sigma_{\rm l}^{(1)} &=& \sum_{s}\dfrac{s\sigma_{1}}{6R_{s}^{4}}{\rm sgn}(R_{s})\Big[(11-9R_s^2)(\gamma-1) - 6\gamma (3R_s^2-1)
\\
&&-3\delta_{s}^{1} \{(R_s^2 - 1) (1- \gamma) -2\gamma R_s^2\}
\Big]~.
\eea
The linear-$B$ correction to $\sigma_{zz}$ is given by 
\bea\nn
\sigma_{\rm l z}^{(1)} &=& \sum_{s}\dfrac{s\sigma_{1}}{3R_{s}^{4}}{\rm sgn}(R_{s})\Big[\left(17 - 27R_s^2 + 6R_s^4\right)(1-\gamma )
\\
&& - 6(1 - 2R_s^2 + 2R_s^4) + 3\delta_s^1\mathcal{F}^2 (\gamma -1)
\Big]~.
\eea
The quadratic-$B$ correction to $\sigma_{xx}$ and $\sigma_{yy}$ is given by 
\be
\sigma_{\rm l}^{(2)}=\sum_{s}\dfrac{\sigma_{0}}{8|R_{s}|^{5}}
\left[\left(-2+5R_{s}^{2}+5R_{s}^{6}\right)+6\gamma\left(1-5R_{s}^2\right)\right].
\ee
The corresponding term for the $\sigma_{zz}$ component is given by 
\bea
\sigma_{\rm l z}^{(2)}&=&\sum_{s}\dfrac{\sigma_{0}}{2|R_{s}|^{5}}
\Big[\left(1 - 5R_{s}^{2} + 15R_{s}^{4} + 5R_{s}^{6}\right)\\
&&+\gamma \left(-3 + 10R_s^2 - 20R_s^4 + 15R_s^6\right)\Big].
\eea 

\bibliography{PNE_Weyl_v1.bib}

\begin{thebibliography}{83}%
\makeatletter
\providecommand \@ifxundefined [1]{%
 \@ifx{#1\undefined}
}%
\providecommand \@ifnum [1]{%
 \ifnum #1\expandafter \@firstoftwo
 \else \expandafter \@secondoftwo
 \fi
}%
\providecommand \@ifx [1]{%
 \ifx #1\expandafter \@firstoftwo
 \else \expandafter \@secondoftwo
 \fi
}%
\providecommand \natexlab [1]{#1}%
\providecommand \enquote  [1]{``#1''}%
\providecommand \bibnamefont  [1]{#1}%
\providecommand \bibfnamefont [1]{#1}%
\providecommand \citenamefont [1]{#1}%
\providecommand \href@noop [0]{\@secondoftwo}%
\providecommand \href [0]{\begingroup \@sanitize@url \@href}%
\providecommand \@href[1]{\@@startlink{#1}\@@href}%
\providecommand \@@href[1]{\endgroup#1\@@endlink}%
\providecommand \@sanitize@url [0]{\catcode `\\12\catcode `\$12\catcode
  `\&12\catcode `\#12\catcode `\^12\catcode `\_12\catcode `\%12\relax}%
\providecommand \@@startlink[1]{}%
\providecommand \@@endlink[0]{}%
\providecommand \url  [0]{\begingroup\@sanitize@url \@url }%
\providecommand \@url [1]{\endgroup\@href {#1}{\urlprefix }}%
\providecommand \urlprefix  [0]{URL }%
\providecommand \Eprint [0]{\href }%
\providecommand \doibase [0]{http://dx.doi.org/}%
\providecommand \selectlanguage [0]{\@gobble}%
\providecommand \bibinfo  [0]{\@secondoftwo}%
\providecommand \bibfield  [0]{\@secondoftwo}%
\providecommand \translation [1]{[#1]}%
\providecommand \BibitemOpen [0]{}%
\providecommand \bibitemStop [0]{}%
\providecommand \bibitemNoStop [0]{.\EOS\space}%
\providecommand \EOS [0]{\spacefactor3000\relax}%
\providecommand \BibitemShut  [1]{\csname bibitem#1\endcsname}%
\let\auto@bib@innerbib\@empty
\bibitem [{\citenamefont {Nielsen}\ and\ \citenamefont
  {Ninomiya}(1981)}]{Nielsen81}%
  \BibitemOpen
  \bibfield  {author} {\bibinfo {author} {\bibfnamefont {H.}~\bibnamefont
  {Nielsen}}\ and\ \bibinfo {author} {\bibfnamefont {M.}~\bibnamefont
  {Ninomiya}},\ }\href {\doibase 10.1016/0370-2693(81)91026-1} {\bibfield
  {journal} {\bibinfo  {journal} {Physics Letters B}\ }\textbf {\bibinfo
  {volume} {105}},\ \bibinfo {pages} {219 } (\bibinfo {year}
  {1981})}\BibitemShut {NoStop}%
\bibitem [{\citenamefont {Armitage}\ \emph {et~al.}(2018)\citenamefont
  {Armitage}, \citenamefont {Mele},\ and\ \citenamefont
  {Vishwanath}}]{Ashvin18}%
  \BibitemOpen
  \bibfield  {author} {\bibinfo {author} {\bibfnamefont {N.~P.}\ \bibnamefont
  {Armitage}}, \bibinfo {author} {\bibfnamefont {E.~J.}\ \bibnamefont {Mele}},
  \ and\ \bibinfo {author} {\bibfnamefont {A.}~\bibnamefont {Vishwanath}},\
  }\href {\doibase 10.1103/RevModPhys.90.015001} {\bibfield  {journal}
  {\bibinfo  {journal} {Rev. Mod. Phys.}\ }\textbf {\bibinfo {volume} {90}},\
  \bibinfo {pages} {015001} (\bibinfo {year} {2018})}\BibitemShut {NoStop}%
\bibitem [{\citenamefont {Bansil}\ \emph {et~al.}(2016)\citenamefont {Bansil},
  \citenamefont {Lin},\ and\ \citenamefont {Das}}]{Bansil16}%
  \BibitemOpen
  \bibfield  {author} {\bibinfo {author} {\bibfnamefont {A.}~\bibnamefont
  {Bansil}}, \bibinfo {author} {\bibfnamefont {H.}~\bibnamefont {Lin}}, \ and\
  \bibinfo {author} {\bibfnamefont {T.}~\bibnamefont {Das}},\ }\href {\doibase
  10.1103/RevModPhys.88.021004} {\bibfield  {journal} {\bibinfo  {journal}
  {Rev. Mod. Phys.}\ }\textbf {\bibinfo {volume} {88}},\ \bibinfo {pages}
  {021004} (\bibinfo {year} {2016})}\BibitemShut {NoStop}%
\bibitem [{\citenamefont {Yan}\ and\ \citenamefont
  {Felser}(2017)}]{Claudia_rev17}%
  \BibitemOpen
  \bibfield  {author} {\bibinfo {author} {\bibfnamefont {B.}~\bibnamefont
  {Yan}}\ and\ \bibinfo {author} {\bibfnamefont {C.}~\bibnamefont {Felser}},\
  }\href {\doibase 10.1146/annurev-conmatphys-031016-025458} {\bibfield
  {journal} {\bibinfo  {journal} {Annual Review of Condensed Matter Physics}\
  }\textbf {\bibinfo {volume} {8}},\ \bibinfo {pages} {337} (\bibinfo {year}
  {2017})}\BibitemShut {NoStop}%
\bibitem [{\citenamefont {Kim}\ \emph {et~al.}(2013)\citenamefont {Kim},
  \citenamefont {Kim}, \citenamefont {Wang}, \citenamefont {Sasaki},
  \citenamefont {Satoh}, \citenamefont {Ohnishi}, \citenamefont {Kitaura},
  \citenamefont {Yang},\ and\ \citenamefont {Li}}]{Kim13}%
  \BibitemOpen
  \bibfield  {author} {\bibinfo {author} {\bibfnamefont {H.-J.}\ \bibnamefont
  {Kim}}, \bibinfo {author} {\bibfnamefont {K.-S.}\ \bibnamefont {Kim}},
  \bibinfo {author} {\bibfnamefont {J.-F.}\ \bibnamefont {Wang}}, \bibinfo
  {author} {\bibfnamefont {M.}~\bibnamefont {Sasaki}}, \bibinfo {author}
  {\bibfnamefont {N.}~\bibnamefont {Satoh}}, \bibinfo {author} {\bibfnamefont
  {A.}~\bibnamefont {Ohnishi}}, \bibinfo {author} {\bibfnamefont
  {M.}~\bibnamefont {Kitaura}}, \bibinfo {author} {\bibfnamefont
  {M.}~\bibnamefont {Yang}}, \ and\ \bibinfo {author} {\bibfnamefont
  {L.}~\bibnamefont {Li}},\ }\href {\doibase 10.1103/PhysRevLett.111.246603}
  {\bibfield  {journal} {\bibinfo  {journal} {Phys. Rev. Lett.}\ }\textbf
  {\bibinfo {volume} {111}},\ \bibinfo {pages} {246603} (\bibinfo {year}
  {2013})}\BibitemShut {NoStop}%
\bibitem [{\citenamefont {Xu}\ \emph {et~al.}(2015{\natexlab{a}})\citenamefont
  {Xu}, \citenamefont {Belopolski}, \citenamefont {Alidoust}, \citenamefont
  {Neupane}, \citenamefont {Bian}, \citenamefont {Zhang}, \citenamefont
  {Sankar}, \citenamefont {Chang}, \citenamefont {Yuan}, \citenamefont {Lee},
  \citenamefont {Huang}, \citenamefont {Zheng}, \citenamefont {Ma},
  \citenamefont {Sanchez}, \citenamefont {Wang}, \citenamefont {Bansil},
  \citenamefont {Chou}, \citenamefont {Shibayev}, \citenamefont {Lin},
  \citenamefont {Jia},\ and\ \citenamefont {Hasan}}]{xu15a}%
  \BibitemOpen
  \bibfield  {author} {\bibinfo {author} {\bibfnamefont {S.-Y.}\ \bibnamefont
  {Xu}}, \bibinfo {author} {\bibfnamefont {I.}~\bibnamefont {Belopolski}},
  \bibinfo {author} {\bibfnamefont {N.}~\bibnamefont {Alidoust}}, \bibinfo
  {author} {\bibfnamefont {M.}~\bibnamefont {Neupane}}, \bibinfo {author}
  {\bibfnamefont {G.}~\bibnamefont {Bian}}, \bibinfo {author} {\bibfnamefont
  {C.}~\bibnamefont {Zhang}}, \bibinfo {author} {\bibfnamefont
  {R.}~\bibnamefont {Sankar}}, \bibinfo {author} {\bibfnamefont
  {G.}~\bibnamefont {Chang}}, \bibinfo {author} {\bibfnamefont
  {Z.}~\bibnamefont {Yuan}}, \bibinfo {author} {\bibfnamefont {C.-C.}\
  \bibnamefont {Lee}}, \bibinfo {author} {\bibfnamefont {S.-M.}\ \bibnamefont
  {Huang}}, \bibinfo {author} {\bibfnamefont {H.}~\bibnamefont {Zheng}},
  \bibinfo {author} {\bibfnamefont {J.}~\bibnamefont {Ma}}, \bibinfo {author}
  {\bibfnamefont {D.~S.}\ \bibnamefont {Sanchez}}, \bibinfo {author}
  {\bibfnamefont {B.}~\bibnamefont {Wang}}, \bibinfo {author} {\bibfnamefont
  {A.}~\bibnamefont {Bansil}}, \bibinfo {author} {\bibfnamefont
  {F.}~\bibnamefont {Chou}}, \bibinfo {author} {\bibfnamefont {P.~P.}\
  \bibnamefont {Shibayev}}, \bibinfo {author} {\bibfnamefont {H.}~\bibnamefont
  {Lin}}, \bibinfo {author} {\bibfnamefont {S.}~\bibnamefont {Jia}}, \ and\
  \bibinfo {author} {\bibfnamefont {M.~Z.}\ \bibnamefont {Hasan}},\ }\href
  {\doibase 10.1126/science.aaa9297} {\bibfield  {journal} {\bibinfo  {journal}
  {Science}\ }\textbf {\bibinfo {volume} {349}},\ \bibinfo {pages} {613}
  (\bibinfo {year} {2015}{\natexlab{a}})}\BibitemShut {NoStop}%
\bibitem [{\citenamefont {Xiong}\ \emph {et~al.}(2015)\citenamefont {Xiong},
  \citenamefont {Kushwaha}, \citenamefont {Liang}, \citenamefont {Krizan},
  \citenamefont {Hirschberger}, \citenamefont {Wang}, \citenamefont {Cava},\
  and\ \citenamefont {Ong}}]{Xiong15}%
  \BibitemOpen
  \bibfield  {author} {\bibinfo {author} {\bibfnamefont {J.}~\bibnamefont
  {Xiong}}, \bibinfo {author} {\bibfnamefont {S.~K.}\ \bibnamefont {Kushwaha}},
  \bibinfo {author} {\bibfnamefont {T.}~\bibnamefont {Liang}}, \bibinfo
  {author} {\bibfnamefont {J.~W.}\ \bibnamefont {Krizan}}, \bibinfo {author}
  {\bibfnamefont {M.}~\bibnamefont {Hirschberger}}, \bibinfo {author}
  {\bibfnamefont {W.}~\bibnamefont {Wang}}, \bibinfo {author} {\bibfnamefont
  {R.~J.}\ \bibnamefont {Cava}}, \ and\ \bibinfo {author} {\bibfnamefont
  {N.~P.}\ \bibnamefont {Ong}},\ }\href {\doibase 10.1126/science.aac6089}
  {\bibfield  {journal} {\bibinfo  {journal} {Science}\ }\textbf {\bibinfo
  {volume} {350}},\ \bibinfo {pages} {413} (\bibinfo {year}
  {2015})}\BibitemShut {NoStop}%
\bibitem [{\citenamefont {Lv}\ \emph {et~al.}(2015)\citenamefont {Lv},
  \citenamefont {Weng}, \citenamefont {Fu}, \citenamefont {Wang}, \citenamefont
  {Miao}, \citenamefont {Ma}, \citenamefont {Richard}, \citenamefont {Huang},
  \citenamefont {Zhao}, \citenamefont {Chen}, \citenamefont {Fang},
  \citenamefont {Dai}, \citenamefont {Qian},\ and\ \citenamefont
  {Ding}}]{Lv15}%
  \BibitemOpen
  \bibfield  {author} {\bibinfo {author} {\bibfnamefont {B.~Q.}\ \bibnamefont
  {Lv}}, \bibinfo {author} {\bibfnamefont {H.~M.}\ \bibnamefont {Weng}},
  \bibinfo {author} {\bibfnamefont {B.~B.}\ \bibnamefont {Fu}}, \bibinfo
  {author} {\bibfnamefont {X.~P.}\ \bibnamefont {Wang}}, \bibinfo {author}
  {\bibfnamefont {H.}~\bibnamefont {Miao}}, \bibinfo {author} {\bibfnamefont
  {J.}~\bibnamefont {Ma}}, \bibinfo {author} {\bibfnamefont {P.}~\bibnamefont
  {Richard}}, \bibinfo {author} {\bibfnamefont {X.~C.}\ \bibnamefont {Huang}},
  \bibinfo {author} {\bibfnamefont {L.~X.}\ \bibnamefont {Zhao}}, \bibinfo
  {author} {\bibfnamefont {G.~F.}\ \bibnamefont {Chen}}, \bibinfo {author}
  {\bibfnamefont {Z.}~\bibnamefont {Fang}}, \bibinfo {author} {\bibfnamefont
  {X.}~\bibnamefont {Dai}}, \bibinfo {author} {\bibfnamefont {T.}~\bibnamefont
  {Qian}}, \ and\ \bibinfo {author} {\bibfnamefont {H.}~\bibnamefont {Ding}},\
  }\href {\doibase 10.1103/PhysRevX.5.031013} {\bibfield  {journal} {\bibinfo
  {journal} {Phys. Rev. X}\ }\textbf {\bibinfo {volume} {5}},\ \bibinfo {pages}
  {031013} (\bibinfo {year} {2015})}\BibitemShut {NoStop}%
\bibitem [{\citenamefont {Xu}\ \emph {et~al.}(2015{\natexlab{b}})\citenamefont
  {Xu}, \citenamefont {Alidoust}, \citenamefont {Belopolski}, \citenamefont
  {Yuan}, \citenamefont {Bian}, \citenamefont {Chang}, \citenamefont {Zheng},
  \citenamefont {Strocov}, \citenamefont {Sanchez}, \citenamefont {Chang},
  \citenamefont {Zhang}, \citenamefont {Mou}, \citenamefont {Wu}, \citenamefont
  {Huang}, \citenamefont {Lee}, \citenamefont {Huang}, \citenamefont {Wang},
  \citenamefont {Bansil}, \citenamefont {Jeng}, \citenamefont {Neupert},
  \citenamefont {Kaminski}, \citenamefont {Lin}, \citenamefont {Jia},\ and\
  \citenamefont {Zahid~Hasan}}]{xu15b}%
  \BibitemOpen
  \bibfield  {author} {\bibinfo {author} {\bibfnamefont {S.-Y.}\ \bibnamefont
  {Xu}}, \bibinfo {author} {\bibfnamefont {N.}~\bibnamefont {Alidoust}},
  \bibinfo {author} {\bibfnamefont {I.}~\bibnamefont {Belopolski}}, \bibinfo
  {author} {\bibfnamefont {Z.}~\bibnamefont {Yuan}}, \bibinfo {author}
  {\bibfnamefont {G.}~\bibnamefont {Bian}}, \bibinfo {author} {\bibfnamefont
  {T.-R.}\ \bibnamefont {Chang}}, \bibinfo {author} {\bibfnamefont
  {H.}~\bibnamefont {Zheng}}, \bibinfo {author} {\bibfnamefont {V.~N.}\
  \bibnamefont {Strocov}}, \bibinfo {author} {\bibfnamefont {D.~S.}\
  \bibnamefont {Sanchez}}, \bibinfo {author} {\bibfnamefont {G.}~\bibnamefont
  {Chang}}, \bibinfo {author} {\bibfnamefont {C.}~\bibnamefont {Zhang}},
  \bibinfo {author} {\bibfnamefont {D.}~\bibnamefont {Mou}}, \bibinfo {author}
  {\bibfnamefont {Y.}~\bibnamefont {Wu}}, \bibinfo {author} {\bibfnamefont
  {L.}~\bibnamefont {Huang}}, \bibinfo {author} {\bibfnamefont {C.-C.}\
  \bibnamefont {Lee}}, \bibinfo {author} {\bibfnamefont {S.-M.}\ \bibnamefont
  {Huang}}, \bibinfo {author} {\bibfnamefont {B.}~\bibnamefont {Wang}},
  \bibinfo {author} {\bibfnamefont {A.}~\bibnamefont {Bansil}}, \bibinfo
  {author} {\bibfnamefont {H.-T.}\ \bibnamefont {Jeng}}, \bibinfo {author}
  {\bibfnamefont {T.}~\bibnamefont {Neupert}}, \bibinfo {author} {\bibfnamefont
  {A.}~\bibnamefont {Kaminski}}, \bibinfo {author} {\bibfnamefont
  {H.}~\bibnamefont {Lin}}, \bibinfo {author} {\bibfnamefont {S.}~\bibnamefont
  {Jia}}, \ and\ \bibinfo {author} {\bibfnamefont {M.}~\bibnamefont
  {Zahid~Hasan}},\ }\href {http://dx.doi.org/10.1038/nphys3437} {\bibfield
  {journal} {\bibinfo  {journal} {Nature Physics}\ }\textbf {\bibinfo {volume}
  {11}},\ \bibinfo {pages} {748} (\bibinfo {year}
  {2015}{\natexlab{b}})}\BibitemShut {NoStop}%
\bibitem [{\citenamefont {Soluyanov}\ \emph {et~al.}(2015)\citenamefont
  {Soluyanov}, \citenamefont {Gresch}, \citenamefont {Wang}, \citenamefont
  {Wu}, \citenamefont {Troyer}, \citenamefont {Dai},\ and\ \citenamefont
  {Bernevig}}]{Soluyanov15}%
  \BibitemOpen
  \bibfield  {author} {\bibinfo {author} {\bibfnamefont {A.~A.}\ \bibnamefont
  {Soluyanov}}, \bibinfo {author} {\bibfnamefont {D.}~\bibnamefont {Gresch}},
  \bibinfo {author} {\bibfnamefont {Z.}~\bibnamefont {Wang}}, \bibinfo {author}
  {\bibfnamefont {Q.}~\bibnamefont {Wu}}, \bibinfo {author} {\bibfnamefont
  {M.}~\bibnamefont {Troyer}}, \bibinfo {author} {\bibfnamefont
  {X.}~\bibnamefont {Dai}}, \ and\ \bibinfo {author} {\bibfnamefont {B.~A.}\
  \bibnamefont {Bernevig}},\ }\href {http://dx.doi.org/10.1038/nature15768}
  {\bibfield  {journal} {\bibinfo  {journal} {Nature}\ }\textbf {\bibinfo
  {volume} {527}},\ \bibinfo {pages} {495} (\bibinfo {year}
  {2015})}\BibitemShut {NoStop}%
\bibitem [{\citenamefont {Aut\`es}\ \emph {et~al.}(2016)\citenamefont
  {Aut\`es}, \citenamefont {Gresch}, \citenamefont {Troyer}, \citenamefont
  {Soluyanov},\ and\ \citenamefont {Yazyev}}]{Yazyev16}%
  \BibitemOpen
  \bibfield  {author} {\bibinfo {author} {\bibfnamefont {G.}~\bibnamefont
  {Aut\`es}}, \bibinfo {author} {\bibfnamefont {D.}~\bibnamefont {Gresch}},
  \bibinfo {author} {\bibfnamefont {M.}~\bibnamefont {Troyer}}, \bibinfo
  {author} {\bibfnamefont {A.~A.}\ \bibnamefont {Soluyanov}}, \ and\ \bibinfo
  {author} {\bibfnamefont {O.~V.}\ \bibnamefont {Yazyev}},\ }\href {\doibase
  10.1103/PhysRevLett.117.066402} {\bibfield  {journal} {\bibinfo  {journal}
  {Phys. Rev. Lett.}\ }\textbf {\bibinfo {volume} {117}},\ \bibinfo {pages}
  {066402} (\bibinfo {year} {2016})}\BibitemShut {NoStop}%
\bibitem [{\citenamefont {Chang}\ \emph {et~al.}(2016)\citenamefont {Chang},
  \citenamefont {Xu}, \citenamefont {Chang}, \citenamefont {Lee}, \citenamefont
  {Huang}, \citenamefont {Wang}, \citenamefont {Bian}, \citenamefont {Zheng},
  \citenamefont {Sanchez}, \citenamefont {Belopolski}, \citenamefont
  {Alidoust}, \citenamefont {Neupane}, \citenamefont {Bansil}, \citenamefont
  {Jeng}, \citenamefont {Lin},\ and\ \citenamefont {Zahid~Hasan}}]{Chang16}%
  \BibitemOpen
  \bibfield  {author} {\bibinfo {author} {\bibfnamefont {T.-R.}\ \bibnamefont
  {Chang}}, \bibinfo {author} {\bibfnamefont {S.-Y.}\ \bibnamefont {Xu}},
  \bibinfo {author} {\bibfnamefont {G.}~\bibnamefont {Chang}}, \bibinfo
  {author} {\bibfnamefont {C.-C.}\ \bibnamefont {Lee}}, \bibinfo {author}
  {\bibfnamefont {S.-M.}\ \bibnamefont {Huang}}, \bibinfo {author}
  {\bibfnamefont {B.}~\bibnamefont {Wang}}, \bibinfo {author} {\bibfnamefont
  {G.}~\bibnamefont {Bian}}, \bibinfo {author} {\bibfnamefont {H.}~\bibnamefont
  {Zheng}}, \bibinfo {author} {\bibfnamefont {D.~S.}\ \bibnamefont {Sanchez}},
  \bibinfo {author} {\bibfnamefont {I.}~\bibnamefont {Belopolski}}, \bibinfo
  {author} {\bibfnamefont {N.}~\bibnamefont {Alidoust}}, \bibinfo {author}
  {\bibfnamefont {M.}~\bibnamefont {Neupane}}, \bibinfo {author} {\bibfnamefont
  {A.}~\bibnamefont {Bansil}}, \bibinfo {author} {\bibfnamefont {H.-T.}\
  \bibnamefont {Jeng}}, \bibinfo {author} {\bibfnamefont {H.}~\bibnamefont
  {Lin}}, \ and\ \bibinfo {author} {\bibfnamefont {M.}~\bibnamefont
  {Zahid~Hasan}},\ }\href {http://dx.doi.org/10.1038/ncomms10639} {\bibfield
  {journal} {\bibinfo  {journal} {Nature Communications}\ }\textbf {\bibinfo
  {volume} {7}},\ \bibinfo {pages} {10639} (\bibinfo {year}
  {2016})}\BibitemShut {NoStop}%
\bibitem [{\citenamefont {Li}\ \emph {et~al.}(2017{\natexlab{a}})\citenamefont
  {Li}, \citenamefont {Wen}, \citenamefont {He}, \citenamefont {Zhang},
  \citenamefont {Xia}, \citenamefont {Yu}, \citenamefont {Yang}, \citenamefont
  {Zhu}, \citenamefont {Alshareef},\ and\ \citenamefont {Zhang}}]{Pengli17}%
  \BibitemOpen
  \bibfield  {author} {\bibinfo {author} {\bibfnamefont {P.}~\bibnamefont
  {Li}}, \bibinfo {author} {\bibfnamefont {Y.}~\bibnamefont {Wen}}, \bibinfo
  {author} {\bibfnamefont {X.}~\bibnamefont {He}}, \bibinfo {author}
  {\bibfnamefont {Q.}~\bibnamefont {Zhang}}, \bibinfo {author} {\bibfnamefont
  {C.}~\bibnamefont {Xia}}, \bibinfo {author} {\bibfnamefont {Z.-M.}\
  \bibnamefont {Yu}}, \bibinfo {author} {\bibfnamefont {S.~A.}\ \bibnamefont
  {Yang}}, \bibinfo {author} {\bibfnamefont {Z.}~\bibnamefont {Zhu}}, \bibinfo
  {author} {\bibfnamefont {H.~N.}\ \bibnamefont {Alshareef}}, \ and\ \bibinfo
  {author} {\bibfnamefont {X.-X.}\ \bibnamefont {Zhang}},\ }\href {\doibase
  10.1038/s41467-017-02237-1} {\bibfield  {journal} {\bibinfo  {journal}
  {Nature Communications}\ }\textbf {\bibinfo {volume} {8}},\ \bibinfo {pages}
  {2150} (\bibinfo {year} {2017}{\natexlab{a}})}\BibitemShut {NoStop}%
\bibitem [{\citenamefont {Jiang}\ \emph {et~al.}(2017)\citenamefont {Jiang},
  \citenamefont {Liu}, \citenamefont {Sun}, \citenamefont {Yang}, \citenamefont
  {Rajamathi}, \citenamefont {Qi}, \citenamefont {Yang}, \citenamefont {Chen},
  \citenamefont {Peng}, \citenamefont {Hwang}, \citenamefont {Sun},
  \citenamefont {Mo}, \citenamefont {Vobornik}, \citenamefont {Fujii},
  \citenamefont {Parkin}, \citenamefont {Felser}, \citenamefont {Yan},\ and\
  \citenamefont {Chen}}]{Jiang17}%
  \BibitemOpen
  \bibfield  {author} {\bibinfo {author} {\bibfnamefont {J.}~\bibnamefont
  {Jiang}}, \bibinfo {author} {\bibfnamefont {Z.~K.}\ \bibnamefont {Liu}},
  \bibinfo {author} {\bibfnamefont {Y.}~\bibnamefont {Sun}}, \bibinfo {author}
  {\bibfnamefont {H.~F.}\ \bibnamefont {Yang}}, \bibinfo {author}
  {\bibfnamefont {C.~R.}\ \bibnamefont {Rajamathi}}, \bibinfo {author}
  {\bibfnamefont {Y.~P.}\ \bibnamefont {Qi}}, \bibinfo {author} {\bibfnamefont
  {L.~X.}\ \bibnamefont {Yang}}, \bibinfo {author} {\bibfnamefont
  {C.}~\bibnamefont {Chen}}, \bibinfo {author} {\bibfnamefont {H.}~\bibnamefont
  {Peng}}, \bibinfo {author} {\bibfnamefont {C.-C.}\ \bibnamefont {Hwang}},
  \bibinfo {author} {\bibfnamefont {S.~Z.}\ \bibnamefont {Sun}}, \bibinfo
  {author} {\bibfnamefont {S.-K.}\ \bibnamefont {Mo}}, \bibinfo {author}
  {\bibfnamefont {I.}~\bibnamefont {Vobornik}}, \bibinfo {author}
  {\bibfnamefont {J.}~\bibnamefont {Fujii}}, \bibinfo {author} {\bibfnamefont
  {S.~S.~P.}\ \bibnamefont {Parkin}}, \bibinfo {author} {\bibfnamefont
  {C.}~\bibnamefont {Felser}}, \bibinfo {author} {\bibfnamefont {B.~H.}\
  \bibnamefont {Yan}}, \ and\ \bibinfo {author} {\bibfnamefont {Y.~L.}\
  \bibnamefont {Chen}},\ }\href {http://dx.doi.org/10.1038/ncomms13973}
  {\bibfield  {journal} {\bibinfo  {journal} {Nature Communications}\ }\textbf
  {\bibinfo {volume} {8}},\ \bibinfo {pages} {13973} (\bibinfo {year}
  {2017})}\BibitemShut {NoStop}%
\bibitem [{\citenamefont {Xu}\ \emph {et~al.}(2017)\citenamefont {Xu},
  \citenamefont {Alidoust}, \citenamefont {Chang}, \citenamefont {Lu},
  \citenamefont {Singh}, \citenamefont {Belopolski}, \citenamefont {Sanchez},
  \citenamefont {Zhang}, \citenamefont {Bian}, \citenamefont {Zheng},
  \citenamefont {Husanu}, \citenamefont {Bian}, \citenamefont {Huang},
  \citenamefont {Hsu}, \citenamefont {Chang}, \citenamefont {Jeng},
  \citenamefont {Bansil}, \citenamefont {Neupert}, \citenamefont {Strocov},
  \citenamefont {Lin}, \citenamefont {Jia},\ and\ \citenamefont
  {Hasan}}]{Xu17}%
  \BibitemOpen
  \bibfield  {author} {\bibinfo {author} {\bibfnamefont {S.-Y.}\ \bibnamefont
  {Xu}}, \bibinfo {author} {\bibfnamefont {N.}~\bibnamefont {Alidoust}},
  \bibinfo {author} {\bibfnamefont {G.}~\bibnamefont {Chang}}, \bibinfo
  {author} {\bibfnamefont {H.}~\bibnamefont {Lu}}, \bibinfo {author}
  {\bibfnamefont {B.}~\bibnamefont {Singh}}, \bibinfo {author} {\bibfnamefont
  {I.}~\bibnamefont {Belopolski}}, \bibinfo {author} {\bibfnamefont {D.~S.}\
  \bibnamefont {Sanchez}}, \bibinfo {author} {\bibfnamefont {X.}~\bibnamefont
  {Zhang}}, \bibinfo {author} {\bibfnamefont {G.}~\bibnamefont {Bian}},
  \bibinfo {author} {\bibfnamefont {H.}~\bibnamefont {Zheng}}, \bibinfo
  {author} {\bibfnamefont {M.-A.}\ \bibnamefont {Husanu}}, \bibinfo {author}
  {\bibfnamefont {Y.}~\bibnamefont {Bian}}, \bibinfo {author} {\bibfnamefont
  {S.-M.}\ \bibnamefont {Huang}}, \bibinfo {author} {\bibfnamefont {C.-H.}\
  \bibnamefont {Hsu}}, \bibinfo {author} {\bibfnamefont {T.-R.}\ \bibnamefont
  {Chang}}, \bibinfo {author} {\bibfnamefont {H.-T.}\ \bibnamefont {Jeng}},
  \bibinfo {author} {\bibfnamefont {A.}~\bibnamefont {Bansil}}, \bibinfo
  {author} {\bibfnamefont {T.}~\bibnamefont {Neupert}}, \bibinfo {author}
  {\bibfnamefont {V.~N.}\ \bibnamefont {Strocov}}, \bibinfo {author}
  {\bibfnamefont {H.}~\bibnamefont {Lin}}, \bibinfo {author} {\bibfnamefont
  {S.}~\bibnamefont {Jia}}, \ and\ \bibinfo {author} {\bibfnamefont {M.~Z.}\
  \bibnamefont {Hasan}},\ }\href {\doibase 10.1126/sciadv.1603266} {\bibfield
  {journal} {\bibinfo  {journal} {Science Advances}\ }\textbf {\bibinfo
  {volume} {3}},\ \bibinfo {pages} {1603266} (\bibinfo {year}
  {2017})}\BibitemShut {NoStop}%
\bibitem [{\citenamefont {Pal}\ \emph {et~al.}(2018)\citenamefont {Pal},
  \citenamefont {Chinotti}, \citenamefont {Degiorgi}, \citenamefont {Ren},\
  and\ \citenamefont {Petrovic}}]{Pal18}%
  \BibitemOpen
  \bibfield  {author} {\bibinfo {author} {\bibfnamefont {A.}~\bibnamefont
  {Pal}}, \bibinfo {author} {\bibfnamefont {M.}~\bibnamefont {Chinotti}},
  \bibinfo {author} {\bibfnamefont {L.}~\bibnamefont {Degiorgi}}, \bibinfo
  {author} {\bibfnamefont {W.}~\bibnamefont {Ren}}, \ and\ \bibinfo {author}
  {\bibfnamefont {C.}~\bibnamefont {Petrovic}},\ }\href {\doibase
  10.1016/j.physb.2017.09.079} {\bibfield  {journal} {\bibinfo  {journal}
  {Physica B: Condensed Matter}\ }\textbf {\bibinfo {volume} {536}},\ \bibinfo
  {pages} {64 } (\bibinfo {year} {2018})}\BibitemShut {NoStop}%
\bibitem [{\citenamefont {Zhang}\ \emph {et~al.}(2018)\citenamefont {Zhang},
  \citenamefont {Yang},\ and\ \citenamefont {Wang}}]{Zhang18}%
  \BibitemOpen
  \bibfield  {author} {\bibinfo {author} {\bibfnamefont {M.}~\bibnamefont
  {Zhang}}, \bibinfo {author} {\bibfnamefont {Z.}~\bibnamefont {Yang}}, \ and\
  \bibinfo {author} {\bibfnamefont {G.}~\bibnamefont {Wang}},\ }\href {\doibase
  10.1021/acs.jpcc.8b00920} {\bibfield  {journal} {\bibinfo  {journal} {The
  Journal of Physical Chemistry C}\ }\textbf {\bibinfo {volume} {122}},\
  \bibinfo {pages} {3533} (\bibinfo {year} {2018})}\BibitemShut {NoStop}%
\bibitem [{\citenamefont {Chang}\ \emph {et~al.}(2018)\citenamefont {Chang},
  \citenamefont {Singh}, \citenamefont {Xu}, \citenamefont {Bian},
  \citenamefont {Huang}, \citenamefont {Hsu}, \citenamefont {Belopolski},
  \citenamefont {Alidoust}, \citenamefont {Sanchez}, \citenamefont {Zheng},
  \citenamefont {Lu}, \citenamefont {Zhang}, \citenamefont {Bian},
  \citenamefont {Chang}, \citenamefont {Jeng}, \citenamefont {Bansil},
  \citenamefont {Hsu}, \citenamefont {Jia}, \citenamefont {Neupert},
  \citenamefont {Lin},\ and\ \citenamefont {Hasan}}]{Chang18}%
  \BibitemOpen
  \bibfield  {author} {\bibinfo {author} {\bibfnamefont {G.}~\bibnamefont
  {Chang}}, \bibinfo {author} {\bibfnamefont {B.}~\bibnamefont {Singh}},
  \bibinfo {author} {\bibfnamefont {S.-Y.}\ \bibnamefont {Xu}}, \bibinfo
  {author} {\bibfnamefont {G.}~\bibnamefont {Bian}}, \bibinfo {author}
  {\bibfnamefont {S.-M.}\ \bibnamefont {Huang}}, \bibinfo {author}
  {\bibfnamefont {C.-H.}\ \bibnamefont {Hsu}}, \bibinfo {author} {\bibfnamefont
  {I.}~\bibnamefont {Belopolski}}, \bibinfo {author} {\bibfnamefont
  {N.}~\bibnamefont {Alidoust}}, \bibinfo {author} {\bibfnamefont {D.~S.}\
  \bibnamefont {Sanchez}}, \bibinfo {author} {\bibfnamefont {H.}~\bibnamefont
  {Zheng}}, \bibinfo {author} {\bibfnamefont {H.}~\bibnamefont {Lu}}, \bibinfo
  {author} {\bibfnamefont {X.}~\bibnamefont {Zhang}}, \bibinfo {author}
  {\bibfnamefont {Y.}~\bibnamefont {Bian}}, \bibinfo {author} {\bibfnamefont
  {T.-R.}\ \bibnamefont {Chang}}, \bibinfo {author} {\bibfnamefont {H.-T.}\
  \bibnamefont {Jeng}}, \bibinfo {author} {\bibfnamefont {A.}~\bibnamefont
  {Bansil}}, \bibinfo {author} {\bibfnamefont {H.}~\bibnamefont {Hsu}},
  \bibinfo {author} {\bibfnamefont {S.}~\bibnamefont {Jia}}, \bibinfo {author}
  {\bibfnamefont {T.}~\bibnamefont {Neupert}}, \bibinfo {author} {\bibfnamefont
  {H.}~\bibnamefont {Lin}}, \ and\ \bibinfo {author} {\bibfnamefont {M.~Z.}\
  \bibnamefont {Hasan}},\ }\href {\doibase 10.1103/PhysRevB.97.041104}
  {\bibfield  {journal} {\bibinfo  {journal} {Phys. Rev. B}\ }\textbf {\bibinfo
  {volume} {97}},\ \bibinfo {pages} {041104} (\bibinfo {year}
  {2018})}\BibitemShut {NoStop}%
\bibitem [{\citenamefont {Berry}(1984)}]{Berry84}%
  \BibitemOpen
  \bibfield  {author} {\bibinfo {author} {\bibfnamefont {M.~V.}\ \bibnamefont
  {Berry}},\ }\href {http://www.jstor.org/stable/2397741} {\bibfield  {journal}
  {\bibinfo  {journal} {Proceedings of the Royal Society of London. Series A,
  Mathematical and Physical Sciences}\ }\textbf {\bibinfo {volume} {392}},\
  \bibinfo {pages} {45} (\bibinfo {year} {1984})}\BibitemShut {NoStop}%
\bibitem [{\citenamefont {Xiao}\ \emph {et~al.}(2010)\citenamefont {Xiao},
  \citenamefont {Chang},\ and\ \citenamefont {Niu}}]{Xiao_Niu_rev10}%
  \BibitemOpen
  \bibfield  {author} {\bibinfo {author} {\bibfnamefont {D.}~\bibnamefont
  {Xiao}}, \bibinfo {author} {\bibfnamefont {M.-C.}\ \bibnamefont {Chang}}, \
  and\ \bibinfo {author} {\bibfnamefont {Q.}~\bibnamefont {Niu}},\ }\href
  {\doibase 10.1103/RevModPhys.82.1959} {\bibfield  {journal} {\bibinfo
  {journal} {Rev. Mod. Phys.}\ }\textbf {\bibinfo {volume} {82}},\ \bibinfo
  {pages} {1959} (\bibinfo {year} {2010})}\BibitemShut {NoStop}%
\bibitem [{\citenamefont {Nielsen}\ and\ \citenamefont
  {Ninomiya}(1983)}]{Nielsen83}%
  \BibitemOpen
  \bibfield  {author} {\bibinfo {author} {\bibfnamefont {H.~B.}\ \bibnamefont
  {Nielsen}}\ and\ \bibinfo {author} {\bibfnamefont {M.}~\bibnamefont
  {Ninomiya}},\ }\href
  {http://www.sciencedirect.com/science/article/pii/0370269383915290}
  {\bibfield  {journal} {\bibinfo  {journal} {Physics Letters B}\ }\textbf
  {\bibinfo {volume} {130}},\ \bibinfo {pages} {389} (\bibinfo {year}
  {1983})}\BibitemShut {NoStop}%
\bibitem [{\citenamefont {Son}\ and\ \citenamefont {Yamamoto}(2012)}]{Son12}%
  \BibitemOpen
  \bibfield  {author} {\bibinfo {author} {\bibfnamefont {D.~T.}\ \bibnamefont
  {Son}}\ and\ \bibinfo {author} {\bibfnamefont {N.}~\bibnamefont {Yamamoto}},\
  }\href {\doibase 10.1103/PhysRevLett.109.181602} {\bibfield  {journal}
  {\bibinfo  {journal} {Phys. Rev. Lett.}\ }\textbf {\bibinfo {volume} {109}},\
  \bibinfo {pages} {181602} (\bibinfo {year} {2012})}\BibitemShut {NoStop}%
\bibitem [{\citenamefont {Zyuzin}\ and\ \citenamefont
  {Burkov}(2012)}]{Zyuzin_Burkov12}%
  \BibitemOpen
  \bibfield  {author} {\bibinfo {author} {\bibfnamefont {A.~A.}\ \bibnamefont
  {Zyuzin}}\ and\ \bibinfo {author} {\bibfnamefont {A.~A.}\ \bibnamefont
  {Burkov}},\ }\href {\doibase 10.1103/PhysRevB.86.115133} {\bibfield
  {journal} {\bibinfo  {journal} {Phys. Rev. B}\ }\textbf {\bibinfo {volume}
  {86}},\ \bibinfo {pages} {115133} (\bibinfo {year} {2012})}\BibitemShut
  {NoStop}%
\bibitem [{\citenamefont {Son}\ and\ \citenamefont
  {Spivak}(2013)}]{Son_Spivak13}%
  \BibitemOpen
  \bibfield  {author} {\bibinfo {author} {\bibfnamefont {D.~T.}\ \bibnamefont
  {Son}}\ and\ \bibinfo {author} {\bibfnamefont {B.~Z.}\ \bibnamefont
  {Spivak}},\ }\href {\doibase 10.1103/PhysRevB.88.104412} {\bibfield
  {journal} {\bibinfo  {journal} {Phys. Rev. B}\ }\textbf {\bibinfo {volume}
  {88}},\ \bibinfo {pages} {104412} (\bibinfo {year} {2013})}\BibitemShut
  {NoStop}%
\bibitem [{\citenamefont {Hosur}\ and\ \citenamefont {Qi}(2013)}]{Hosur13}%
  \BibitemOpen
  \bibfield  {author} {\bibinfo {author} {\bibfnamefont {P.}~\bibnamefont
  {Hosur}}\ and\ \bibinfo {author} {\bibfnamefont {X.}~\bibnamefont {Qi}},\
  }\href {http://www.sciencedirect.com/science/article/pii/S1631070513001710}
  {\bibfield  {journal} {\bibinfo  {journal} {Comptes Rendus Physique}\
  }\textbf {\bibinfo {volume} {14}},\ \bibinfo {pages} {857} (\bibinfo {year}
  {2013})}\BibitemShut {NoStop}%
\bibitem [{\citenamefont {Kim}\ \emph {et~al.}(2014)\citenamefont {Kim},
  \citenamefont {Kim},\ and\ \citenamefont {Sasaki}}]{Kim14}%
  \BibitemOpen
  \bibfield  {author} {\bibinfo {author} {\bibfnamefont {K.-S.}\ \bibnamefont
  {Kim}}, \bibinfo {author} {\bibfnamefont {H.-J.}\ \bibnamefont {Kim}}, \ and\
  \bibinfo {author} {\bibfnamefont {M.}~\bibnamefont {Sasaki}},\ }\href
  {\doibase 10.1103/PhysRevB.89.195137} {\bibfield  {journal} {\bibinfo
  {journal} {Phys. Rev. B}\ }\textbf {\bibinfo {volume} {89}},\ \bibinfo
  {pages} {195137} (\bibinfo {year} {2014})}\BibitemShut {NoStop}%
\bibitem [{\citenamefont {Lundgren}\ \emph {et~al.}(2014)\citenamefont
  {Lundgren}, \citenamefont {Laurell},\ and\ \citenamefont {Fiete}}]{Fiete14}%
  \BibitemOpen
  \bibfield  {author} {\bibinfo {author} {\bibfnamefont {R.}~\bibnamefont
  {Lundgren}}, \bibinfo {author} {\bibfnamefont {P.}~\bibnamefont {Laurell}}, \
  and\ \bibinfo {author} {\bibfnamefont {G.~A.}\ \bibnamefont {Fiete}},\ }\href
  {\doibase 10.1103/PhysRevB.90.165115} {\bibfield  {journal} {\bibinfo
  {journal} {Phys. Rev. B}\ }\textbf {\bibinfo {volume} {90}},\ \bibinfo
  {pages} {165115} (\bibinfo {year} {2014})}\BibitemShut {NoStop}%
\bibitem [{\citenamefont {Zyuzin}\ and\ \citenamefont
  {Tiwari}(2016)}]{Zyuzin_A_A16}%
  \BibitemOpen
  \bibfield  {author} {\bibinfo {author} {\bibfnamefont {A.~A.}\ \bibnamefont
  {Zyuzin}}\ and\ \bibinfo {author} {\bibfnamefont {R.~P.}\ \bibnamefont
  {Tiwari}},\ }\href {\doibase 10.1134/S002136401611014X} {\bibfield  {journal}
  {\bibinfo  {journal} {JETP Letters}\ }\textbf {\bibinfo {volume} {103}},\
  \bibinfo {pages} {717} (\bibinfo {year} {2016})}\BibitemShut {NoStop}%
\bibitem [{\citenamefont {Sharma}\ \emph {et~al.}(2016)\citenamefont {Sharma},
  \citenamefont {Goswami},\ and\ \citenamefont {Tewari}}]{G_Sharma16}%
  \BibitemOpen
  \bibfield  {author} {\bibinfo {author} {\bibfnamefont {G.}~\bibnamefont
  {Sharma}}, \bibinfo {author} {\bibfnamefont {P.}~\bibnamefont {Goswami}}, \
  and\ \bibinfo {author} {\bibfnamefont {S.}~\bibnamefont {Tewari}},\ }\href
  {\doibase 10.1103/PhysRevB.93.035116} {\bibfield  {journal} {\bibinfo
  {journal} {Phys. Rev. B}\ }\textbf {\bibinfo {volume} {93}},\ \bibinfo
  {pages} {035116} (\bibinfo {year} {2016})}\BibitemShut {NoStop}%
\bibitem [{\citenamefont {van~der Wurff}\ and\ \citenamefont
  {Stoof}(2017)}]{Wurff17}%
  \BibitemOpen
  \bibfield  {author} {\bibinfo {author} {\bibfnamefont {E.~C.~I.}\
  \bibnamefont {van~der Wurff}}\ and\ \bibinfo {author} {\bibfnamefont
  {H.~T.~C.}\ \bibnamefont {Stoof}},\ }\href {\doibase
  10.1103/PhysRevB.96.121116} {\bibfield  {journal} {\bibinfo  {journal} {Phys.
  Rev. B}\ }\textbf {\bibinfo {volume} {96}},\ \bibinfo {pages} {121116}
  (\bibinfo {year} {2017})}\BibitemShut {NoStop}%
\bibitem [{\citenamefont {McCormick}\ \emph {et~al.}(2017)\citenamefont
  {McCormick}, \citenamefont {McKay},\ and\ \citenamefont
  {Trivedi}}]{Nandini17}%
  \BibitemOpen
  \bibfield  {author} {\bibinfo {author} {\bibfnamefont {T.~M.}\ \bibnamefont
  {McCormick}}, \bibinfo {author} {\bibfnamefont {R.~C.}\ \bibnamefont
  {McKay}}, \ and\ \bibinfo {author} {\bibfnamefont {N.}~\bibnamefont
  {Trivedi}},\ }\href {\doibase 10.1103/PhysRevB.96.235116} {\bibfield
  {journal} {\bibinfo  {journal} {Phys. Rev. B}\ }\textbf {\bibinfo {volume}
  {96}},\ \bibinfo {pages} {235116} (\bibinfo {year} {2017})}\BibitemShut
  {NoStop}%
\bibitem [{\citenamefont {Sekine}\ \emph {et~al.}(2017)\citenamefont {Sekine},
  \citenamefont {Culcer},\ and\ \citenamefont {MacDonald}}]{Sekine17}%
  \BibitemOpen
  \bibfield  {author} {\bibinfo {author} {\bibfnamefont {A.}~\bibnamefont
  {Sekine}}, \bibinfo {author} {\bibfnamefont {D.}~\bibnamefont {Culcer}}, \
  and\ \bibinfo {author} {\bibfnamefont {A.~H.}\ \bibnamefont {MacDonald}},\
  }\href {\doibase 10.1103/PhysRevB.96.235134} {\bibfield  {journal} {\bibinfo
  {journal} {Phys. Rev. B}\ }\textbf {\bibinfo {volume} {96}},\ \bibinfo
  {pages} {235134} (\bibinfo {year} {2017})}\BibitemShut {NoStop}%
\bibitem [{\citenamefont {Ferreiros}\ \emph {et~al.}(2017)\citenamefont
  {Ferreiros}, \citenamefont {Zyuzin},\ and\ \citenamefont
  {Bardarson}}]{Ferreiros17}%
  \BibitemOpen
  \bibfield  {author} {\bibinfo {author} {\bibfnamefont {Y.}~\bibnamefont
  {Ferreiros}}, \bibinfo {author} {\bibfnamefont {A.~A.}\ \bibnamefont
  {Zyuzin}}, \ and\ \bibinfo {author} {\bibfnamefont {J.~H.}\ \bibnamefont
  {Bardarson}},\ }\href {\doibase 10.1103/PhysRevB.96.115202} {\bibfield
  {journal} {\bibinfo  {journal} {Phys. Rev. B}\ }\textbf {\bibinfo {volume}
  {96}},\ \bibinfo {pages} {115202} (\bibinfo {year} {2017})}\BibitemShut
  {NoStop}%
\bibitem [{\citenamefont {Steiner}\ \emph {et~al.}(2017)\citenamefont
  {Steiner}, \citenamefont {Andreev},\ and\ \citenamefont {Pesin}}]{Steiner17}%
  \BibitemOpen
  \bibfield  {author} {\bibinfo {author} {\bibfnamefont {J.~F.}\ \bibnamefont
  {Steiner}}, \bibinfo {author} {\bibfnamefont {A.~V.}\ \bibnamefont
  {Andreev}}, \ and\ \bibinfo {author} {\bibfnamefont {D.~A.}\ \bibnamefont
  {Pesin}},\ }\href {\doibase 10.1103/PhysRevLett.119.036601} {\bibfield
  {journal} {\bibinfo  {journal} {Phys. Rev. Lett.}\ }\textbf {\bibinfo
  {volume} {119}},\ \bibinfo {pages} {036601} (\bibinfo {year}
  {2017})}\BibitemShut {NoStop}%
\bibitem [{\citenamefont {Watzman}\ \emph {et~al.}(2018)\citenamefont
  {Watzman}, \citenamefont {McCormick}, \citenamefont {Shekhar}, \citenamefont
  {Wu}, \citenamefont {Sun}, \citenamefont {Prakash}, \citenamefont {Felser},
  \citenamefont {Trivedi},\ and\ \citenamefont {Heremans}}]{Watzman18}%
  \BibitemOpen
  \bibfield  {author} {\bibinfo {author} {\bibfnamefont {S.~J.}\ \bibnamefont
  {Watzman}}, \bibinfo {author} {\bibfnamefont {T.~M.}\ \bibnamefont
  {McCormick}}, \bibinfo {author} {\bibfnamefont {C.}~\bibnamefont {Shekhar}},
  \bibinfo {author} {\bibfnamefont {S.-C.}\ \bibnamefont {Wu}}, \bibinfo
  {author} {\bibfnamefont {Y.}~\bibnamefont {Sun}}, \bibinfo {author}
  {\bibfnamefont {A.}~\bibnamefont {Prakash}}, \bibinfo {author} {\bibfnamefont
  {C.}~\bibnamefont {Felser}}, \bibinfo {author} {\bibfnamefont
  {N.}~\bibnamefont {Trivedi}}, \ and\ \bibinfo {author} {\bibfnamefont
  {J.~P.}\ \bibnamefont {Heremans}},\ }\href {\doibase
  10.1103/PhysRevB.97.161404} {\bibfield  {journal} {\bibinfo  {journal} {Phys.
  Rev. B}\ }\textbf {\bibinfo {volume} {97}},\ \bibinfo {pages} {161404}
  (\bibinfo {year} {2018})}\BibitemShut {NoStop}%
\bibitem [{\citenamefont {Desrat}\ \emph {et~al.}(2015)\citenamefont {Desrat},
  \citenamefont {Consejo}, \citenamefont {Teppe}, \citenamefont {Contreras},
  \citenamefont {Marcinkiewicz}, \citenamefont {Knap}, \citenamefont
  {Nateprov},\ and\ \citenamefont {Arushanov}}]{Desrat15}%
  \BibitemOpen
  \bibfield  {author} {\bibinfo {author} {\bibfnamefont {W.}~\bibnamefont
  {Desrat}}, \bibinfo {author} {\bibfnamefont {C.}~\bibnamefont {Consejo}},
  \bibinfo {author} {\bibfnamefont {F.}~\bibnamefont {Teppe}}, \bibinfo
  {author} {\bibfnamefont {S.}~\bibnamefont {Contreras}}, \bibinfo {author}
  {\bibfnamefont {M.}~\bibnamefont {Marcinkiewicz}}, \bibinfo {author}
  {\bibfnamefont {W.}~\bibnamefont {Knap}}, \bibinfo {author} {\bibfnamefont
  {A.}~\bibnamefont {Nateprov}}, \ and\ \bibinfo {author} {\bibfnamefont
  {E.}~\bibnamefont {Arushanov}},\ }\href
  {http://stacks.iop.org/1742-6596/647/i=1/a=012064} {\bibfield  {journal}
  {\bibinfo  {journal} {Journal of Physics: Conference Series}\ }\textbf
  {\bibinfo {volume} {647}},\ \bibinfo {pages} {012064} (\bibinfo {year}
  {2015})}\BibitemShut {NoStop}%
\bibitem [{\citenamefont {Huang}\ \emph {et~al.}(2015)\citenamefont {Huang},
  \citenamefont {Zhao}, \citenamefont {Long}, \citenamefont {Wang},
  \citenamefont {Chen}, \citenamefont {Yang}, \citenamefont {Liang},
  \citenamefont {Xue}, \citenamefont {Weng}, \citenamefont {Fang},
  \citenamefont {Dai},\ and\ \citenamefont {Chen}}]{Huang15b}%
  \BibitemOpen
  \bibfield  {author} {\bibinfo {author} {\bibfnamefont {X.}~\bibnamefont
  {Huang}}, \bibinfo {author} {\bibfnamefont {L.}~\bibnamefont {Zhao}},
  \bibinfo {author} {\bibfnamefont {Y.}~\bibnamefont {Long}}, \bibinfo {author}
  {\bibfnamefont {P.}~\bibnamefont {Wang}}, \bibinfo {author} {\bibfnamefont
  {D.}~\bibnamefont {Chen}}, \bibinfo {author} {\bibfnamefont {Z.}~\bibnamefont
  {Yang}}, \bibinfo {author} {\bibfnamefont {H.}~\bibnamefont {Liang}},
  \bibinfo {author} {\bibfnamefont {M.}~\bibnamefont {Xue}}, \bibinfo {author}
  {\bibfnamefont {H.}~\bibnamefont {Weng}}, \bibinfo {author} {\bibfnamefont
  {Z.}~\bibnamefont {Fang}}, \bibinfo {author} {\bibfnamefont {X.}~\bibnamefont
  {Dai}}, \ and\ \bibinfo {author} {\bibfnamefont {G.}~\bibnamefont {Chen}},\
  }\href {\doibase 10.1103/PhysRevX.5.031023} {\bibfield  {journal} {\bibinfo
  {journal} {Phys. Rev. X}\ }\textbf {\bibinfo {volume} {5}},\ \bibinfo {pages}
  {031023} (\bibinfo {year} {2015})}\BibitemShut {NoStop}%
\bibitem [{\citenamefont {Hu}\ \emph {et~al.}(2016)\citenamefont {Hu},
  \citenamefont {Liu}, \citenamefont {Graf}, \citenamefont {Radmanesh},
  \citenamefont {Adams}, \citenamefont {Chuang}, \citenamefont {Wang},
  \citenamefont {Chiorescu}, \citenamefont {Wei}, \citenamefont {Spinu},\ and\
  \citenamefont {Mao}}]{Hu16}%
  \BibitemOpen
  \bibfield  {author} {\bibinfo {author} {\bibfnamefont {J.}~\bibnamefont
  {Hu}}, \bibinfo {author} {\bibfnamefont {J.~Y.}\ \bibnamefont {Liu}},
  \bibinfo {author} {\bibfnamefont {D.}~\bibnamefont {Graf}}, \bibinfo {author}
  {\bibfnamefont {S.~M.~A.}\ \bibnamefont {Radmanesh}}, \bibinfo {author}
  {\bibfnamefont {D.~J.}\ \bibnamefont {Adams}}, \bibinfo {author}
  {\bibfnamefont {A.}~\bibnamefont {Chuang}}, \bibinfo {author} {\bibfnamefont
  {Y.}~\bibnamefont {Wang}}, \bibinfo {author} {\bibfnamefont {I.}~\bibnamefont
  {Chiorescu}}, \bibinfo {author} {\bibfnamefont {J.}~\bibnamefont {Wei}},
  \bibinfo {author} {\bibfnamefont {L.}~\bibnamefont {Spinu}}, \ and\ \bibinfo
  {author} {\bibfnamefont {Z.~Q.}\ \bibnamefont {Mao}},\ }\href
  {http://dx.doi.org/10.1038/srep18674} {\bibfield  {journal} {\bibinfo
  {journal} {Scientific Reports}\ }\textbf {\bibinfo {volume} {6}},\ \bibinfo
  {pages} {18674} (\bibinfo {year} {2016})}\BibitemShut {NoStop}%
\bibitem [{\citenamefont {Arnold}\ \emph {et~al.}(2016)\citenamefont {Arnold},
  \citenamefont {Shekhar}, \citenamefont {Wu}, \citenamefont {Sun},
  \citenamefont {dos Reis}, \citenamefont {Kumar}, \citenamefont {Naumann},
  \citenamefont {Ajeesh}, \citenamefont {Schmidt}, \citenamefont {Grushin},
  \citenamefont {Bardarson}, \citenamefont {Baenitz}, \citenamefont {Sokolov},
  \citenamefont {Borrmann}, \citenamefont {Nicklas}, \citenamefont {Felser},
  \citenamefont {Hassinger},\ and\ \citenamefont {Yan}}]{Arnold16}%
  \BibitemOpen
  \bibfield  {author} {\bibinfo {author} {\bibfnamefont {F.}~\bibnamefont
  {Arnold}}, \bibinfo {author} {\bibfnamefont {C.}~\bibnamefont {Shekhar}},
  \bibinfo {author} {\bibfnamefont {S.-C.}\ \bibnamefont {Wu}}, \bibinfo
  {author} {\bibfnamefont {Y.}~\bibnamefont {Sun}}, \bibinfo {author}
  {\bibfnamefont {R.~D.}\ \bibnamefont {dos Reis}}, \bibinfo {author}
  {\bibfnamefont {N.}~\bibnamefont {Kumar}}, \bibinfo {author} {\bibfnamefont
  {M.}~\bibnamefont {Naumann}}, \bibinfo {author} {\bibfnamefont {M.~O.}\
  \bibnamefont {Ajeesh}}, \bibinfo {author} {\bibfnamefont {M.}~\bibnamefont
  {Schmidt}}, \bibinfo {author} {\bibfnamefont {A.~G.}\ \bibnamefont
  {Grushin}}, \bibinfo {author} {\bibfnamefont {J.~H.}\ \bibnamefont
  {Bardarson}}, \bibinfo {author} {\bibfnamefont {M.}~\bibnamefont {Baenitz}},
  \bibinfo {author} {\bibfnamefont {D.}~\bibnamefont {Sokolov}}, \bibinfo
  {author} {\bibfnamefont {H.}~\bibnamefont {Borrmann}}, \bibinfo {author}
  {\bibfnamefont {M.}~\bibnamefont {Nicklas}}, \bibinfo {author} {\bibfnamefont
  {C.}~\bibnamefont {Felser}}, \bibinfo {author} {\bibfnamefont
  {E.}~\bibnamefont {Hassinger}}, \ and\ \bibinfo {author} {\bibfnamefont
  {B.}~\bibnamefont {Yan}},\ }\href {http://dx.doi.org/10.1038/ncomms11615}
  {\bibfield  {journal} {\bibinfo  {journal} {Nature Communications}\ }\textbf
  {\bibinfo {volume} {7}},\ \bibinfo {pages} {11615} (\bibinfo {year}
  {2016})}\BibitemShut {NoStop}%
\bibitem [{\citenamefont {Li}\ \emph {et~al.}(2016{\natexlab{a}})\citenamefont
  {Li}, \citenamefont {He}, \citenamefont {Lu}, \citenamefont {Zhang},
  \citenamefont {Liu}, \citenamefont {Ma}, \citenamefont {Fan}, \citenamefont
  {Shen},\ and\ \citenamefont {Wang}}]{Li16}%
  \BibitemOpen
  \bibfield  {author} {\bibinfo {author} {\bibfnamefont {H.}~\bibnamefont
  {Li}}, \bibinfo {author} {\bibfnamefont {H.}~\bibnamefont {He}}, \bibinfo
  {author} {\bibfnamefont {H.-Z.}\ \bibnamefont {Lu}}, \bibinfo {author}
  {\bibfnamefont {H.}~\bibnamefont {Zhang}}, \bibinfo {author} {\bibfnamefont
  {H.}~\bibnamefont {Liu}}, \bibinfo {author} {\bibfnamefont {R.}~\bibnamefont
  {Ma}}, \bibinfo {author} {\bibfnamefont {Z.}~\bibnamefont {Fan}}, \bibinfo
  {author} {\bibfnamefont {S.-Q.}\ \bibnamefont {Shen}}, \ and\ \bibinfo
  {author} {\bibfnamefont {J.}~\bibnamefont {Wang}},\ }\href
  {http://dx.doi.org/10.1038/ncomms10301} {\bibfield  {journal} {\bibinfo
  {journal} {Nature Communications}\ }\textbf {\bibinfo {volume} {7}},\
  \bibinfo {pages} {10301} (\bibinfo {year} {2016}{\natexlab{a}})}\BibitemShut
  {NoStop}%
\bibitem [{\citenamefont {Li}\ \emph {et~al.}(2017{\natexlab{b}})\citenamefont
  {Li}, \citenamefont {Wang}, \citenamefont {Li}, \citenamefont {Yang},
  \citenamefont {Shen}, \citenamefont {Sheng}, \citenamefont {Li},
  \citenamefont {Lu}, \citenamefont {Zheng},\ and\ \citenamefont {Xu}}]{Li17}%
  \BibitemOpen
  \bibfield  {author} {\bibinfo {author} {\bibfnamefont {Y.}~\bibnamefont
  {Li}}, \bibinfo {author} {\bibfnamefont {Z.}~\bibnamefont {Wang}}, \bibinfo
  {author} {\bibfnamefont {P.}~\bibnamefont {Li}}, \bibinfo {author}
  {\bibfnamefont {X.}~\bibnamefont {Yang}}, \bibinfo {author} {\bibfnamefont
  {Z.}~\bibnamefont {Shen}}, \bibinfo {author} {\bibfnamefont {F.}~\bibnamefont
  {Sheng}}, \bibinfo {author} {\bibfnamefont {X.}~\bibnamefont {Li}}, \bibinfo
  {author} {\bibfnamefont {Y.}~\bibnamefont {Lu}}, \bibinfo {author}
  {\bibfnamefont {Y.}~\bibnamefont {Zheng}}, \ and\ \bibinfo {author}
  {\bibfnamefont {Z.-A.}\ \bibnamefont {Xu}},\ }\href {\doibase
  10.1007/s11467-016-0636-8} {\bibfield  {journal} {\bibinfo  {journal}
  {Frontiers of Physics}\ }\textbf {\bibinfo {volume} {12}},\ \bibinfo {pages}
  {127205} (\bibinfo {year} {2017}{\natexlab{b}})}\BibitemShut {NoStop}%
\bibitem [{\citenamefont {Liang}\ \emph {et~al.}(2017)\citenamefont {Liang},
  \citenamefont {Lin}, \citenamefont {Gibson}, \citenamefont {Gao},
  \citenamefont {Hirschberger}, \citenamefont {Liu}, \citenamefont {Cava},\
  and\ \citenamefont {Ong}}]{Liang17}%
  \BibitemOpen
  \bibfield  {author} {\bibinfo {author} {\bibfnamefont {T.}~\bibnamefont
  {Liang}}, \bibinfo {author} {\bibfnamefont {J.}~\bibnamefont {Lin}}, \bibinfo
  {author} {\bibfnamefont {Q.}~\bibnamefont {Gibson}}, \bibinfo {author}
  {\bibfnamefont {T.}~\bibnamefont {Gao}}, \bibinfo {author} {\bibfnamefont
  {M.}~\bibnamefont {Hirschberger}}, \bibinfo {author} {\bibfnamefont
  {M.}~\bibnamefont {Liu}}, \bibinfo {author} {\bibfnamefont {R.~J.}\
  \bibnamefont {Cava}}, \ and\ \bibinfo {author} {\bibfnamefont {N.~P.}\
  \bibnamefont {Ong}},\ }\href {\doibase 10.1103/PhysRevLett.118.136601}
  {\bibfield  {journal} {\bibinfo  {journal} {Phys. Rev. Lett.}\ }\textbf
  {\bibinfo {volume} {118}},\ \bibinfo {pages} {136601} (\bibinfo {year}
  {2017})}\BibitemShut {NoStop}%
\bibitem [{\citenamefont {Sudesh}\ \emph {et~al.}(2017)\citenamefont {Sudesh},
  \citenamefont {Kumar}, \citenamefont {Neha}, \citenamefont {Das},\ and\
  \citenamefont {Patnaik}}]{Sudesh17}%
  \BibitemOpen
  \bibfield  {author} {\bibinfo {author} {\bibnamefont {Sudesh}}, \bibinfo
  {author} {\bibfnamefont {P.}~\bibnamefont {Kumar}}, \bibinfo {author}
  {\bibfnamefont {P.}~\bibnamefont {Neha}}, \bibinfo {author} {\bibfnamefont
  {T.}~\bibnamefont {Das}}, \ and\ \bibinfo {author} {\bibfnamefont
  {S.}~\bibnamefont {Patnaik}},\ }\href {http://dx.doi.org/10.1038/srep46062}
  {\bibfield  {journal} {\bibinfo  {journal} {Scientific Reports}\ }\textbf
  {\bibinfo {volume} {7}},\ \bibinfo {pages} {46062} (\bibinfo {year}
  {2017})}\BibitemShut {NoStop}%
\bibitem [{\citenamefont {Niemann}\ \emph {et~al.}(2017)\citenamefont
  {Niemann}, \citenamefont {Gooth}, \citenamefont {Wu}, \citenamefont
  {B{\"a}{\ss}ler}, \citenamefont {Sergelius}, \citenamefont {H{\"u}hne},
  \citenamefont {Rellinghaus}, \citenamefont {Shekhar}, \citenamefont
  {S{\"u}{\ss}}, \citenamefont {Schmidt}, \citenamefont {Felser}, \citenamefont
  {Yan},\ and\ \citenamefont {Nielsch}}]{Niemann17}%
  \BibitemOpen
  \bibfield  {author} {\bibinfo {author} {\bibfnamefont {A.~C.}\ \bibnamefont
  {Niemann}}, \bibinfo {author} {\bibfnamefont {J.}~\bibnamefont {Gooth}},
  \bibinfo {author} {\bibfnamefont {S.-C.}\ \bibnamefont {Wu}}, \bibinfo
  {author} {\bibfnamefont {S.}~\bibnamefont {B{\"a}{\ss}ler}}, \bibinfo
  {author} {\bibfnamefont {P.}~\bibnamefont {Sergelius}}, \bibinfo {author}
  {\bibfnamefont {R.}~\bibnamefont {H{\"u}hne}}, \bibinfo {author}
  {\bibfnamefont {B.}~\bibnamefont {Rellinghaus}}, \bibinfo {author}
  {\bibfnamefont {C.}~\bibnamefont {Shekhar}}, \bibinfo {author} {\bibfnamefont
  {V.}~\bibnamefont {S{\"u}{\ss}}}, \bibinfo {author} {\bibfnamefont
  {M.}~\bibnamefont {Schmidt}}, \bibinfo {author} {\bibfnamefont
  {C.}~\bibnamefont {Felser}}, \bibinfo {author} {\bibfnamefont
  {B.}~\bibnamefont {Yan}}, \ and\ \bibinfo {author} {\bibfnamefont
  {K.}~\bibnamefont {Nielsch}},\ }\href {http://dx.doi.org/10.1038/srep43394}
  {\bibfield  {journal} {\bibinfo  {journal} {Scientific Reports}\ }\textbf
  {\bibinfo {volume} {7}},\ \bibinfo {pages} {43394} (\bibinfo {year}
  {2017})}\BibitemShut {NoStop}%
\bibitem [{\citenamefont {Liang}\ \emph {et~al.}(2018)\citenamefont {Liang},
  \citenamefont {Lin}, \citenamefont {Gibson}, \citenamefont {Kushwaha},
  \citenamefont {Liu}, \citenamefont {Wang}, \citenamefont {Xiong},
  \citenamefont {Sobota}, \citenamefont {Hashimoto}, \citenamefont {Kirchmann},
  \citenamefont {Shen}, \citenamefont {Cava},\ and\ \citenamefont
  {Ong}}]{Liang18}%
  \BibitemOpen
  \bibfield  {author} {\bibinfo {author} {\bibfnamefont {T.}~\bibnamefont
  {Liang}}, \bibinfo {author} {\bibfnamefont {J.}~\bibnamefont {Lin}}, \bibinfo
  {author} {\bibfnamefont {Q.}~\bibnamefont {Gibson}}, \bibinfo {author}
  {\bibfnamefont {S.}~\bibnamefont {Kushwaha}}, \bibinfo {author}
  {\bibfnamefont {M.}~\bibnamefont {Liu}}, \bibinfo {author} {\bibfnamefont
  {W.}~\bibnamefont {Wang}}, \bibinfo {author} {\bibfnamefont {H.}~\bibnamefont
  {Xiong}}, \bibinfo {author} {\bibfnamefont {J.~A.}\ \bibnamefont {Sobota}},
  \bibinfo {author} {\bibfnamefont {M.}~\bibnamefont {Hashimoto}}, \bibinfo
  {author} {\bibfnamefont {P.~S.}\ \bibnamefont {Kirchmann}}, \bibinfo {author}
  {\bibfnamefont {Z.-X.}\ \bibnamefont {Shen}}, \bibinfo {author}
  {\bibfnamefont {R.~J.}\ \bibnamefont {Cava}}, \ and\ \bibinfo {author}
  {\bibfnamefont {N.~P.}\ \bibnamefont {Ong}},\ }\href {\doibase
  10.1038/s41567-018-0078-z} {\bibfield  {journal} {\bibinfo  {journal} {Nature
  Physics}\ }\textbf {\bibinfo {volume} {14}},\ \bibinfo {pages} {451}
  (\bibinfo {year} {2018})}\BibitemShut {NoStop}%
\bibitem [{\citenamefont {Kumar}\ \emph {et~al.}(2018)\citenamefont {Kumar},
  \citenamefont {Guin}, \citenamefont {Felser},\ and\ \citenamefont
  {Shekhar}}]{Nitesh18}%
  \BibitemOpen
  \bibfield  {author} {\bibinfo {author} {\bibfnamefont {N.}~\bibnamefont
  {Kumar}}, \bibinfo {author} {\bibfnamefont {S.~N.}\ \bibnamefont {Guin}},
  \bibinfo {author} {\bibfnamefont {C.}~\bibnamefont {Felser}}, \ and\ \bibinfo
  {author} {\bibfnamefont {C.}~\bibnamefont {Shekhar}},\ }\href {\doibase
  10.1103/PhysRevB.98.041103} {\bibfield  {journal} {\bibinfo  {journal} {Phys.
  Rev. B}\ }\textbf {\bibinfo {volume} {98}},\ \bibinfo {pages} {041103}
  (\bibinfo {year} {2018})}\BibitemShut {NoStop}%
\bibitem [{\citenamefont {Noky}\ \emph {et~al.}(2018)\citenamefont {Noky},
  \citenamefont {Gayles}, \citenamefont {Felser},\ and\ \citenamefont
  {Sun}}]{Noky18}%
  \BibitemOpen
  \bibfield  {author} {\bibinfo {author} {\bibfnamefont {J.}~\bibnamefont
  {Noky}}, \bibinfo {author} {\bibfnamefont {J.}~\bibnamefont {Gayles}},
  \bibinfo {author} {\bibfnamefont {C.}~\bibnamefont {Felser}}, \ and\ \bibinfo
  {author} {\bibfnamefont {Y.}~\bibnamefont {Sun}},\ }\href {\doibase
  10.1103/PhysRevB.97.220405} {\bibfield  {journal} {\bibinfo  {journal} {Phys.
  Rev. B}\ }\textbf {\bibinfo {volume} {97}},\ \bibinfo {pages} {220405}
  (\bibinfo {year} {2018})}\BibitemShut {NoStop}%
\bibitem [{\citenamefont {Yang}\ \emph {et~al.}(2019)\citenamefont {Yang},
  \citenamefont {Zhen}, \citenamefont {Liang}, \citenamefont {Wang},
  \citenamefont {Yan}, \citenamefont {Weng}, \citenamefont {Wang},
  \citenamefont {Tong}, \citenamefont {Pi}, \citenamefont {Zhu},\ and\
  \citenamefont {Zhang}}]{Yang19}%
  \BibitemOpen
  \bibfield  {author} {\bibinfo {author} {\bibfnamefont {J.}~\bibnamefont
  {Yang}}, \bibinfo {author} {\bibfnamefont {W.~L.}\ \bibnamefont {Zhen}},
  \bibinfo {author} {\bibfnamefont {D.~D.}\ \bibnamefont {Liang}}, \bibinfo
  {author} {\bibfnamefont {Y.~J.}\ \bibnamefont {Wang}}, \bibinfo {author}
  {\bibfnamefont {X.}~\bibnamefont {Yan}}, \bibinfo {author} {\bibfnamefont
  {S.~R.}\ \bibnamefont {Weng}}, \bibinfo {author} {\bibfnamefont {J.~R.}\
  \bibnamefont {Wang}}, \bibinfo {author} {\bibfnamefont {W.}~\bibnamefont
  {Tong}}, \bibinfo {author} {\bibfnamefont {L.}~\bibnamefont {Pi}}, \bibinfo
  {author} {\bibfnamefont {W.~K.}\ \bibnamefont {Zhu}}, \ and\ \bibinfo
  {author} {\bibfnamefont {C.~J.}\ \bibnamefont {Zhang}},\ }\href {\doibase
  10.1103/PhysRevMaterials.3.014201} {\bibfield  {journal} {\bibinfo  {journal}
  {Phys. Rev. Materials}\ }\textbf {\bibinfo {volume} {3}},\ \bibinfo {pages}
  {014201} (\bibinfo {year} {2019})}\BibitemShut {NoStop}%
\bibitem [{\citenamefont {Haldane}(2004)}]{Haldane04}%
  \BibitemOpen
  \bibfield  {author} {\bibinfo {author} {\bibfnamefont {F.~D.~M.}\
  \bibnamefont {Haldane}},\ }\href {\doibase 10.1103/PhysRevLett.93.206602}
  {\bibfield  {journal} {\bibinfo  {journal} {Phys. Rev. Lett.}\ }\textbf
  {\bibinfo {volume} {93}},\ \bibinfo {pages} {206602} (\bibinfo {year}
  {2004})}\BibitemShut {NoStop}%
\bibitem [{\citenamefont {Burkov}(2014)}]{Burkov14}%
  \BibitemOpen
  \bibfield  {author} {\bibinfo {author} {\bibfnamefont {A.~A.}\ \bibnamefont
  {Burkov}},\ }\href {\doibase 10.1103/PhysRevLett.113.187202} {\bibfield
  {journal} {\bibinfo  {journal} {Phys. Rev. Lett.}\ }\textbf {\bibinfo
  {volume} {113}},\ \bibinfo {pages} {187202} (\bibinfo {year}
  {2014})}\BibitemShut {NoStop}%
\bibitem [{\citenamefont {Li}\ \emph {et~al.}(2016{\natexlab{b}})\citenamefont
  {Li}, \citenamefont {Kharzeev}, \citenamefont {Zhang}, \citenamefont {Huang},
  \citenamefont {Pletikosic}, \citenamefont {Fedorov}, \citenamefont {Zhong},
  \citenamefont {Schneeloch}, \citenamefont {Gu},\ and\ \citenamefont
  {Valla}}]{Qiang_Li16}%
  \BibitemOpen
  \bibfield  {author} {\bibinfo {author} {\bibfnamefont {Q.}~\bibnamefont
  {Li}}, \bibinfo {author} {\bibfnamefont {D.~E.}\ \bibnamefont {Kharzeev}},
  \bibinfo {author} {\bibfnamefont {C.}~\bibnamefont {Zhang}}, \bibinfo
  {author} {\bibfnamefont {Y.}~\bibnamefont {Huang}}, \bibinfo {author}
  {\bibfnamefont {I.}~\bibnamefont {Pletikosic}}, \bibinfo {author}
  {\bibfnamefont {A.~.~V.}\ \bibnamefont {Fedorov}}, \bibinfo {author}
  {\bibfnamefont {R.~.~D.}\ \bibnamefont {Zhong}}, \bibinfo {author}
  {\bibfnamefont {J.~.~A.}\ \bibnamefont {Schneeloch}}, \bibinfo {author}
  {\bibfnamefont {G.~.~D.}\ \bibnamefont {Gu}}, \ and\ \bibinfo {author}
  {\bibfnamefont {T.}~\bibnamefont {Valla}},\ }\href
  {http://dx.doi.org/10.1038/nphys3648} {\bibfield  {journal} {\bibinfo
  {journal} {Nature Physics}\ }\textbf {\bibinfo {volume} {12}},\ \bibinfo
  {pages} {550 EP } (\bibinfo {year} {2016}{\natexlab{b}})}\BibitemShut
  {NoStop}%
\bibitem [{\citenamefont {Nandy}\ \emph {et~al.}(2017)\citenamefont {Nandy},
  \citenamefont {Sharma}, \citenamefont {Taraphder},\ and\ \citenamefont
  {Tewari}}]{Nandy17}%
  \BibitemOpen
  \bibfield  {author} {\bibinfo {author} {\bibfnamefont {S.}~\bibnamefont
  {Nandy}}, \bibinfo {author} {\bibfnamefont {G.}~\bibnamefont {Sharma}},
  \bibinfo {author} {\bibfnamefont {A.}~\bibnamefont {Taraphder}}, \ and\
  \bibinfo {author} {\bibfnamefont {S.}~\bibnamefont {Tewari}},\ }\href
  {\doibase 10.1103/PhysRevLett.119.176804} {\bibfield  {journal} {\bibinfo
  {journal} {Phys. Rev. Lett.}\ }\textbf {\bibinfo {volume} {119}},\ \bibinfo
  {pages} {176804} (\bibinfo {year} {2017})}\BibitemShut {NoStop}%
\bibitem [{\citenamefont {Burkov}(2017)}]{Burkov17}%
  \BibitemOpen
  \bibfield  {author} {\bibinfo {author} {\bibfnamefont {A.~A.}\ \bibnamefont
  {Burkov}},\ }\href {\doibase 10.1103/PhysRevB.96.041110} {\bibfield
  {journal} {\bibinfo  {journal} {Phys. Rev. B}\ }\textbf {\bibinfo {volume}
  {96}},\ \bibinfo {pages} {041110} (\bibinfo {year} {2017})}\BibitemShut
  {NoStop}%
\bibitem [{\citenamefont {Dantas}\ \emph {et~al.}(2018)\citenamefont {Dantas},
  \citenamefont {Pe{\~{n}}a-Benitez}, \citenamefont {Roy},\ and\ \citenamefont
  {Sur{\'o}wka}}]{Dantas18}%
  \BibitemOpen
  \bibfield  {author} {\bibinfo {author} {\bibfnamefont {R.~M.~A.}\
  \bibnamefont {Dantas}}, \bibinfo {author} {\bibfnamefont {F.}~\bibnamefont
  {Pe{\~{n}}a-Benitez}}, \bibinfo {author} {\bibfnamefont {B.}~\bibnamefont
  {Roy}}, \ and\ \bibinfo {author} {\bibfnamefont {P.}~\bibnamefont
  {Sur{\'o}wka}},\ }\href {\doibase 10.1007/JHEP12(2018)069} {\bibfield
  {journal} {\bibinfo  {journal} {Journal of High Energy Physics}\ }\textbf
  {\bibinfo {volume} {2018}},\ \bibinfo {pages} {69} (\bibinfo {year}
  {2018})}\BibitemShut {NoStop}%
\bibitem [{\citenamefont {{Nag}}\ and\ \citenamefont {{Nandy}}(2018)}]{Nag18}%
  \BibitemOpen
  \bibfield  {author} {\bibinfo {author} {\bibfnamefont {T.}~\bibnamefont
  {{Nag}}}\ and\ \bibinfo {author} {\bibfnamefont {S.}~\bibnamefont
  {{Nandy}}},\ }\href@noop {} {\bibfield  {journal} {\bibinfo  {journal} {arXiv
  e-prints}\ ,\ \bibinfo {eid} {arXiv:1812.08322}} (\bibinfo {year} {2018})},\
  \Eprint {http://arxiv.org/abs/1812.08322} {arXiv:1812.08322} \BibitemShut
  {NoStop}%
\bibitem [{\citenamefont {Zyuzin}(2017)}]{Zyuzin_V_A17}%
  \BibitemOpen
  \bibfield  {author} {\bibinfo {author} {\bibfnamefont {V.~A.}\ \bibnamefont
  {Zyuzin}},\ }\href {\doibase 10.1103/PhysRevB.95.245128} {\bibfield
  {journal} {\bibinfo  {journal} {Phys. Rev. B}\ }\textbf {\bibinfo {volume}
  {95}},\ \bibinfo {pages} {245128} (\bibinfo {year} {2017})}\BibitemShut
  {NoStop}%
\bibitem [{\citenamefont {Sharma}\ \emph {et~al.}(2017)\citenamefont {Sharma},
  \citenamefont {Goswami},\ and\ \citenamefont {Tewari}}]{G_Sharma17a}%
  \BibitemOpen
  \bibfield  {author} {\bibinfo {author} {\bibfnamefont {G.}~\bibnamefont
  {Sharma}}, \bibinfo {author} {\bibfnamefont {P.}~\bibnamefont {Goswami}}, \
  and\ \bibinfo {author} {\bibfnamefont {S.}~\bibnamefont {Tewari}},\ }\href
  {\doibase 10.1103/PhysRevB.96.045112} {\bibfield  {journal} {\bibinfo
  {journal} {Phys. Rev. B}\ }\textbf {\bibinfo {volume} {96}},\ \bibinfo
  {pages} {045112} (\bibinfo {year} {2017})}\BibitemShut {NoStop}%
\bibitem [{\citenamefont {Das}\ and\ \citenamefont {Agarwal}(2019)}]{Kamal19}%
  \BibitemOpen
  \bibfield  {author} {\bibinfo {author} {\bibfnamefont {K.}~\bibnamefont
  {Das}}\ and\ \bibinfo {author} {\bibfnamefont {A.}~\bibnamefont {Agarwal}},\
  }\href {\doibase 10.1103/PhysRevB.99.085405} {\bibfield  {journal} {\bibinfo
  {journal} {Phys. Rev. B}\ }\textbf {\bibinfo {volume} {99}},\ \bibinfo
  {pages} {085405} (\bibinfo {year} {2019})}\BibitemShut {NoStop}%
\bibitem [{\citenamefont {{Dong}}\ \emph {et~al.}(2018)\citenamefont {{Dong}},
  \citenamefont {{Xiao}},\ and\ \citenamefont {{Niu}}}]{Dong2018}%
  \BibitemOpen
  \bibfield  {author} {\bibinfo {author} {\bibfnamefont {L.}~\bibnamefont
  {{Dong}}}, \bibinfo {author} {\bibfnamefont {C.}~\bibnamefont {{Xiao}}}, \
  and\ \bibinfo {author} {\bibfnamefont {Q.}~\bibnamefont {{Niu}}},\
  }\href@noop {} {\bibfield  {journal} {\bibinfo  {journal} {arXiv e-prints}\
  ,\ \bibinfo {eid} {arXiv:1812.11721}} (\bibinfo {year} {2018})},\ \Eprint
  {http://arxiv.org/abs/1812.11721} {arXiv:1812.11721} \BibitemShut {NoStop}%
\bibitem [{\citenamefont {Bui}\ and\ \citenamefont {Rivadulla}(2014)}]{Bui14}%
  \BibitemOpen
  \bibfield  {author} {\bibinfo {author} {\bibfnamefont {C.~T.}\ \bibnamefont
  {Bui}}\ and\ \bibinfo {author} {\bibfnamefont {F.}~\bibnamefont
  {Rivadulla}},\ }\href {\doibase 10.1103/PhysRevB.90.100403} {\bibfield
  {journal} {\bibinfo  {journal} {Phys. Rev. B}\ }\textbf {\bibinfo {volume}
  {90}},\ \bibinfo {pages} {100403} (\bibinfo {year} {2014})}\BibitemShut
  {NoStop}%
\bibitem [{\citenamefont {Bui}\ \emph {et~al.}(2018)\citenamefont {Bui},
  \citenamefont {Garcia}, \citenamefont {Tu}, \citenamefont {Tanaka},\ and\
  \citenamefont {Hai}}]{Bui18}%
  \BibitemOpen
  \bibfield  {author} {\bibinfo {author} {\bibfnamefont {C.~T.}\ \bibnamefont
  {Bui}}, \bibinfo {author} {\bibfnamefont {C.~A.~C.}\ \bibnamefont {Garcia}},
  \bibinfo {author} {\bibfnamefont {N.~T.}\ \bibnamefont {Tu}}, \bibinfo
  {author} {\bibfnamefont {M.}~\bibnamefont {Tanaka}}, \ and\ \bibinfo {author}
  {\bibfnamefont {P.~N.}\ \bibnamefont {Hai}},\ }\href {\doibase
  10.1063/1.5026452} {\bibfield  {journal} {\bibinfo  {journal} {Journal of
  Applied Physics}\ }\textbf {\bibinfo {volume} {123}},\ \bibinfo {pages}
  {175102} (\bibinfo {year} {2018})}\BibitemShut {NoStop}%
\bibitem [{\citenamefont {Wesenberg}\ \emph {et~al.}(2018)\citenamefont
  {Wesenberg}, \citenamefont {Hojem}, \citenamefont {Bennet},\ and\
  \citenamefont {Zink}}]{Wesenberg18}%
  \BibitemOpen
  \bibfield  {author} {\bibinfo {author} {\bibfnamefont {D.}~\bibnamefont
  {Wesenberg}}, \bibinfo {author} {\bibfnamefont {A.}~\bibnamefont {Hojem}},
  \bibinfo {author} {\bibfnamefont {R.~K.}\ \bibnamefont {Bennet}}, \ and\
  \bibinfo {author} {\bibfnamefont {B.~L.}\ \bibnamefont {Zink}},\ }\href
  {http://stacks.iop.org/0022-3727/51/i=24/a=244005} {\bibfield  {journal}
  {\bibinfo  {journal} {Journal of Physics D: Applied Physics}\ }\textbf
  {\bibinfo {volume} {51}},\ \bibinfo {pages} {244005} (\bibinfo {year}
  {2018})}\BibitemShut {NoStop}%
\bibitem [{\citenamefont {Ashcroft}\ and\ \citenamefont
  {Mermin}(1976)}]{Ashcroft76}%
  \BibitemOpen
  \bibfield  {author} {\bibinfo {author} {\bibfnamefont {N.}~\bibnamefont
  {Ashcroft}}\ and\ \bibinfo {author} {\bibfnamefont {N.}~\bibnamefont
  {Mermin}},\ }\href {https://books.google.co.in/books?id=1C9HAQAAIAAJ} {\emph
  {\bibinfo {title} {Solid State Physics}}},\ HRW international editions\
  (\bibinfo  {publisher} {Holt, Rinehart and Winston},\ \bibinfo {year}
  {1976})\BibitemShut {NoStop}%
\bibitem [{\citenamefont {Xiao}\ \emph {et~al.}(2006)\citenamefont {Xiao},
  \citenamefont {Yao}, \citenamefont {Fang},\ and\ \citenamefont
  {Niu}}]{Xiao_Niu06}%
  \BibitemOpen
  \bibfield  {author} {\bibinfo {author} {\bibfnamefont {D.}~\bibnamefont
  {Xiao}}, \bibinfo {author} {\bibfnamefont {Y.}~\bibnamefont {Yao}}, \bibinfo
  {author} {\bibfnamefont {Z.}~\bibnamefont {Fang}}, \ and\ \bibinfo {author}
  {\bibfnamefont {Q.}~\bibnamefont {Niu}},\ }\href {\doibase
  10.1103/PhysRevLett.97.026603} {\bibfield  {journal} {\bibinfo  {journal}
  {Phys. Rev. Lett.}\ }\textbf {\bibinfo {volume} {97}},\ \bibinfo {pages}
  {026603} (\bibinfo {year} {2006})}\BibitemShut {NoStop}%
\bibitem [{\citenamefont {Carbotte}(2016)}]{Carbotte16}%
  \BibitemOpen
  \bibfield  {author} {\bibinfo {author} {\bibfnamefont {J.~P.}\ \bibnamefont
  {Carbotte}},\ }\href {\doibase 10.1103/PhysRevB.94.165111} {\bibfield
  {journal} {\bibinfo  {journal} {Phys. Rev. B}\ }\textbf {\bibinfo {volume}
  {94}},\ \bibinfo {pages} {165111} (\bibinfo {year} {2016})}\BibitemShut
  {NoStop}%
\bibitem [{\citenamefont {Hayata}\ \emph {et~al.}(2017)\citenamefont {Hayata},
  \citenamefont {Kikuchi},\ and\ \citenamefont {Tanizaki}}]{Hayata17}%
  \BibitemOpen
  \bibfield  {author} {\bibinfo {author} {\bibfnamefont {T.}~\bibnamefont
  {Hayata}}, \bibinfo {author} {\bibfnamefont {Y.}~\bibnamefont {Kikuchi}}, \
  and\ \bibinfo {author} {\bibfnamefont {Y.}~\bibnamefont {Tanizaki}},\ }\href
  {\doibase 10.1103/PhysRevB.96.085112} {\bibfield  {journal} {\bibinfo
  {journal} {Phys. Rev. B}\ }\textbf {\bibinfo {volume} {96}},\ \bibinfo
  {pages} {085112} (\bibinfo {year} {2017})}\BibitemShut {NoStop}%
\bibitem [{\citenamefont {Jia}\ \emph {et~al.}(2016)\citenamefont {Jia},
  \citenamefont {Li}, \citenamefont {Li}, \citenamefont {Shi}, \citenamefont
  {Liao}, \citenamefont {Yu},\ and\ \citenamefont {Wu}}]{Jia16}%
  \BibitemOpen
  \bibfield  {author} {\bibinfo {author} {\bibfnamefont {Z.}~\bibnamefont
  {Jia}}, \bibinfo {author} {\bibfnamefont {C.}~\bibnamefont {Li}}, \bibinfo
  {author} {\bibfnamefont {X.}~\bibnamefont {Li}}, \bibinfo {author}
  {\bibfnamefont {J.}~\bibnamefont {Shi}}, \bibinfo {author} {\bibfnamefont
  {Z.}~\bibnamefont {Liao}}, \bibinfo {author} {\bibfnamefont {D.}~\bibnamefont
  {Yu}}, \ and\ \bibinfo {author} {\bibfnamefont {X.}~\bibnamefont {Wu}},\
  }\href {http://dx.doi.org/10.1038/ncomms13013} {\bibfield  {journal}
  {\bibinfo  {journal} {Nature Communications}\ }\textbf {\bibinfo {volume}
  {7}},\ \bibinfo {pages} {13013} (\bibinfo {year} {2016})}\BibitemShut
  {NoStop}%
\bibitem [{\citenamefont {Stockert}\ \emph {et~al.}(2017)\citenamefont
  {Stockert}, \citenamefont {dos Reis}, \citenamefont {Ajeesh}, \citenamefont
  {Watzman}, \citenamefont {Schmidt}, \citenamefont {Shekhar}, \citenamefont
  {Heremans}, \citenamefont {Felser}, \citenamefont {Baenitz},\ and\
  \citenamefont {Nicklas}}]{Stockert17}%
  \BibitemOpen
  \bibfield  {author} {\bibinfo {author} {\bibfnamefont {U.}~\bibnamefont
  {Stockert}}, \bibinfo {author} {\bibfnamefont {R.~D.}\ \bibnamefont {dos
  Reis}}, \bibinfo {author} {\bibfnamefont {M.~O.}\ \bibnamefont {Ajeesh}},
  \bibinfo {author} {\bibfnamefont {S.~J.}\ \bibnamefont {Watzman}}, \bibinfo
  {author} {\bibfnamefont {M.}~\bibnamefont {Schmidt}}, \bibinfo {author}
  {\bibfnamefont {C.}~\bibnamefont {Shekhar}}, \bibinfo {author} {\bibfnamefont
  {J.~P.}\ \bibnamefont {Heremans}}, \bibinfo {author} {\bibfnamefont
  {C.}~\bibnamefont {Felser}}, \bibinfo {author} {\bibfnamefont
  {M.}~\bibnamefont {Baenitz}}, \ and\ \bibinfo {author} {\bibfnamefont
  {M.}~\bibnamefont {Nicklas}},\ }\href
  {http://stacks.iop.org/0953-8984/29/i=32/a=325701} {\bibfield  {journal}
  {\bibinfo  {journal} {Journal of Physics: Condensed Matter}\ }\textbf
  {\bibinfo {volume} {29}},\ \bibinfo {pages} {325701} (\bibinfo {year}
  {2017})}\BibitemShut {NoStop}%
\bibitem [{\citenamefont {Thakur}\ \emph {et~al.}(2018)\citenamefont {Thakur},
  \citenamefont {Sadhukhan},\ and\ \citenamefont
  {Agarwal}}]{PhysRevB.97.035403}%
  \BibitemOpen
  \bibfield  {author} {\bibinfo {author} {\bibfnamefont {A.}~\bibnamefont
  {Thakur}}, \bibinfo {author} {\bibfnamefont {K.}~\bibnamefont {Sadhukhan}}, \
  and\ \bibinfo {author} {\bibfnamefont {A.}~\bibnamefont {Agarwal}},\ }\href
  {\doibase 10.1103/PhysRevB.97.035403} {\bibfield  {journal} {\bibinfo
  {journal} {Phys. Rev. B}\ }\textbf {\bibinfo {volume} {97}},\ \bibinfo
  {pages} {035403} (\bibinfo {year} {2018})}\BibitemShut {NoStop}%
\bibitem [{\citenamefont {Zhang}\ \emph {et~al.}(2017)\citenamefont {Zhang},
  \citenamefont {Yuan}, \citenamefont {Jiang}, \citenamefont {Tong},
  \citenamefont {Zhang}, \citenamefont {Xie},\ and\ \citenamefont
  {Jia}}]{Zhang17}%
  \BibitemOpen
  \bibfield  {author} {\bibinfo {author} {\bibfnamefont {C.-L.}\ \bibnamefont
  {Zhang}}, \bibinfo {author} {\bibfnamefont {Z.}~\bibnamefont {Yuan}},
  \bibinfo {author} {\bibfnamefont {Q.-D.}\ \bibnamefont {Jiang}}, \bibinfo
  {author} {\bibfnamefont {B.}~\bibnamefont {Tong}}, \bibinfo {author}
  {\bibfnamefont {C.}~\bibnamefont {Zhang}}, \bibinfo {author} {\bibfnamefont
  {X.~C.}\ \bibnamefont {Xie}}, \ and\ \bibinfo {author} {\bibfnamefont
  {S.}~\bibnamefont {Jia}},\ }\href {\doibase 10.1103/PhysRevB.95.085202}
  {\bibfield  {journal} {\bibinfo  {journal} {Phys. Rev. B}\ }\textbf {\bibinfo
  {volume} {95}},\ \bibinfo {pages} {085202} (\bibinfo {year}
  {2017})}\BibitemShut {NoStop}%
\bibitem [{\citenamefont {Morimoto}\ \emph {et~al.}(2016)\citenamefont
  {Morimoto}, \citenamefont {Zhong}, \citenamefont {Orenstein},\ and\
  \citenamefont {Moore}}]{Morimoto16}%
  \BibitemOpen
  \bibfield  {author} {\bibinfo {author} {\bibfnamefont {T.}~\bibnamefont
  {Morimoto}}, \bibinfo {author} {\bibfnamefont {S.}~\bibnamefont {Zhong}},
  \bibinfo {author} {\bibfnamefont {J.}~\bibnamefont {Orenstein}}, \ and\
  \bibinfo {author} {\bibfnamefont {J.~E.}\ \bibnamefont {Moore}},\ }\href
  {\doibase 10.1103/PhysRevB.94.245121} {\bibfield  {journal} {\bibinfo
  {journal} {Phys. Rev. B}\ }\textbf {\bibinfo {volume} {94}},\ \bibinfo
  {pages} {245121} (\bibinfo {year} {2016})}\BibitemShut {NoStop}%
\bibitem [{\citenamefont {Marder}(2010)}]{Marder10}%
  \BibitemOpen
  \bibfield  {author} {\bibinfo {author} {\bibfnamefont {M.}~\bibnamefont
  {Marder}},\ }\href@noop {} {\emph {\bibinfo {title} {Condensed matter
  physics}}}\ (\bibinfo  {publisher} {Wiley},\ \bibinfo {address} {Hoboken,
  N.J},\ \bibinfo {year} {2010})\BibitemShut {NoStop}%
\bibitem [{\citenamefont {Sinitsyn}(2008)}]{Sinitsyn08}%
  \BibitemOpen
  \bibfield  {author} {\bibinfo {author} {\bibfnamefont {N.~A.}\ \bibnamefont
  {Sinitsyn}},\ }\href {http://stacks.iop.org/0953-8984/20/i=2/a=023201}
  {\bibfield  {journal} {\bibinfo  {journal} {Journal of Physics: Condensed
  Matter}\ }\textbf {\bibinfo {volume} {20}},\ \bibinfo {pages} {023201}
  (\bibinfo {year} {2008})}\BibitemShut {NoStop}%
\bibitem [{\citenamefont {Xiao}\ \emph {et~al.}(2005)\citenamefont {Xiao},
  \citenamefont {Shi},\ and\ \citenamefont {Niu}}]{Xiao_Niu05}%
  \BibitemOpen
  \bibfield  {author} {\bibinfo {author} {\bibfnamefont {D.}~\bibnamefont
  {Xiao}}, \bibinfo {author} {\bibfnamefont {J.}~\bibnamefont {Shi}}, \ and\
  \bibinfo {author} {\bibfnamefont {Q.}~\bibnamefont {Niu}},\ }\href {\doibase
  10.1103/PhysRevLett.95.137204} {\bibfield  {journal} {\bibinfo  {journal}
  {Phys. Rev. Lett.}\ }\textbf {\bibinfo {volume} {95}},\ \bibinfo {pages}
  {137204} (\bibinfo {year} {2005})}\BibitemShut {NoStop}%
\bibitem [{\citenamefont {Duval}\ \emph {et~al.}(2006)\citenamefont {Duval},
  \citenamefont {Horvath}, \citenamefont {Horvathy}, \citenamefont {Martina},\
  and\ \citenamefont {Stichel}}]{Duval06}%
  \BibitemOpen
  \bibfield  {author} {\bibinfo {author} {\bibfnamefont {C.}~\bibnamefont
  {Duval}}, \bibinfo {author} {\bibfnamefont {Z.}~\bibnamefont {Horvath}},
  \bibinfo {author} {\bibfnamefont {P.~A.}\ \bibnamefont {Horvathy}}, \bibinfo
  {author} {\bibfnamefont {L.}~\bibnamefont {Martina}}, \ and\ \bibinfo
  {author} {\bibfnamefont {P.~C.}\ \bibnamefont {Stichel}},\ }\href {\doibase
  10.1142/S0217984906010573} {\bibfield  {journal} {\bibinfo  {journal} {Modern
  Physics Letters B}\ }\textbf {\bibinfo {volume} {20}},\ \bibinfo {pages}
  {373} (\bibinfo {year} {2006})}\BibitemShut {NoStop}%
\bibitem [{\citenamefont {Chang}\ and\ \citenamefont {Niu}(1996)}]{Chang96}%
  \BibitemOpen
  \bibfield  {author} {\bibinfo {author} {\bibfnamefont {M.-C.}\ \bibnamefont
  {Chang}}\ and\ \bibinfo {author} {\bibfnamefont {Q.}~\bibnamefont {Niu}},\
  }\href {\doibase 10.1103/PhysRevB.53.7010} {\bibfield  {journal} {\bibinfo
  {journal} {Phys. Rev. B}\ }\textbf {\bibinfo {volume} {53}},\ \bibinfo
  {pages} {7010} (\bibinfo {year} {1996})}\BibitemShut {NoStop}%
\bibitem [{\citenamefont {Dai}\ \emph {et~al.}(2017)\citenamefont {Dai},
  \citenamefont {Du},\ and\ \citenamefont {Lu}}]{Dai17}%
  \BibitemOpen
  \bibfield  {author} {\bibinfo {author} {\bibfnamefont {X.}~\bibnamefont
  {Dai}}, \bibinfo {author} {\bibfnamefont {Z.~Z.}\ \bibnamefont {Du}}, \ and\
  \bibinfo {author} {\bibfnamefont {H.-Z.}\ \bibnamefont {Lu}},\ }\href
  {\doibase 10.1103/PhysRevLett.119.166601} {\bibfield  {journal} {\bibinfo
  {journal} {Phys. Rev. Lett.}\ }\textbf {\bibinfo {volume} {119}},\ \bibinfo
  {pages} {166601} (\bibinfo {year} {2017})}\BibitemShut {NoStop}%
\bibitem [{\citenamefont {Jacoboni}(2010)}]{Jacoboni10}%
  \BibitemOpen
  \bibfield  {author} {\bibinfo {author} {\bibfnamefont {C.}~\bibnamefont
  {Jacoboni}},\ }\href@noop {} {\emph {\bibinfo {title} {Theory of electron
  transport in semiconductors : a pathway from elementary physics to
  nonequilibrium green functions}}}\ (\bibinfo  {publisher} {Springer},\
  \bibinfo {address} {Berlin},\ \bibinfo {year} {2010})\BibitemShut {NoStop}%
\bibitem [{\citenamefont {Imran}\ and\ \citenamefont
  {Hershfield}(2018)}]{PhysRevB.98.205139}%
  \BibitemOpen
  \bibfield  {author} {\bibinfo {author} {\bibfnamefont {M.}~\bibnamefont
  {Imran}}\ and\ \bibinfo {author} {\bibfnamefont {S.}~\bibnamefont
  {Hershfield}},\ }\href {\doibase 10.1103/PhysRevB.98.205139} {\bibfield
  {journal} {\bibinfo  {journal} {Phys. Rev. B}\ }\textbf {\bibinfo {volume}
  {98}},\ \bibinfo {pages} {205139} (\bibinfo {year} {2018})}\BibitemShut
  {NoStop}%
\bibitem [{\citenamefont {Cooper}\ \emph {et~al.}(1997)\citenamefont {Cooper},
  \citenamefont {Halperin},\ and\ \citenamefont {Ruzin}}]{Cooper97}%
  \BibitemOpen
  \bibfield  {author} {\bibinfo {author} {\bibfnamefont {N.~R.}\ \bibnamefont
  {Cooper}}, \bibinfo {author} {\bibfnamefont {B.~I.}\ \bibnamefont
  {Halperin}}, \ and\ \bibinfo {author} {\bibfnamefont {I.~M.}\ \bibnamefont
  {Ruzin}},\ }\href {\doibase 10.1103/PhysRevB.55.2344} {\bibfield  {journal}
  {\bibinfo  {journal} {Phys. Rev. B}\ }\textbf {\bibinfo {volume} {55}},\
  \bibinfo {pages} {2344} (\bibinfo {year} {1997})}\BibitemShut {NoStop}%
\bibitem [{\citenamefont {Goswami}\ and\ \citenamefont
  {Tewari}(2013)}]{Goswami13}%
  \BibitemOpen
  \bibfield  {author} {\bibinfo {author} {\bibfnamefont {P.}~\bibnamefont
  {Goswami}}\ and\ \bibinfo {author} {\bibfnamefont {S.}~\bibnamefont
  {Tewari}},\ }\href {\doibase 10.1103/PhysRevB.88.245107} {\bibfield
  {journal} {\bibinfo  {journal} {Phys. Rev. B}\ }\textbf {\bibinfo {volume}
  {88}},\ \bibinfo {pages} {245107} (\bibinfo {year} {2013})}\BibitemShut
  {NoStop}%
\bibitem [{\citenamefont {Gorbar}\ \emph {et~al.}(2017)\citenamefont {Gorbar},
  \citenamefont {Miransky}, \citenamefont {Shovkovy},\ and\ \citenamefont
  {Sukhachov}}]{Gorbar17}%
  \BibitemOpen
  \bibfield  {author} {\bibinfo {author} {\bibfnamefont {E.~V.}\ \bibnamefont
  {Gorbar}}, \bibinfo {author} {\bibfnamefont {V.~A.}\ \bibnamefont
  {Miransky}}, \bibinfo {author} {\bibfnamefont {I.~A.}\ \bibnamefont
  {Shovkovy}}, \ and\ \bibinfo {author} {\bibfnamefont {P.~O.}\ \bibnamefont
  {Sukhachov}},\ }\href {\doibase 10.1103/PhysRevB.96.155138} {\bibfield
  {journal} {\bibinfo  {journal} {Phys. Rev. B}\ }\textbf {\bibinfo {volume}
  {96}},\ \bibinfo {pages} {155138} (\bibinfo {year} {2017})}\BibitemShut
  {NoStop}%
\bibitem [{\citenamefont {Pellegrino}\ \emph {et~al.}(2015)\citenamefont
  {Pellegrino}, \citenamefont {Katsnelson},\ and\ \citenamefont
  {Polini}}]{Marco15}%
  \BibitemOpen
  \bibfield  {author} {\bibinfo {author} {\bibfnamefont {F.~M.~D.}\
  \bibnamefont {Pellegrino}}, \bibinfo {author} {\bibfnamefont {M.~I.}\
  \bibnamefont {Katsnelson}}, \ and\ \bibinfo {author} {\bibfnamefont
  {M.}~\bibnamefont {Polini}},\ }\href {\doibase 10.1103/PhysRevB.92.201407}
  {\bibfield  {journal} {\bibinfo  {journal} {Phys. Rev. B}\ }\textbf {\bibinfo
  {volume} {92}},\ \bibinfo {pages} {201407} (\bibinfo {year}
  {2015})}\BibitemShut {NoStop}%
\end{thebibliography}%

\end{document}